\documentclass[12pt,preprint]{aastex}
\usepackage{natbib}
\def\aa#1#2#3#4#5{\bibitem[#1]{#2}#3, {A\&A}, {#4}, #5}
\def\aasup#1#2#3#4#5{\bibitem[#1]{#2}#3, {A\&AS}, {#4}, #5}
\def\aj#1#2#3#4#5{\bibitem[#1]{#2}#3, {AJ}, {#4}, #5}
\def\apj#1#2#3#4#5{\bibitem[#1]{#2}#3, {Ap. J.}, {#4}, #5}

\def\apss#1#2#3#4#5{\bibitem[#1]{#2}#3, {Ap\&SS}, {#4}, #5}
\def\apjsup#1#2#3#4#5{\bibitem[#1]{#2}#3, ApJS, #4, #5}
\def\araa#1#2#3#4#5{\bibitem[#1]{#2}#3, ARA\&A, #4, #5}

\def\mnras#1#2#3#4#5{\bibitem[#1]{#2}#3, {M.N.R.A.S.}, {#4}, #5}
\def\nature#1#2#3#4#5{\bibitem[#1]{#2}#3, {Nature}, {#4}, #5}

\def\pasp#1#2#3#4#5{\bibitem[#1]{#2}#3, {PASP}, {#4}, #5}

\def\spie#1#2#3#4#5{\bibitem[#1]{#2}#3, {Proc. SPIE}, {#4}, #5}

\begin{document}
\title{Resolved Inner Disks around Herbig Ae/Be Stars}

\author{J.A. Eisner\altaffilmark{1}, B.F. Lane\altaffilmark{2}, 
L.A. Hillenbrand\altaffilmark{1}, R.L. Akeson\altaffilmark{3}, 
\& A.I. Sargent\altaffilmark{1}}
\email{jae@astro.caltech.edu}
\altaffiltext{1}{California Institute of Technology,
Department of Astronomy MC 105-24,
Pasadena, CA 91125}
\altaffiltext{2}{Center for Space Research,
MIT Department of Physics, 70 Vassar Street, Cambridge, MA 02139}
\altaffiltext{3}{California Institute of Technology, 
Michelson Science Center MC 100-22,
Pasadena, CA 91125}


\slugcomment{Draft of {\bf \today}}

\begin{abstract}
We have observed 14 Herbig Ae/Be sources
with the long-baseline near-IR Palomar Testbed Interferometer.
All except two sources are resolved at 2.2 $\mu$m, with
angular sizes generally $\la 5$ mas.
We determine the size scales and orientations of the 2.2 $\mu$m
emission using various models: uniform disks, Gaussians, uniform rings, flat
accretion disks with inner holes, and flared disks with puffed-up inner rims.
Although it is difficult to distinguish different radial distributions,
we are able to place firm constraints on the inclinations of most
sources; 7 objects display significantly inclined morphologies.
The inner disk inclinations derived from our near-IR data are 
generally compatible
with the outer disk geometries inferred from millimeter interferometric
observations, implying that HAEBE disks are not significantly warped.
Using the derived inner disk sizes and inclinations, we compute the
spectral energy distributions for two simple physical disk models, 
and compare these with 
observed SEDs compiled from the
literature and new near-IR photometry.  While geometrically flat accretion
disk models are consistent with the data for the earliest spectral types
in our sample (MWC 297, V1685 Cyg, and MWC 1080),
the later-type sources
are explained better through models incorporating puffed-up inner disk 
walls.  The different inner disk geometries may indicate different accretion 
mechanisms for early and late-type Herbig Ae/Be stars.
\end{abstract}

\section{Introduction \label{sec:intro}}
Herbig Ae/Be \citep[HAEBE;][]{HERBIG60}
stars are intermediate-mass (2--10 M$_{\odot}$) young 
stellar objects
that show broad emission lines, rapid variability, and
excess infrared and millimeter-wavelength emission. 
These properties are consistent with 
the presence of hot and cold circumstellar dust and gas.  
While there is still some debate about the morphology of this circumstellar
material, most evidence supports the hypothesis that in many cases
the dust and gas lie in massive ($\sim 0.01$ M$_{\odot}$) 
circumstellar disks \citep{NGM00,HILLENBRAND+92}.

The strongest evidence for circumstellar disks around HAEBE stars comes
from near-IR and millimeter interferometry.  Flattened structures 
around several sources have been resolved on $\la 1$ AU scales in the
near-IR \citep[][hereafter, ELAHS]{EISNER+03} and on $\sim 100$ AU scales 
at millimeter wavelengths \citep{MS97,MS00,PDK03}, and  
detailed kinematic modeling of one source, MWC 480, is consistent with rotation
in a Keplerian disk \citep{MKS97,SDG01}. 
H$\alpha$ spectropolarimetric observations,
which trace dust on scales of tens of stellar radii \citep{VINK+02},
and modeling of forbidden emission lines that arise in winds and outflows
around HAEBE stars \citep{CR97} provide further evidence for disks.
Finally, disk-like distributions of material around HAEBEs are suggested by
modeling of spectral energy distributions (SEDs).
Various models, including geometrically flat accretion disks
\citep[e.g.,][]{HILLENBRAND+92}, flared outer disks \citep[e.g.,][]{CG97},
and puffed-up inner disk rims \citep[Dullemond, Dominik, \& Natta 2001;
hereafter][]{DDN01}, have been used to fit the data \citep[although the SEDs 
for some sources can also be explained by more spherically distributed
dust; e.g.,][]{HARTMANN+93}.

Previous near-IR interferometric observations probed the inner disks around
several HAEBE sources (Akeson et al. 2000; Millan-Gabet, Schloerb, \&
Traub 2001, hereafter MST; ELAHS; Wilkin \& Akeson 2003; see also
aperture masking results, Danchi, Tuthill, \& Monnier 2001; 
Tuthill et al. 2002).
Here, we expand the
sample and obtain superior $u-v$ coverage, enabling measurements of
size scales and orientations of the inner disks around 14 HAEBEs.
We present results for AB Aur, MWC 480,
MWC 758, CQ Tau, T Ori, MWC 120,
HD 141569, HD 158352, VV Ser, MWC 297, V1295 Aql, 
V1685 Cyg (BD+40$^{\circ}$4124), AS 442, and MWC 1080.  Two sources,
HD 141569 and HD 158352, show no evidence of near-IR circumstellar emission
and appear unresolved in our interferometric observations.  The other
12 sources in our sample show resolved circumstellar emission,
consistent with disks.

We model the structure of circumstellar dust within $\sim 0.1$-1 AU of
these HAEBE stars, fitting three simple geometrical models 
(Gaussians, uniform disks, and uniform rings), and two basic
physical models (flat accretion disks with inner holes and
flared passive disks with puffed-up inner walls).  For each model,
we determine approximate size scales, position angles, and inclinations
of the near-IR circumstellar emission.
Where possible,
we compare our 2.2 $\mu$m interferometry results with previously published
1.6 $\mu$m and 2.2 $\mu$m interferometric data from the 
IOTA interferometer \citep{MST01}.  
In addition, we compare our results with SEDs compiled from
the literature and new data, and millimeter interferometric results
(where available), in order to further constrain simple physical models
of HAEBE disks.

\section{Observations and Calibration \label{sec:obs}}
The Palomar Testbed Interferometer (PTI) 
is a fringe-tracking long-baseline near-IR Michelson interferometer 
located on Palomar Mountain near San Diego, CA \citep{COLAVITA+99}.
PTI combines starlight from two 40-cm aperture telescopes using
a Michelson beam combiner, and the resulting 
fringe visibilities provide a measure of the brightness distribution
on the sky (via the van Cittert-Zernike theorem).  PTI measures
normalized squared visibilities, $V^2$, which provide unbiased estimates of
the visibility amplitudes \citep{COLAVITA99};  $V^2$ is unity for point 
sources and smaller for resolved sources.

We observed 14 HAEBE sources
with PTI between May, 2002 and January, 2004.  Properties of the sample
are included in Table \ref{tab:sources}.
We obtained K-band ($\lambda_0=2.2$ $\mu$m, $\Delta \lambda = 0.4$ $\mu$m) 
measurements
on an 85-m North-West (NW)  baseline for all 14 objects,
on an 86-m South-West (SW) baseline for 8,
and on a 110-m North-South (NS) baseline for 6.
The NW baseline is oriented $109^{\circ}$ west of north and has a fringe 
spacing of $\sim 5$ mas, the SW baseline is $211^{\circ}$ west of north
with a $\sim 5$ mas fringe spacing,
and the NS baseline is $160^{\circ}$ west of
north and has a fringe spacing of $\sim 4$ mas.
A summary of the observations is given in Table \ref{tab:obs}.

For the sample, $V^2$ were measured from a synthetic wide-band
channel formed from five spatially-filtered spectral pixels covering
the K-band \citep{COLAVITA99}.  Since PTI tracks the central (``white light'') 
fringe, and the fringe spacings 
for PTI baselines are much smaller than the interferometric field of view 
($\sim 50$ mas; set by baseline lengths and spectral bandpass), the effects
of spectral smearing on the K-band visibilities are 
negligible.  

The visibilities were calibrated by observing sources of 
known angular size.   Our adopted angular size for a calibrator
is the average of three estimates based on 1) published stellar luminosity 
and distance, 2) a blackbody fit to 
published photometric data with the temperature constrained to
that expected for the published spectral type, and 3) an unconstrained 
blackbody fit to the photometric data.
We propagate the errors on individual estimates to determine the
uncertainty of the average size.
Relevant properties of the calibrators used in these observations are
given in Table \ref{tab:cals}.
For a more thorough discussion of the data measurement and calibration
procedures, see ELAHS.

\subsection{PALAO Observations \label{sec:ao}}
In addition to the PTI data, we obtained
adaptive optics images of the sources in our sample.
As we discuss in 
\S \ref{sec:viscorr}, the adaptive optics data 
help distinguish between components of the PTI visibilities, including 
circumstellar emission, the central stars, and nearby companions.

Using the Palomar 200-inch adaptive optics system \citep[PALAO;][]{TROY+00}
with the PHARO camera \citep{HAYWARD+01}
on November 17-18, 2002, January 13-14, 2003, and July 14-15, 2003,
we imaged our sources
in the JHK bands with a 25 mas plate scale and a $25''$ field of view.
We used 1\% or 0.1\% neutral density filters and short integration times
to prevent saturation of the camera.
After bias correction, flat-fielding, and background subtraction,
the reduced images enabled searches for any companions
$\ga 0\rlap{.}''05$ away from the target sources.  Source counts were
obtained from aperture
photometry with sky subtraction, and
photometric calibration was achieved using nearby sources with measured 
2MASS photometry. The determined JHK magnitudes of our target sources
are listed in Table \ref{tab:sources}.  While the single-night uncertainty of
the photometry is typically $<0.05$ mags, night-to-night variations (due to
variable seeing and small instrumental changes) limit
the photometric accuracy to $\sim 0.15$ mags.  Since AS 442 appears to be
variable with a short timescale (probably days),
Table \ref{tab:sources} quotes the average of two measurements.
For all objects, our photometry is consistent with previously 
published measurements to within $\sim 0.3$ magnitudes
\citep[2MASS;][]{HILLENBRAND+92,MBW98,DEWINTER+01,EIROA+01}.

\section{Modeling \label{sec:modeling}}
From the PTI data,
we measured calibrated squared visibilities for all sources listed in
Table \ref{tab:sources}.
For each source, we constrain the angular size and geometry of the emission
by fitting the visibility data with several simple models: Gaussians,
uniform disks, uniform rings, flat accretion disks with inner holes, and
flared passive disks with puffed-up inner walls.
In this section, we describe the models, as well as corrections to
the visibilities required to separate the circumstellar component
from contributions by the central stars or nearby companions.

\subsection{Visibility Corrections \label{sec:viscorr}}
\subsubsection{Nearby Companions \label{sec:companions}}
Nearby companions that lie outside the interferometric field of view, $\sim 50$
mas, but within the field of view of the detector, $\sim 1''$, will
contribute incoherent light to the visibilities, leading to measured
visibilities smaller than the true values:
\begin{equation}
\frac{V^2_{\rm true}}{V^2_{\rm meas}} = \left(\frac{1}
{1 + 10^{-\Delta K / 2.5}}\right)^2,
\label{eq:companion}
\end{equation}
where $\Delta K$ is the difference in K-band magnitudes between
the two stars. 
Our PALAO Adaptive optics images of the sources in our sample 
(\S \ref{sec:ao}) show that 
most of these have no bright companions ($\Delta K < 5$) at distances between
$\sim 50$ mas and $1''$, and thus no companion corrections are required
($V^2_{\rm true}/V^2_{\rm meas} > 0.98$).
For MWC 1080, we measured $\Delta K = 2.70$ for a companion at 
$0\rlap{.}''78$ separation (consistent with previous
measurements by Corporon 1998)
and Equation \ref{eq:companion} yields
a correction factor of 0.85. V1685 Cyg is also known
to have a faint companion \citep[$\Delta K = 5.50$;][]{CORPORON98},
but its effects on the visibilities are negligible.

\subsubsection{Stellar Emission \label{sec:stellar}}
We account for the effect of the central star
on the visibilities by including it in the models:
\begin{equation}
V^2_{\rm tot} = \left(\frac{F_{\ast} V_{\ast} + F_{\rm x} V_{\rm x}}{F_{\ast} +
F_{\rm x}} \right)^2 \approx 
\left(\frac{F_{\ast} + F_{\rm x} V_{\rm x}}{F_{\ast} +
F_{\rm x}} \right)^2,
\label{eq:vtot}
\end{equation}
where $F_{\ast}$ is the stellar flux, $F_{\rm x}$ is 
the excess flux (both
measured at 2.2 $\mu$m), $V_{\ast} \approx 1$ is the
visibility of the (unresolved) central star,
and $V_{\rm x}$ is the visibility due to the
circumstellar component.  It is reasonable to assume that $V_{\ast} \approx 1$,
since for the stellar radii ($\sim 2$--9 R$_{\odot}$) and distances
($\sim 140$--1000 pc) for our sample, 
the angular diameters of the central stars will be $<0.2$ mas.
In the case of the binary model described below
(\S \ref{sec:binary}), we do not perform any such correction, since the
basic model already includes the stellar component.

Since HAEBE objects are often variable at near-IR wavelengths
\citep[e.g.,][]{SKRUTSKIE+96}, we obtained nearly contemporaneous photometric
K-band measurements for our sample, as described in \S \ref{sec:ao}.
As in ELAHS, we calculate $F_{\ast}$ and $F_{\rm x}$
using our 
K-band photometry (Table \ref{tab:sources}) combined with BVRI photometry,
visual extinctions, stellar radii,
and effective temperatures from the literature
\citep{HS99,HILLENBRAND+92,OUDMAIJER+01,DEWINTER+01,EIROA+02,VIEIRA+03,BG70}.
For several sources where literature values were not available, 
we computed the stellar parameters using the published spectral types
and photometry. The spectral type gives the effective
temperature and expected color, from which we determine the extinction.
We determine the stellar luminosity
(using the adopted distance from Table \ref{tab:sources})
from a bolometric correction applied to the optical photometry, and stellar
radius is given by the Stephan-Boltzmann equation.
Assuming that all of the short-wavelength
flux is due to the central star, we fit a blackbody at 
the assumed effective temperature to the de-reddened
BVRI data.  De-reddening uses the extinction law of Steenman \& Th\'{e} (1991).
The K-band stellar flux, $F_{\ast}$,
is derived from the value of this blackbody curve at 
2.2 $\mu$m, and the excess flux, $F_{\rm x}$,
is given by the difference between the
de-reddened observed flux and the stellar flux. The derived fluxes
are given in Table \ref{tab:sources}.

The uncertainties in our K-band photometric measurements ($\sim 0.15$ mags)
and the uncertainties in the optical photometry used to determine the 
stellar contributions to the K-band fluxes lead to some uncertainties 
in the sizes inferred from the interferometric data.  
This uncertainty decreases for larger $F_{\rm x} / F_{\ast}$
ratios, and is typically $\la 0.1$ mas for the sources discussed here.
CQ Tau, VV Ser, V1685 Cyg, and AS 442 are optically variable by $\Delta V
\ga 1$ magnitudes on timescales of days to months \citep[while the
other sources in our sample show little or no optical variability;][]{HS99},
and thus $F_{\ast}$ is somewhat uncertain.  However,
since $F_{\rm x} / F_{\ast} \gg 1$ for these objects, this uncertainty
is relatively unimportant when modeling the visibilities.  

Uncertainties in the calculation of stellar and circumstellar fluxes can
also have a small effect on the measured inclinations.  However, 
since the fluxes for our sample are dominated by the excess component, only
large errors ($>100\%$) in the flux estimation will produce
measurable effects on the determined inclinations.  Since our photometric
errors are generally $\la 10\%$, the effect on measured inclinations 
is negligible.

\subsubsection{Extended Emission \label{sec:extended}}
Equation \ref{eq:vtot} assumes that the point-like central star 
and compact circumstellar emission contribute 
coherently to the visibilities. 
There may also be an incoherent component due to extended emission
(either thermal or scattered) from tenuous dust.  The
excess flux, $F_{\rm x}$, would then be the sum of the compact circumstellar
emission, $F_{\rm comp}$, and the extended emission, $F_{\rm ext}$.
Including the incoherent contribution in the visibilities would
modify Equation \ref{eq:vtot} to:
\begin{equation}
V^2_{\rm tot} = \left(\frac{F_{\ast} + F_{\rm comp} V_{\rm comp}}
{F_{\ast} + F_{\rm comp} + F_{\rm ext}}\right)^2 = 
\left(\frac{F_{\ast} + [F_{\rm x}-F_{\rm ext}]
 V_{\rm x}}{F_{\ast} + F_{\rm x}} \right)^2,
\label{eq:vscatt}
\end{equation}  
causing a reduction in the measured visibilities. 

The existence of dust on large angular scales around some Herbig Ae/Be stars
is illustrated by optical light scattered from dust grains far from the star
\citep[e.g.,][]{GRADY+99}.  
Furthermore, SED models that include large-scale
optically thin dust, in addition to disks, usually fit the data
well for HAEBE sources \citep[][and references therein]{VINKOVIC+03}.
Nevertheless, these observations do not measure the exact contribution
of extended K-band emission, and our adaptive optics observations
(\S \ref{sec:ao}) do not have sufficient sensitivity to detect faint
emission above the halo of the point spread function within $1''$ 
(the PTI field of view) of the central stars.
Thus, we do not include the effect
of incoherent emission when modeling the visibilities; i.e., we use
Equation \ref{eq:vtot} in the analysis below.  

Because our measured visibilities are found to depend on projected 
baseline (see \S \ref{sec:models}), contrary to
the baseline-independent visibilities expected for
incoherent emission,
the incoherent contribution is probably
insignificant compared to the compact
circumstellar component.  Although small incoherent contributions may
lead to estimated size scales slightly larger 
than true values, we show below in \S \ref{sec:pti+iota} that the
effect is likely to be quite small, perhaps as high as
a few percent in the worst case.

\subsection{Compact Circumstellar Emission \label{sec:mod}}
For each source, we compare the observed visibilities to those derived from
a uniform disk model, a Gaussian model, 
a ring model, a geometrically flat
accretion disk model with an inner hole, and a flared passive disk with a
puffed-up inner wall; all models are 2-D.  
The uniform disk, Gaussian, ring, and flat accretion disk models were
discussed in ELAHS, and the basic equations are merely recalled here.
In addition, we develop and implement a flared disk model with a puffed-up
inner wall, which is a more physically plausible model than 
any of those considered in ELAHS.

As in ELAHS, we consider both face-on and inclined models.
If we assume that the inclination 
of the circumstellar material is zero, then the one remaining
free parameter in the
models is the angular size scale, $\theta$.  When we include inclination
effects, we fit for three parameters: size ($\theta$), 
inclination angle ($\phi$), and position angle ($\psi$).  
Inclination is defined such that a face-on disk has $\phi=0$,
and position angle, $\psi$,  is measured east of north.  
Inclination effects are included in the models
via a simple coordinate transformation:
\begin{equation}
x' = x\sin\psi + y\cos\psi;\: y' = \frac{y\sin\psi - x\cos\psi}{\cos\phi},
\end{equation}
\begin{equation}
u' = u \sin \psi + v \cos \psi;\:
v' = \cos \phi (v \sin \psi - u \cos \psi).
\end{equation}
Here, $(x,y)$,$(u,v)$ are the original sky and $u-v$ plane coordinates, 
and $(x',y')$,$(u',v')$ are the transformed coordinates.  

In addition to these five models, we also examine whether the
data are consistent with a wide binary model, which we approximate
as two stationary point sources.  For this model, the free parameters are
the angular separation ($\theta$), the position angle ($\psi$), and
the brightness ratio of the two components ($R$).  

We fit these models to the PTI data for each source
by searching grids of parameters for the minimum $\chi_r^2$ value.
The grid for face-on disk models was generated
by varying $\theta$ from 0.01 to 10 mas
in increments of 0.01 mas.  
For sources with adequate $u-v$ coverage (Figure \ref{fig:uv}), we also fit
inclined disk models, where
in addition to varying $\theta$, we vary
$\phi$ from 0$^{\circ}$ to $90^{\circ}$ and
$\psi$ from 0$^{\circ}$ to $180^{\circ}$, both in increments of $1^{\circ}$.
Since inclined disk models are symmetric under
reflections through the origin, we do not explore
position angles between $180^{\circ}$ and $360^{\circ}$.
For the binary model, we vary $\theta$ from 1 to 100 mas
in increments of 0.01 mas, $\psi$ from 0$^{\circ}$ to $180^{\circ}$ 
in increments of $1^{\circ}$, and $R=F_2/F_1$ from 0.2 to 1 in increments
of 0.001 (for flux ratios $<0.2$ or angular separations $<1$ mas,
the effects of the companions on the visibilities will be negligible,
and we can rule out angular separations $\ga 100$ mas from adaptive optics
imaging).  Although binary models are not symmetric under reflections through 
the origin, PTI does not measure phase and cannot distinguish 
180$^{\circ}$-rotated models; thus, we only consider position angles from  
0$^{\circ}$ to $180^{\circ}$.

For each point in the parameter grid, we generate
a model for the observed $u-v$ coverage, and calculate the reduced
chi squared, $\chi_r^2$, 
to determine the
``best-fit'' model.  Standard 1-$\sigma$ confidence limits for face-on models
are determined by finding the grid points where the
non-reduced $\chi^2$ equals the minimum 
value plus 1.  For inclined disk or binary models, 
the errors on each parameter correspond to an increment of 3.5 of the
the minimum $\chi^2$ surface.
The quoted uncertainties are not scaled by $\chi_r^2$.

\subsubsection{Simple Geometrical Disk Models \label{sec:gauss}}
The normalized visibilities for a Gaussian, uniform disk, and uniform
ring brightness distribution are, 
\begin{equation}
V_{\rm gauss}(r_{\rm uv}) = \exp \left(-\frac{\pi^2 \theta^2 r_{\rm uv}^2}
{4 \ln 2}\right),
\label{eq:gauss}
\end{equation}
\begin{equation}
V_{\rm uniform}(r_{\rm uv}) = 2 \frac{J_1(\pi \theta r_{\rm uv})}{\pi \theta 
r_{\rm uv}}, 
\label{eq:uniform}
\end{equation}
\begin{equation}
V_{\rm ring} = \frac{2}{\pi \theta r_{\rm uv} (2f + f^2)} \left[
(1+f) J_1([1+f] \pi \theta r_{\rm uv}) - 
J_1(\pi \theta r_{\rm uv}) \right].
\label{eq:ring}
\end{equation}
Here, $(x,y)$ are the angular offsets from the central star, 
$\theta$ is the angular size scale (FWHM, diameter, and inner
diameter for the Gaussian, uniform disk, and ring models,
respectively),
and $r_{\rm uv}$ is the ``uv radius'': 
\begin{equation}
r_{\rm uv} = (u^2+v^2)^{1/2}. 
\label{eq:ruv}
\end{equation}
For the ring model, $f = W/R$, where $R$ is the
radius of the inner edge of the ring, and $W$ is the width of the ring.
In order to facilitate comparison of this model to the more physical
puffed up inner disk models (\S \ref{sec:ddn}), 
we use ring widths derived from radiative transfer modeling by DDN (Table 1);
for stars earlier than spectral type B6, we assume $f = 0.27$, 
and for stars later than B6, we assume $f=0.18$.  The models for the observed
squared visibilities are obtained by substituting Equations \ref{eq:gauss},
\ref{eq:uniform}, or \ref{eq:ring} for $V_{\rm x}$ in Equation \ref{eq:vtot}.

\subsubsection{Geometrically Flat Accretion Disk Model \label{sec:acc}}
We derive the brightness distribution and predicted visibilities
for a geometrically thin irradiated
accretion disk by determining the temperature and
spectral energy distributions for a series of annuli extending from some 
inner radius, $R_{\rm in}$, to some outer radius, $R_{\rm out}$.
We then weight the visibilities for each annulus (which are
given by the ring visibilities described by Equation \ref{eq:ring}) by
their SEDs to determine the visibilities expected for the entire disk.

For a disk heated by stellar 
radiation and accretion \citep{LBP74}, the temperature profile, $T_{\rm R}$,
in the regime where $R_{\ast}/R \ll 1$, is given by 
$T_{\rm 1 AU} \left(R/{\rm AU}\right)^{-3/4}$. Here,
$T_{\rm 1 AU}=T_{\rm in} \left({R_{\rm in}}/{\rm AU}\right)^{3/4}$,
where $T_{\rm in}$ is the temperature at the inner radius.
We consider two values of $T_{\rm in}$: 1500 K and 2000 K.  These are likely
(upper limit) sublimation temperatures for silicate and graphite grains, 
respectively \citep[e.g.,][]{SALPETER77,POLLACK+94},
and it is reasonable to assume that there is little or no dust emission 
interior to $R_{\rm in}$ (although the model does not exclude the 
possibility of optically thin gas interior to $R_{\rm in}$).  
We choose $R_{\rm out}$ to be the lesser of 
100 AU or the radius at which $T_{\rm R} = 10$ K ($R_{\rm out}$ is not 
important in this analysis, since virtually all of the near-IR flux comes
from the hotter inner regions of the disk).

The flux in an annulus specified by inner boundary $R_{\rm i}$ and outer
boundary $R_{\rm f}$ is given by 
\begin{equation}
F_{\rm annulus} = \frac{\pi}{d^2} [B_{\nu}(T_{\rm i}) + B_{\nu}(T_{\rm f})]
R_{\rm i}(R_{\rm f} - R_{\rm i}) \cos(\phi),
\label{eq:acc-flux}
\end{equation}
and the normalized visibilities for this annulus are (following 
Equation \ref{eq:ring}):
\begin{equation}
V_{\rm annulus} = \frac{2}{\pi r_{\rm uv} (\theta^2_{\rm f}-\theta^2_{\rm i})}
\left[\theta_{\rm f} J_1(\pi \theta_{\rm f} r_{\rm uv}) - \theta_{\rm i}
J_1(\pi \theta_{\rm i} r_{\rm uv})\right].
\label{eq:vannulus}
\end{equation}
Here, $d$ is the distance to the source, $\nu$ is the observed frequency,
$B_{\nu}$ is the Planck function,
$T$ is the temperature, $R$ is the physical radius,
$\theta$ is the 
angular size, $\phi$ is the inclination, 
$r_{\rm uv}$ is the ``uv radius'' (Equation \ref{eq:ruv}), 
and $i,f$ indicate the inner
and outer boundaries of the annulus.
To obtain the visibilities for the entire disk, we
sum the flux-weighted visibilities for each annulus 
and normalize by the total flux:
\begin{equation}
V_{\rm disk} = \frac{\sum_{R_{in}}^{R_{out}} F_{\rm annulus} V_{\rm annulus}}{
\sum_{R_{in}}^{R_{out}}{F_{\rm annulus}}}.
\label{eq:vdisk}
\end{equation}
The resultant model visibilities are obtained by substituting this expression
for $V_{\rm x}$ into Equation \ref{eq:vtot}.  

As mentioned above, we compute the visibilities for models using $T_{\rm in}=
1500,2000$ K.  The main effect
of $T_{\rm in}$ is that for lower temperatures, the flux difference between 
the inner several annuli is significantly larger 
(since we are on the Wien tail of
the blackbody curve), leading to a smaller flux-weighted emitting region
probed by the 2.2 $\mu$m visibilities.  
This smaller
effective area for lower $T_{\rm in}$ leads to a slightly larger inner disk
diameter (Equation \ref{eq:ring}).
In practice, $T_{\rm in}$ is not a critical parameter since
most of the 2.2 $\mu$m radiation in the disk comes from the innermost annulus. 


\subsubsection{Puffed-Up Inner Disk Model \label{sec:ddn}}
We consider a two-layer flared disk model \citep{CG97} with a puffed-up inner
disk wall \citep{DDN01}.  The primary difference between this model
and the geometrically flat model discussed above is the angle at which
starlight is incident on the disk.  While starlight tends to hit a flat
disk at grazing angles, for this model the starlight is incident at
larger angles and causes more heating of the disk at a given radius.  
This additional
heating causes the disk to expand in the vertical direction (in order
to maintain hydrostatic equilibrium), which leads to a puffed-up inner
wall as well as flaring in the outer disk.  Since stellar radiation
is incident on the inner wall normally, most of the 2.2 $\mu$m flux is
generated in this region.

As in the case of a geometrically flat disk (\S \ref{sec:acc})
we first calculate the radial
temperature and SED distributions.  The temperature distribution for
this model has been discussed in detail by Chiang \& Goldreich (1997)
and DDN, but we present a brief outline of the equations
here.  For the flared two-layer disk we assume a radial dust surface 
density profile $\Sigma(R)=10^5 (R / 1 {\rm AU})^{-1.5}$ g cm$^{-2}$
and a flaring angle, $\alpha$, which defines
the angle at which starlight impacts the disk: 
\begin{equation}
\alpha = 0.4 \left(\frac{R_{\ast}}{R}\right) + \frac{8}{7}
\left(\frac{T_{\ast}}{T_{\rm c}}\right)^{4/7} \left(\frac{R}{R_{\ast}}
\right)^{2/7}.
\label{eq:ddn-alpha}
\end{equation}
Here, $R$ is the radial coordinate in the disk,
$T_{\rm c} = GM_{\ast}\mu m_{\rm p} / kR_{\ast}$, $M_{\ast}$
is the mass of the star, $\mu$ is the mean molecular weight,
$R_{\ast}$ is the stellar radius, and $T_{\ast}$ is the stellar temperature.
We define a flaring index, $\xi$, which corresponds to the
exponent on the scale height as a function of radius minus one
($\gamma-1$ in DDN).  Following a simple hydrostatic equilibrium calculation 
from Chiang \& Goldreich (1997), we have adopted a value for $\xi$ of 2/7.

Assuming that the opacity in the disk is due to silicate dust \citep{DL84},
we parameterize the opacity of
the surface and interior layers using $\psi_{\rm surf}$ and $\psi_{\rm int}$:
\begin{equation}
\psi_{\rm surf} = \frac{\sum_{\nu=0}^{\infty} B_{\nu}(T_{\rm surf}) 
\kappa_{\nu} \left(1-\exp[-\Sigma(R) \kappa_{\nu}]\right)}
{\sum_{\nu=0}^{\infty} B_{\nu}(T_{\rm surf}) \kappa_{\nu}},
\label{eq:ddn-psis}
\end{equation}
\begin{equation}
\psi_{\rm int} = \frac{\sum_{\nu=0}^{\infty} B_{\nu}(T_{\rm int}) 
\left(1-\exp[-\Sigma(R) \kappa_{\nu}]\right)}
{\sum_{\nu=0}^{\infty} B_{\nu}(T_{\rm int})}.
\label{eq:ddn-psii} 
\end{equation}

The temperature of the surface layer is determined from
\begin{equation}
T_{\rm surf} = \epsilon_{\rm s}^{-0.25} \left(\frac{R_{\ast}}{2R}\right)^{0.5}
T_{\ast}.
\label{eq:ddn-tsurf}
\end{equation}
Here, $\epsilon_{\rm s}=\kappa_p(T_{\rm surf}) / \kappa_p(T_{\ast})$, 
where $\kappa_p$
is the Planck mean opacity.  The interior temperature is then given by
\begin{equation}
T_{\rm int} = \left(\frac{\alpha \psi_{\rm surf}}{2 \psi_{\rm int}}
\right)^{0.25}  \left(\frac{R_{\ast}}{R}\right)^{0.5} T_{\ast}.
\label{eq:ddn-tin}
\end{equation}
Equations \ref{eq:ddn-psis}--\ref{eq:ddn-tin} are solved iteratively.
The SED of an annulus is given by summing the contributions
from the surface and the interior.  We denote the surface temperatures
at the inner and outer radii of the annulus by $T_{\rm si},T_{\rm sf}$,
and the interior temperatures by $T_{\rm ii},T_{\rm if}$.
The surface component of the SED is given by
\begin{equation}
F_{\rm surf} = \frac{\pi}{d^2} [B_{\nu}(T_{\rm si}) + B_{\nu}(T_{\rm sf})]
R_{\rm i}(R_{\rm f} - R_{\rm i}) \left(1+\exp[-\Sigma(R_{\rm i})
\kappa_{\nu}/\cos(\phi)]\right) \kappa_{\nu} \Delta \Sigma,
\label{eq:ddn-fsurf}
\end{equation}
where
\begin{equation}
\Delta \Sigma = \frac{\psi_{\rm int}}{\psi_{\rm surf}} \left(\frac{T_{\rm ii}}
{T_{\rm si}}\right)^4 \frac{1}{2 \kappa_p}.
\label{eq:ddn-dsigma}
\end{equation}
For the interior,
\begin{equation}
F_{\rm int} = \frac{\pi}{d^2} [B_{\nu}(T_{\rm ii}) + B_{\nu}(T_{\rm if})]
R_{\rm i}(R_{\rm f} - R_{\rm i})  \left(1-\exp[-\Sigma(R_{\rm i})
\kappa_{\nu}/\cos(\phi)]\right) \cos(\phi).
\label{eq:ddn-fint}
\end{equation}

In addition to the surface and interior layers of the disk, a puffed-up
inner disk wall is included in the model.  We choose the radius of this
inner rim, $R_{\rm in}$, to be the radius where  $T=T_{\rm in}$ 
($R_{\rm in}$ also defines the inner radius of the flared disk component).
We consider $T_{\rm in}=1500,2000$ K, likely
sublimation temperatures of graphite and silicate grains, respectively.
Because the wall is directly
exposed to stellar radiation (instead of the glancing angles encountered
in a geometrically thin disk model), the inner rim puffs up, attaining
a height given by
\begin{equation}
H_{\rm rim} = \chi_{\rm rim} \sqrt{\frac{k T_{\rm in} R_{\rm in}^3}
{\mu m_{\rm p} G M_{\ast}}}.
\label{eq:ddn-hrim}
\end{equation}
$\chi_{\rm rim}$ is a dimensionless quantity that describes how the disk
height depends on the stellar luminosity.  For the stars in our sample,
we adopt typical values of $\chi_{\rm rim}=5.3$ for stars later
than spectral type B6, and $\chi_{\rm rim}=4.5$ for earlier spectral
types \citep{DDN01}.
The emergent flux from the rim
contributes only to the innermost annulus:
\begin{equation}
F_{\rm rim}(R_{\rm in}) = \frac{4 R_{\rm in} H_{\rm rim}}{d^2} 
B_{\nu}(T_{\rm in}).
\label{eq:ddn-frim}
\end{equation}
While DDN included the effects of inclination, for simplicity we ignore
those here.  In effect, this corresponds to assuming a different shape
for the inner disk wall; DDN assumed a thin cylindrical annulus,
and we adopt a more toroidal shape.  

The total flux for each annulus is given by 
\begin{equation}
F_{\rm annulus}
= F_{\rm int} + F_{\rm surf} + F_{\rm rim},
\label{eq:ddn-ftot}
\end{equation}
where $F_{\rm rim}$ is zero everywhere but the innermost annulus.  
To determine the visibilities for the entire disk, we sum the
flux-weighted visibilities for each annulus, and normalize by the
total flux (Equation \ref{eq:vdisk}).
When calculating the visibilities for each annulus, we retain the
approximation of a geometrically flat inner disk\footnote{This 
approximation is valid for the inner disk (where the K-band
emission arises) because flaring is negligible in the inner regions.
Moreover, the vertical puffing of the
inner rim will cause little deviation between the true and approximate
visibilities.  Even in the edge-on case, the
approximation will not be far-off since the minor axis is essentially
unresolved in either the true or approximate model.}
and use Equation \ref{eq:vannulus}.
Thus, the normalized squared visibilities for this model are computed using
Equations \ref{eq:ddn-ftot}, \ref{eq:vannulus}, \ref{eq:vdisk},
and \ref{eq:vtot}.

As discussed in \S \ref{sec:acc}, the fitted inner disk sizes are not
particularly sensitive to the choice of $T_{\rm in}$.  However, for the
puffed-up inner disk model, fitted sizes will be slightly {\it smaller}
for lower values of $T_{\rm in}$.  The puffed-up
inner rim is much hotter than the flared disk component and, as
the rim temperature decreases, the difference in flux between the rim and
inner disk annuli decreases.  This leads to a larger flux-weighted
effective area, which, in turn, 
leads to smaller fitted sizes (Equation \ref{eq:ring}).
While we have ignored the effect of the shadow cast by the
inner rim onto the disk \citep{DDN01}, this
will not significantly alter the results since the fitted
sizes are relatively insensitive to the temperature difference between
the rim and the disk.  We also note that of the two values of $T_{\rm in}$
considered, models with $T_{\rm in}=2000$ K have a larger temperature
difference between the puffed-up rim and the inner disk annuli, and thus
more closely approximate the effects of shadowing.

\subsubsection{Binary Model \label{sec:binary}}
This model simulates a wide binary, where 
visibilities are effectively due to two stationary point sources.
The squared visibilities for the binary model are,
\begin{equation}
V^2_{\rm binary} = \frac{1 + R^2 + 2R \cos\left(\frac{2\pi}{\lambda}
\vec{B} \cdot \vec{s} \right)}{(1+R)^2},
\label{eq:binary}
\end{equation}
where $(\vec{B} \cdot \vec{s})/\lambda = \theta [u \sin(\psi) + v \cos(\psi)]$,
$\theta$ is the angular separation of the binary, $\psi$ is the
position angle, $R$ is the ratio of the fluxes of the two components,
and $\lambda$ is the observed wavelength.

\section{Results and Analysis \label{sec:res}}

\subsection{PTI Results \label{sec:models}}
Disk models fit the PTI data reasonably well for most
sources in our sample.  All sources except HD 141569 and HD 158352 are 
resolved, with uniform disk diameters between $\sim 2.5$ and 5.8 mas,
and most sources show evidence for non-symmetric circumstellar distributions.
While a nearly circularly symmetric distribution appears
appropriate for AB Aur, the data for
MWC 480, MWC 758, CQ Tau, VV Ser, V1685 Cyg, AS 442, and MWC 1080
show evidence for significantly non-zero inclinations.
A high inclination cannot be ruled out for V1295 Aql, and the data
for T Ori, MWC 120, and MWC 297 are insufficient to constrain the inclinations.

Tables \ref{tab:uniforms}--\ref{tab:ddn} 
list the fitted parameters and $\chi_r^2$ values
for various disk models.  Columns 2 and 3 list the $\chi_r^2$ values and
best-fit angular size scales ($\theta$) for face-on models, and columns
4-7 list the $\chi_r^2$, sizes ($\theta$), position angles ($\psi$), 
and inclinations ($\phi$) for inclined models.
The $u-v$ coverage for T Ori, MWC 120, and MWC 297 is insufficient to
fit inclined disk models, and for these sources we constrain only
the angular size scales of face-on disk models.
Table \ref{tab:ddn}, which lists the fitted parameters for
flared disk models with puffed-up inner walls, does not include
MWC 297, V1685 Cyg, and MWC 1080, the sources with the earliest spectral
types in our sample.  For these objects, the puffed-up inner wall
model cannot fit the visibility data ($\chi_r^2 \gg 100$), since the early-type
central stars lead to hot inner disks at radii much larger than allowed
by the PTI data.

In Table \ref{tab:binary}, we present the angular 
separations ($\theta$), position angles ($\psi$), 
and brightness ratios ($R$) for binary models.  
The $u-v$ coverage for T Ori, MWC 120, and MWC 297 is insufficient to
fit binary models, and these sources are therefore absent from Table
\ref{tab:binary}.  
The best-fit binary separations for all sources in our sample are
$\ga 2.5$ mas.  For the distances and approximate masses of the sources in our 
sample, these separations correspond to orbital periods of $\ga 2$ years.
For most objects, observations span several months, and
our assumption that the two point sources in the binary model
are stationary is reasonable.  This assumption may break down
for short-period binaries in sources with observations spanning more
than 1 year (see Table \ref{tab:obs}).  However, short-period orbits
would produce visibilities that vary with time, and no time-variation is
detected.

\begin{figure}
\plotone{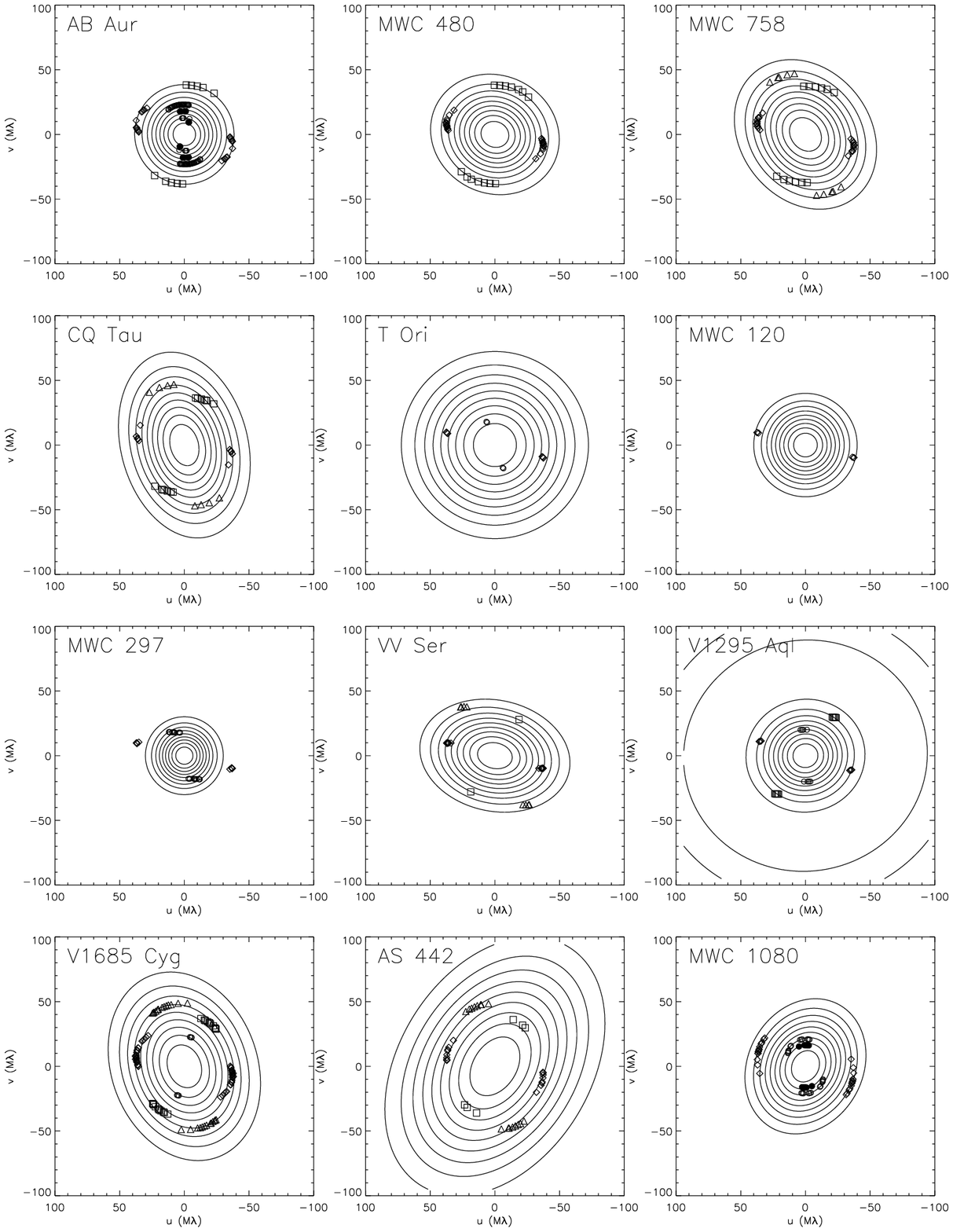}
\caption{Contour plots of best-fit uniform disk models for AB Aur,
MWC 480, MWC 758, CQ Tau, T Ori, MWC 120, MWC 297,
VV Ser, V1295 Aql, V1685 Cyg, AS 442, and MWC 1080, whose parameters
are listed in Table \ref{tab:uniforms}.
We plot the best-fit inclined models except for T Ori, MWC 120, and MWC 297,
where no inclination estimates are available, in which case we use the
best-fit face-on models.  For MWC 297, we have used the model determined
from a combined fit to PTI+IOTA data 
(\S \ref{sec:pti+iota}; Table \ref{tab:pti+iota}), since the PTI visibilities
only provide limits.
The contour increment is
10\% in $V^2$. 
We also plot the $u-v$ points sampled for each source by the PTI NW baseline
(open diamonds), the PTI SW baseline (open squares), the PTI NS baseline 
(open triangles), and by IOTA (MST; H and K-band data, represented by open and
filled circles, respectively).  
\label{fig:uv}}
\end{figure}

\begin{figure}
\plottwo{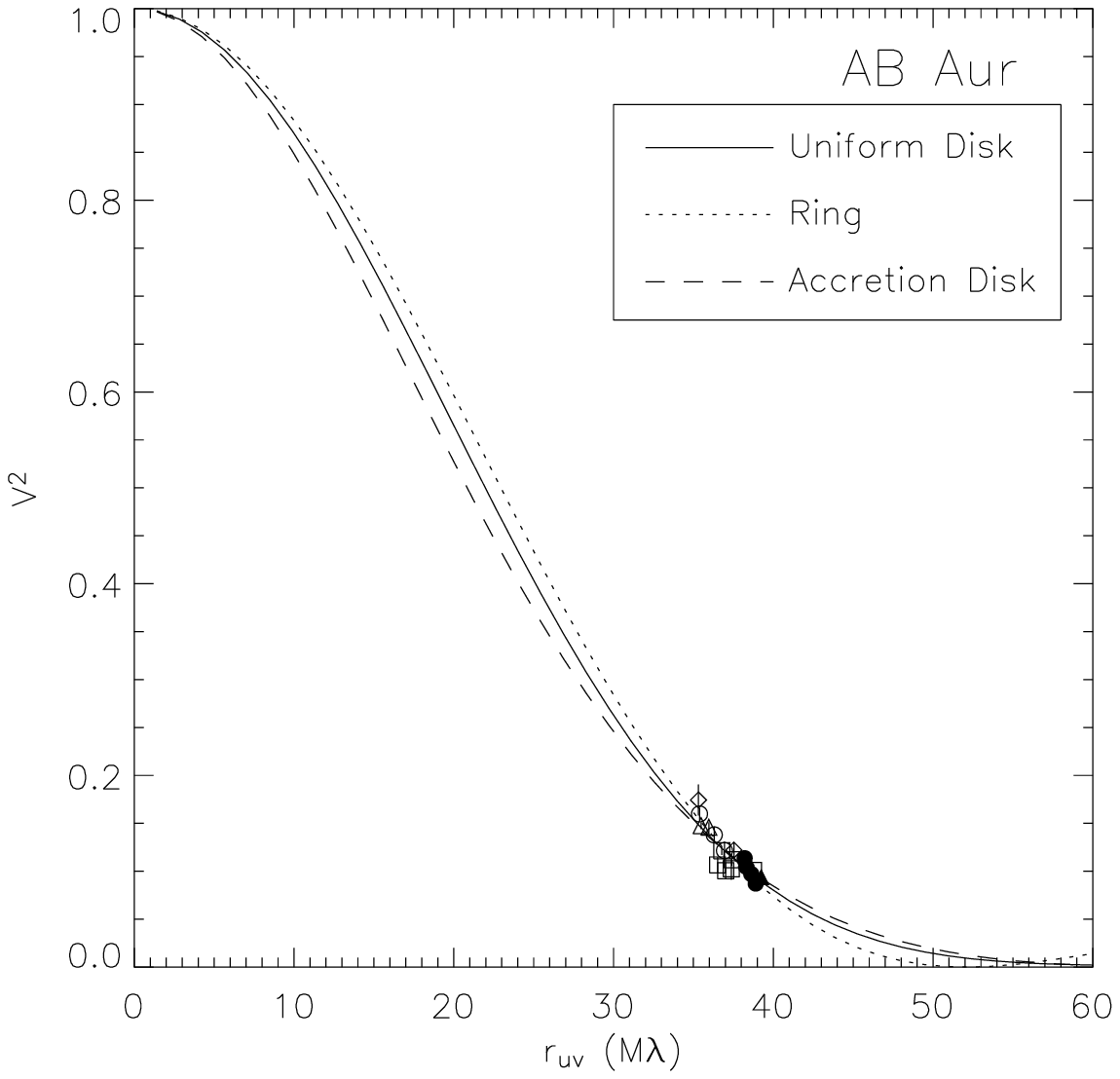}{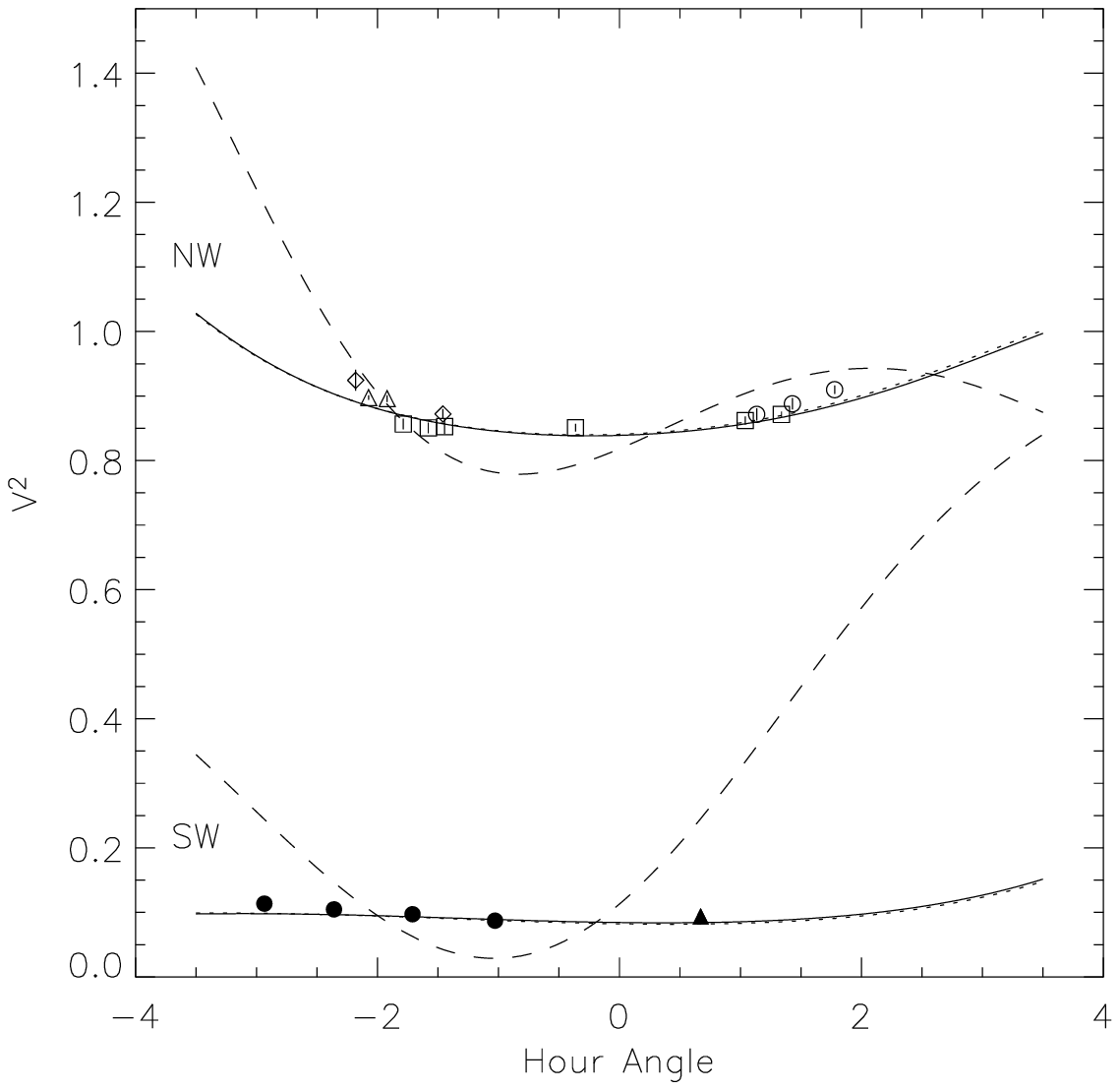}
\caption{$V^2$ PTI data for AB Aur, as a function of 
$r_{\rm uv} = (u^2 + v^2)^{1/2}$ (left panel).  
Data for individual nights are represented by different symbols,
where NW data are plotted with open symbols, and SW data use
filled symbols. Face-on uniform disk (solid line), ring (dotted line), 
and geometrically flat accretion disk (dashed line) models are over-plotted.  
We also plot the data as a function of hour angle (right panel).
For clarity, we have plotted $V^2+0.75$ for the NW data.
The best-fit face-on uniform disk is plotted as a solid line,
the best-fit inclined uniform disk model is represented with a dotted line, and
the best-fit binary model is shown with a dashed line.  We note that
the different linestyles correspond to different models in the
left and right panels.
For this source,
we see that a face-on model provides the best fit to the data.
\label{fig:abaur}}
\end{figure}

\begin{figure}
\plottwo{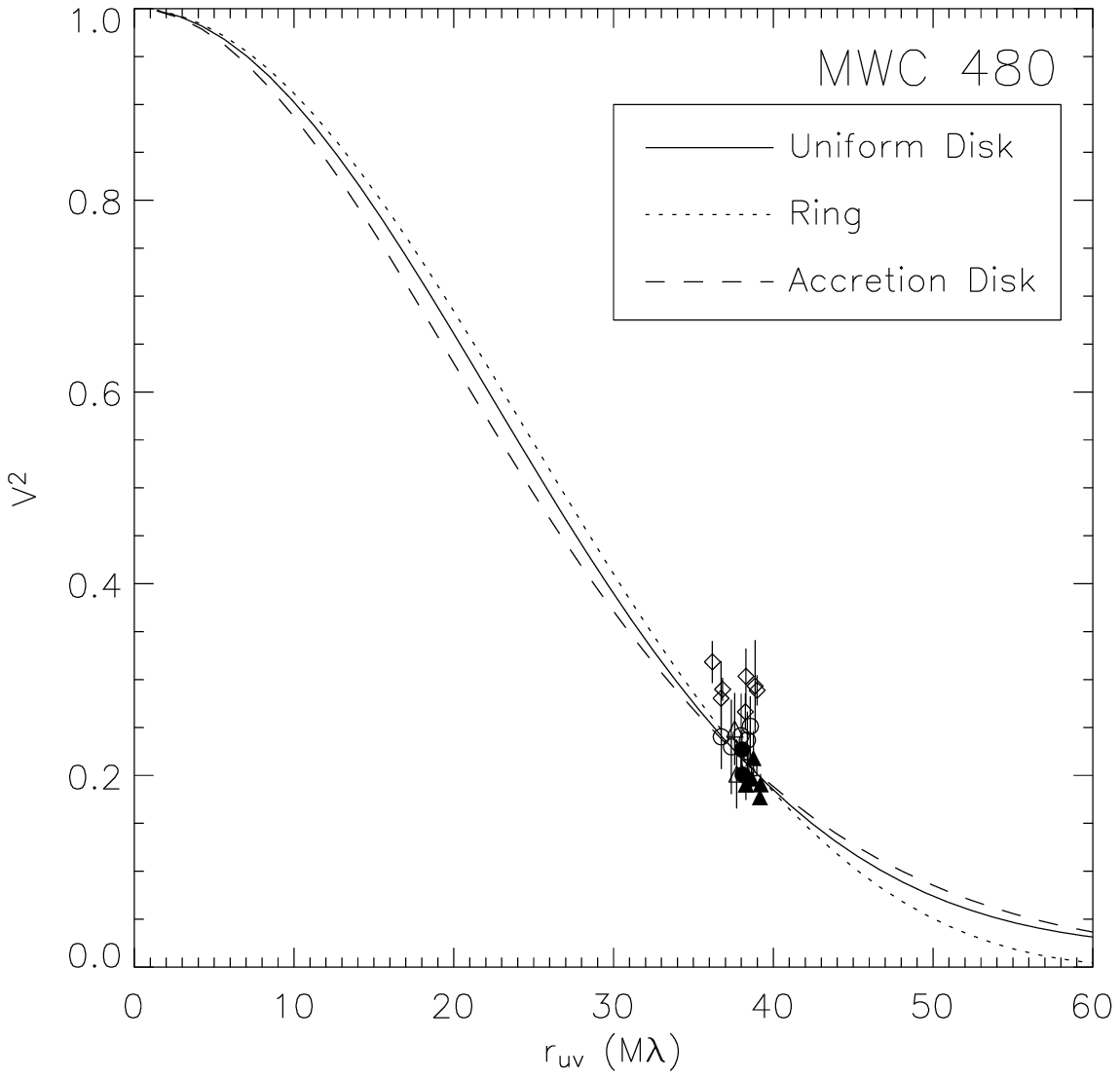}{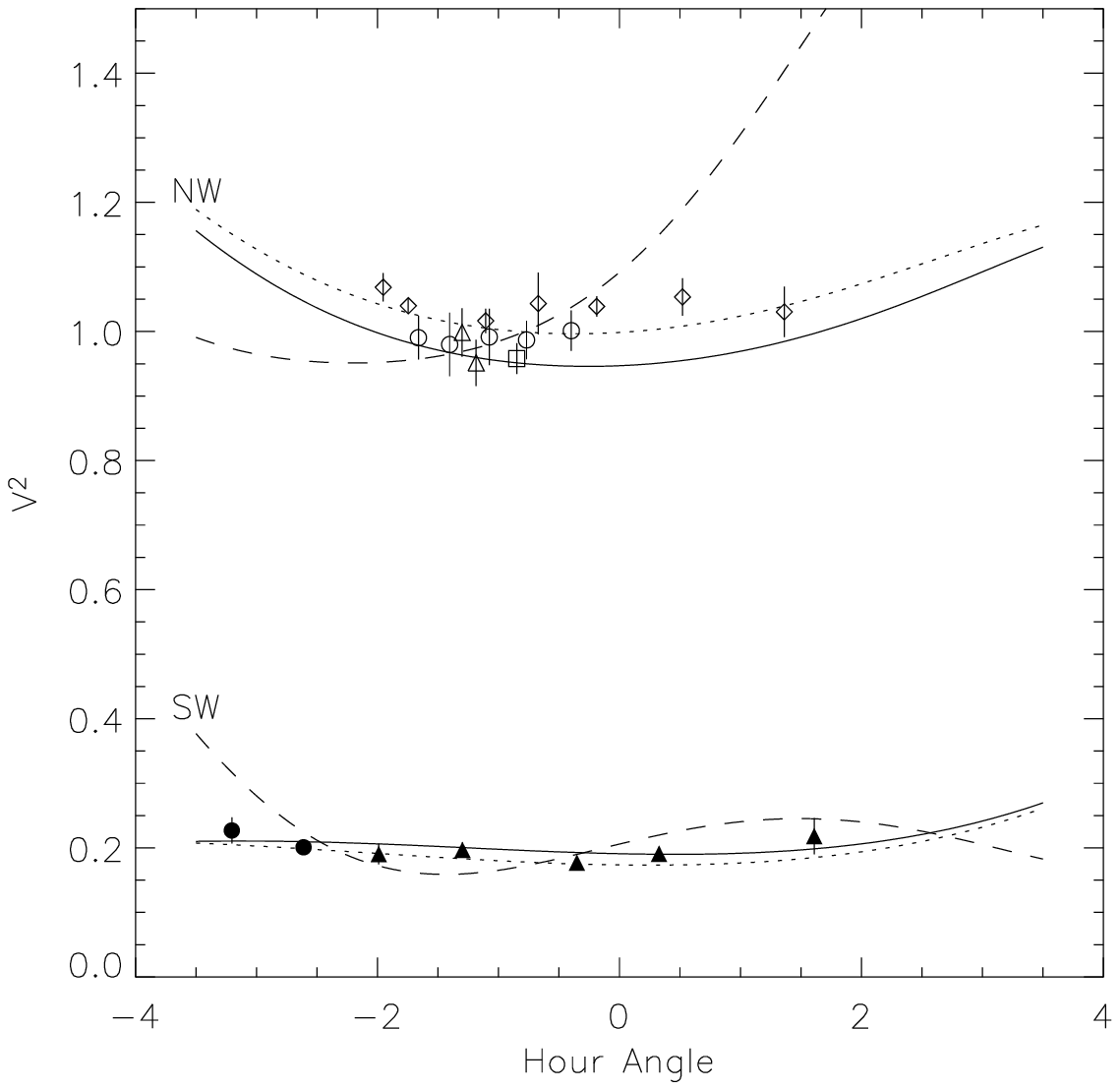}
\caption{$V^2$ PTI data for MWC 480, as a function of 
$r_{\rm uv}$ (left panel) and hour angle (right panel).  
Symbols and models are plotted as in Figure \ref{fig:abaur}.
For this source,
we see that an inclined disk model provides the best fit to the data.
\label{fig:mwc480}}
\end{figure}

\begin{figure}
\plottwo{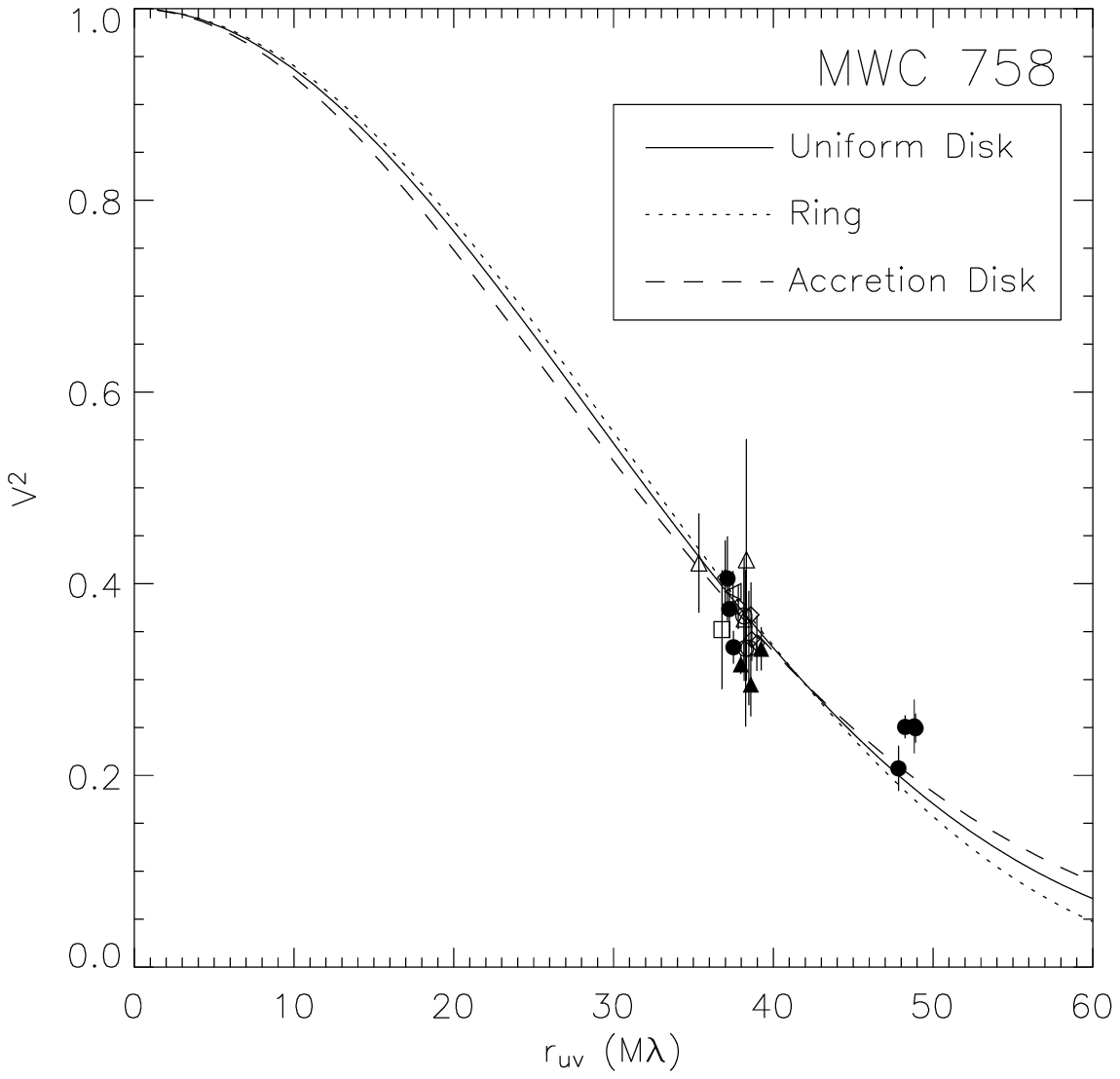}{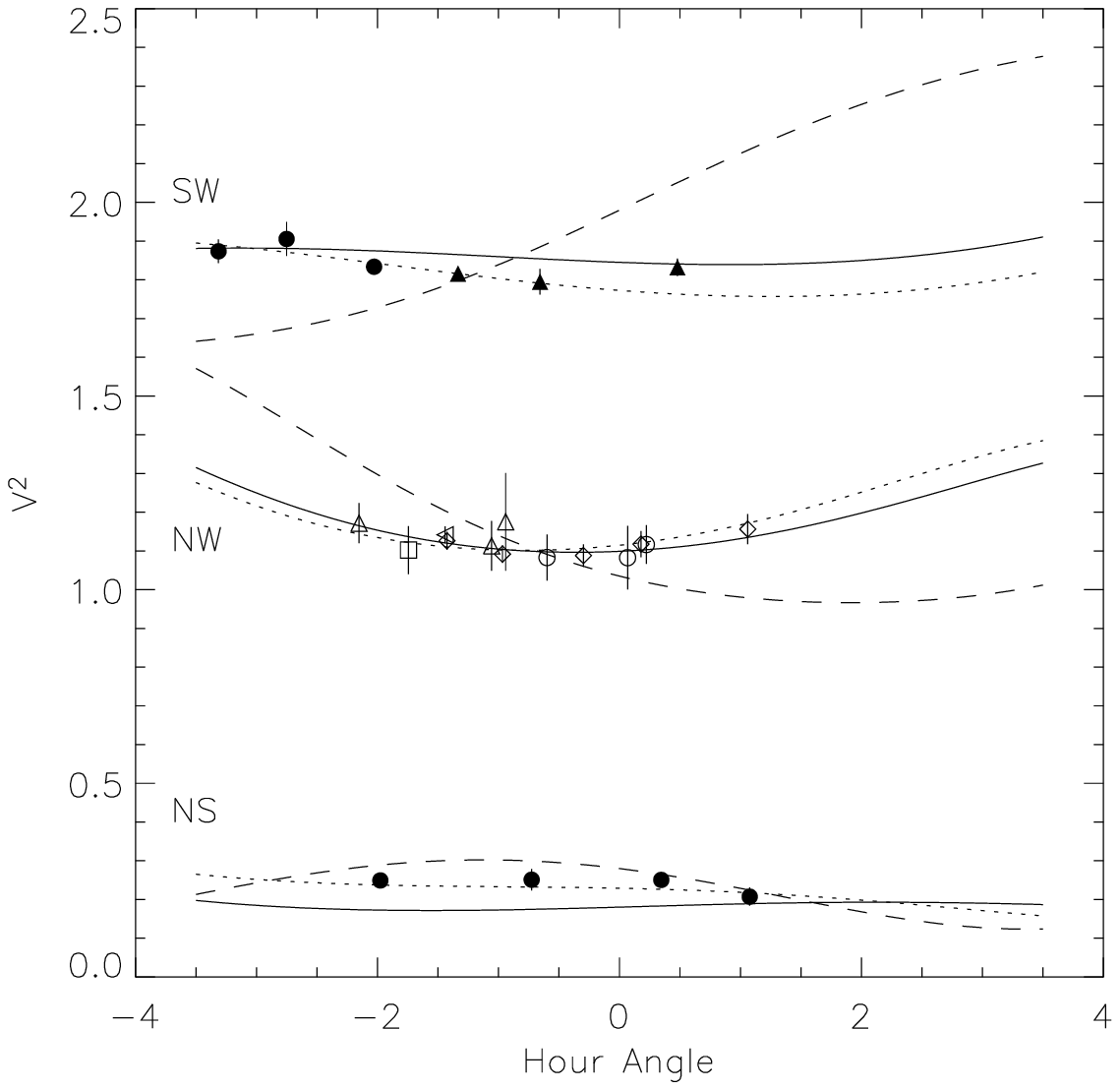}
\caption{$V^2$ PTI data for MWC 758, as a function of 
$r_{\rm uv}$ (left panel) and hour angle (right panel).
Symbols and models are plotted as in Figure \ref{fig:abaur},
except that here, we have three baselines.
NW data are plotted with open symbols, and NS and SW data use
filled symbols.  In the right panel, 
we have plotted $V^2+0.75$ for the NW data and
$V^2+1.5$ for the SW data.  
An inclined disk model provides the best fit to the data.
\label{fig:mwc758}}
\end{figure}

\begin{figure}
\plottwo{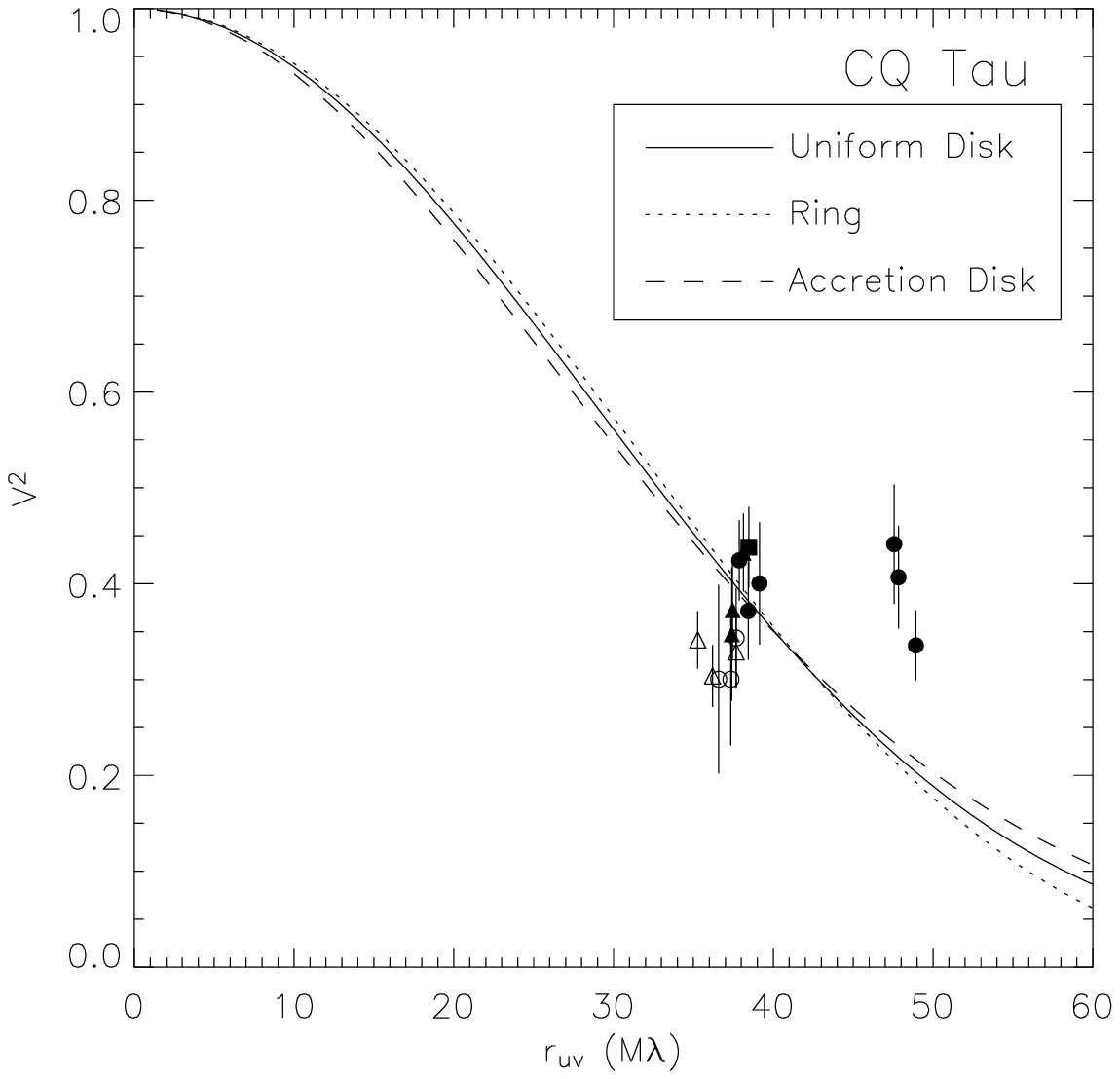}{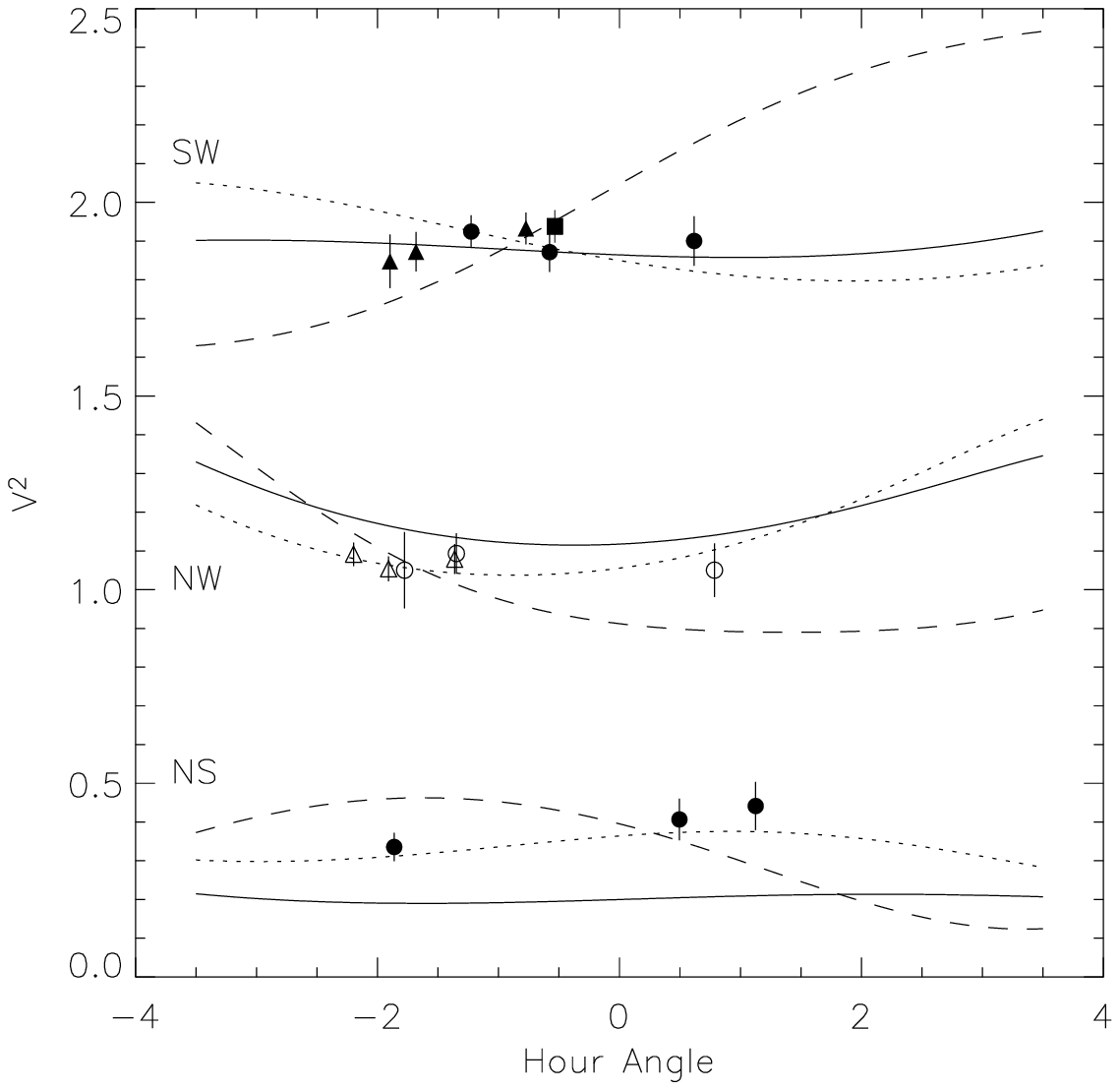}
\caption{$V^2$ PTI data for CQ Tau, as a function of 
$r_{\rm uv}$ (left panel) and hour angle (right panel).  
Symbols and models are plotted as in Figure \ref{fig:mwc758}.
For this source, an inclined disk model provides the best fit to the data.
\label{fig:cqtau}}
\end{figure}

\begin{figure}
\plotone{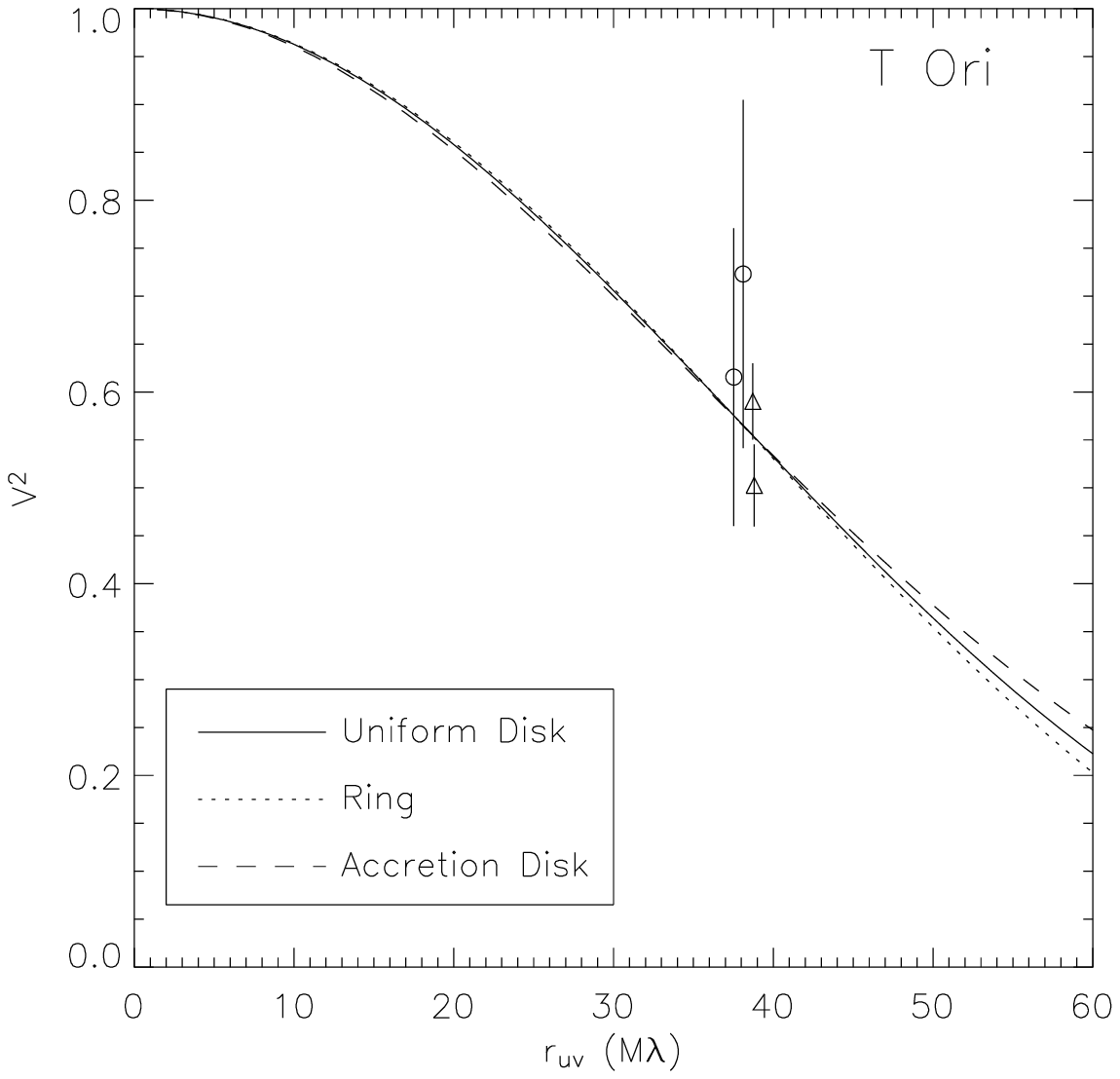}
\caption{$V^2$ PTI data for T Ori, as a function of 
$r_{\rm uv}$.  Symbols and models are plotted as in Figure \ref{fig:abaur}. 
For this source, the limited $u-v$ coverage does not allow an 
estimate of inclination, so we plot only face-on models.
\label{fig:tori}}
\end{figure}

\begin{figure}
\plotone{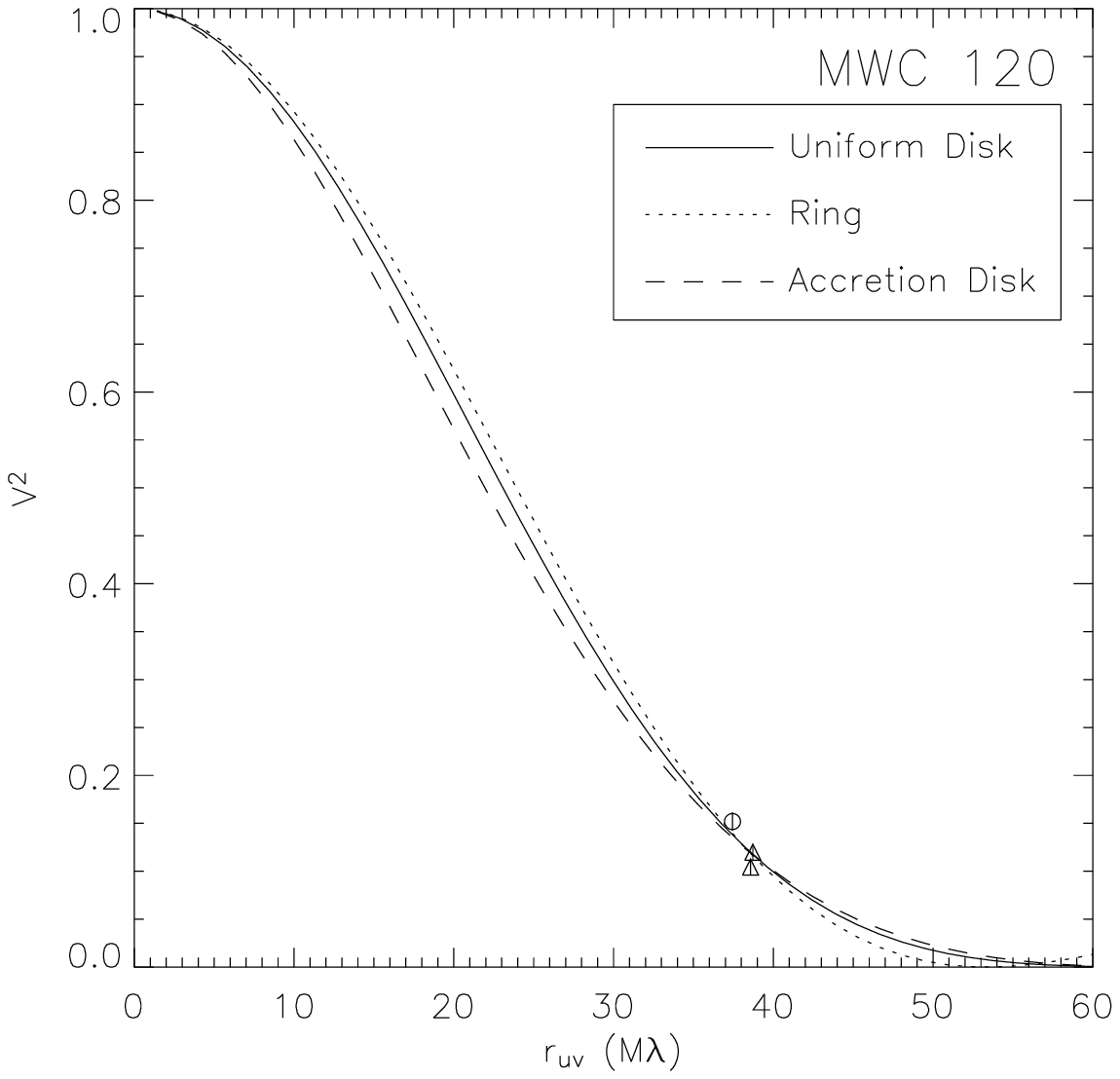}
\caption{$V^2$ PTI data for MWC 120, as a function of 
$r_{\rm uv}$.  Symbols and models are plotted as in Figure \ref{fig:abaur}. 
For this source, the limited $u-v$ coverage does not allow an 
estimate of inclination, so we plot only face-on models.
\label{fig:mwc120}}
\end{figure}

\begin{figure}
\plottwo{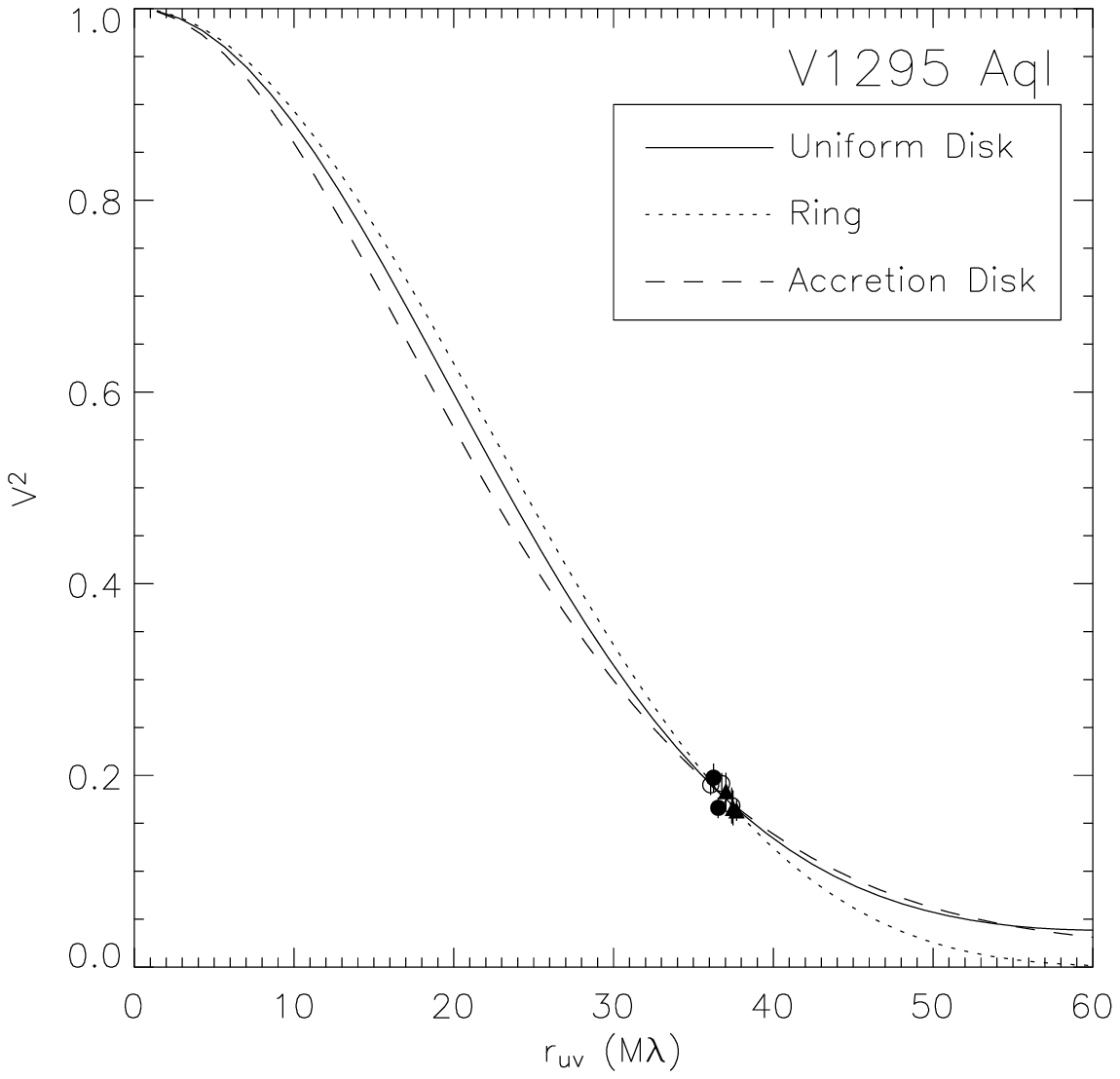}{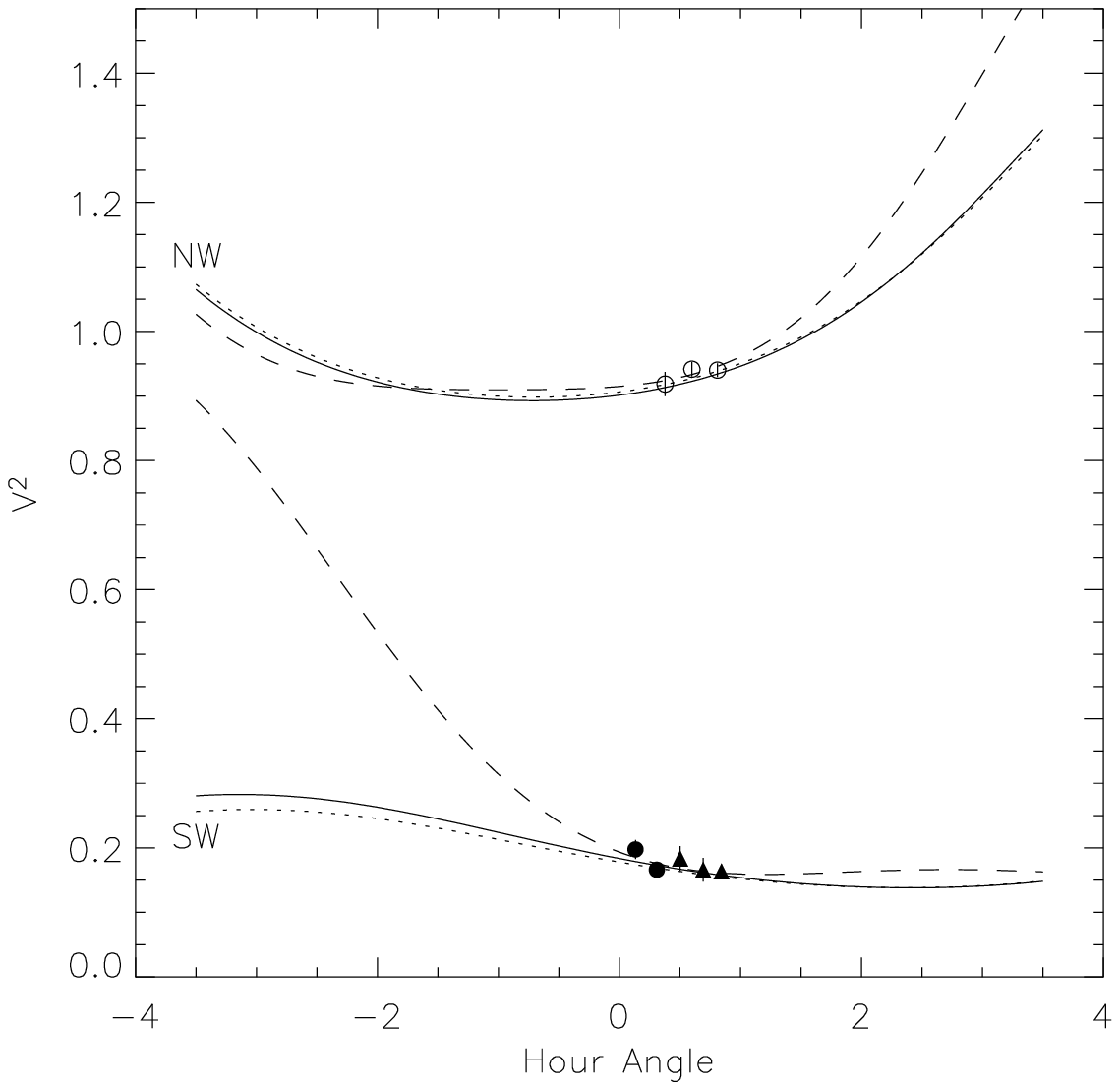}
\caption{$V^2$ PTI data for V1295 Aql, as a function of 
$r_{\rm uv}$ (left panel) and hour angle (right panel).  
Symbols and models are plotted as in Figure \ref{fig:abaur}.
For this source,
it appears that a face-on disk or a binary model provides the best fit 
to the data.
\label{fig:v1295aql}}
\end{figure}

\begin{figure}
\plottwo{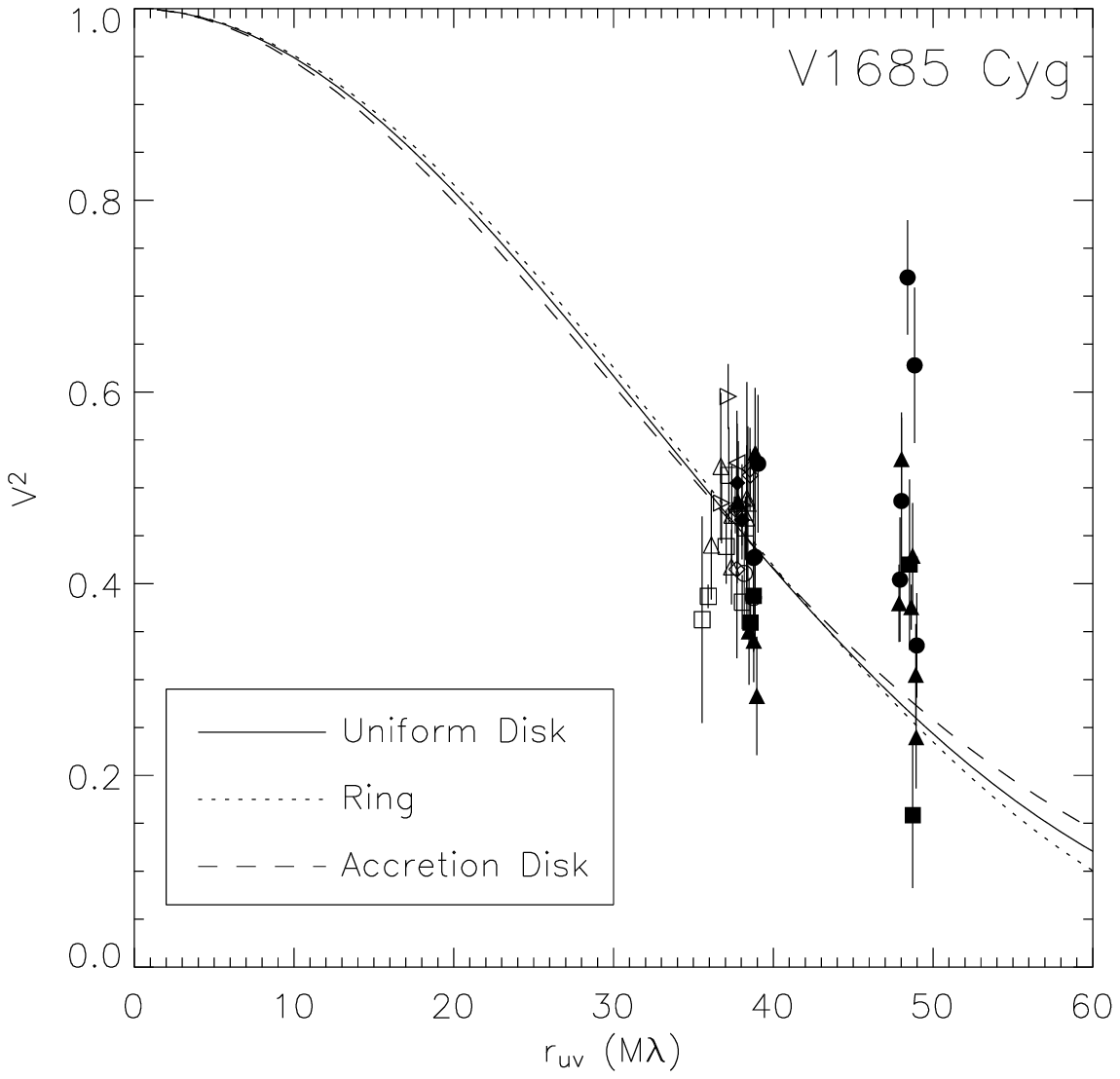}{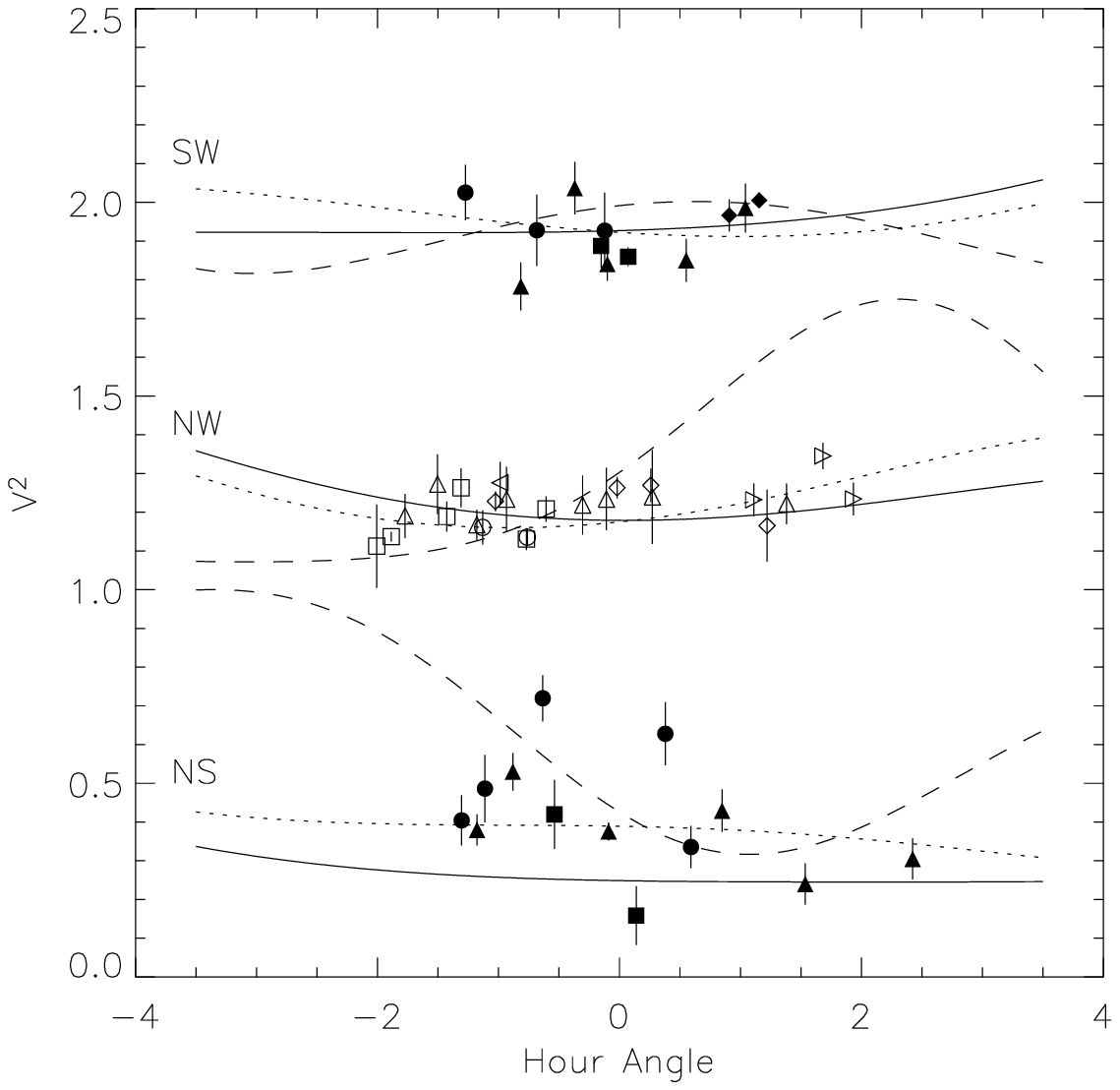}
\caption{$V^2$ PTI data for V1685 Cyg, as a function of 
$r_{\rm uv}$ (left panel) and hour angle (right panel).  
Symbols and models are plotted as in Figure \ref{fig:mwc758}.
For this source,
we see that an inclined disk model provides the best fit to the data,
although no models fit all of the data particularly well.
\label{fig:v1685cyg}}
\end{figure}

\begin{figure}
\plottwo{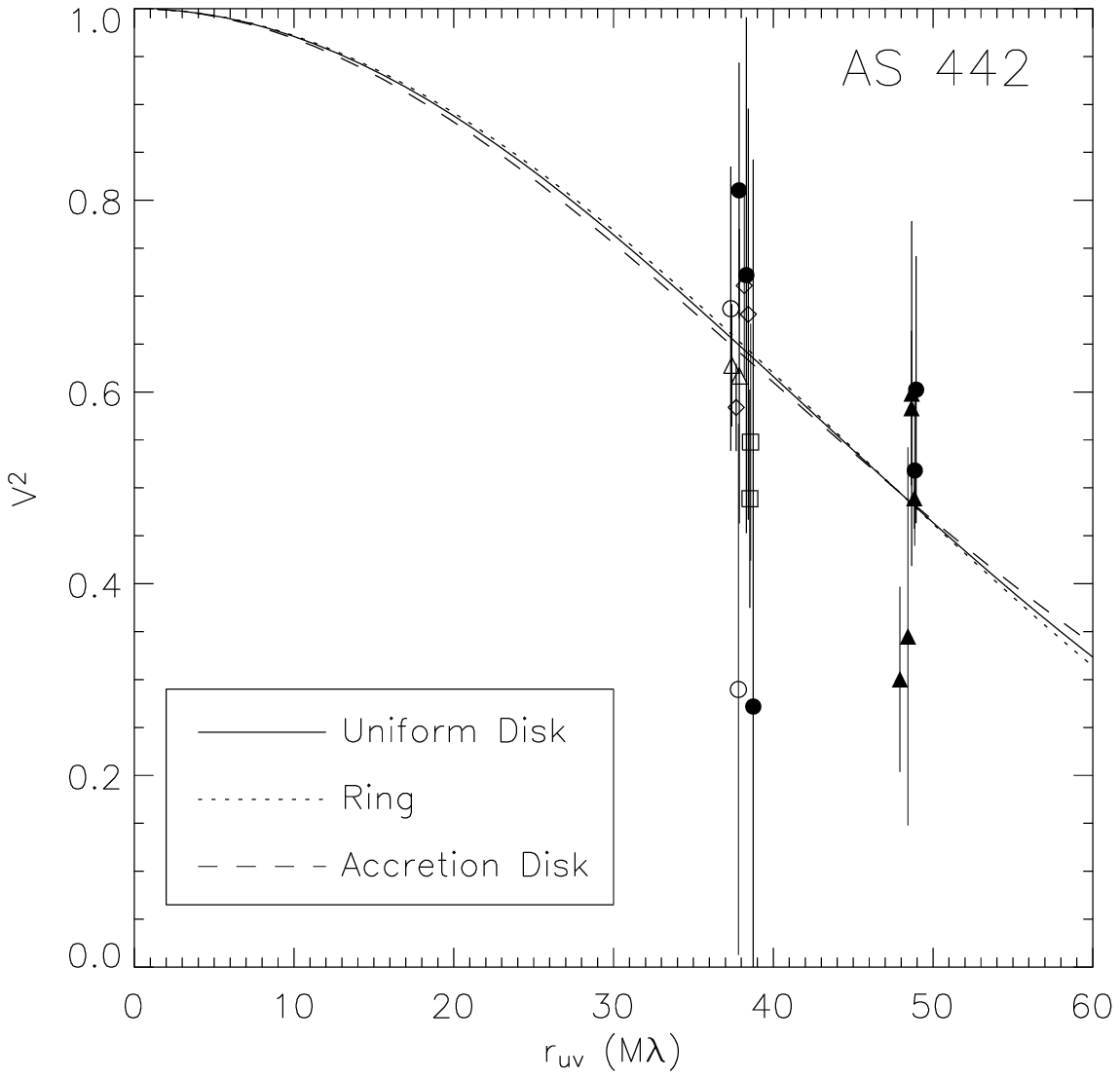}{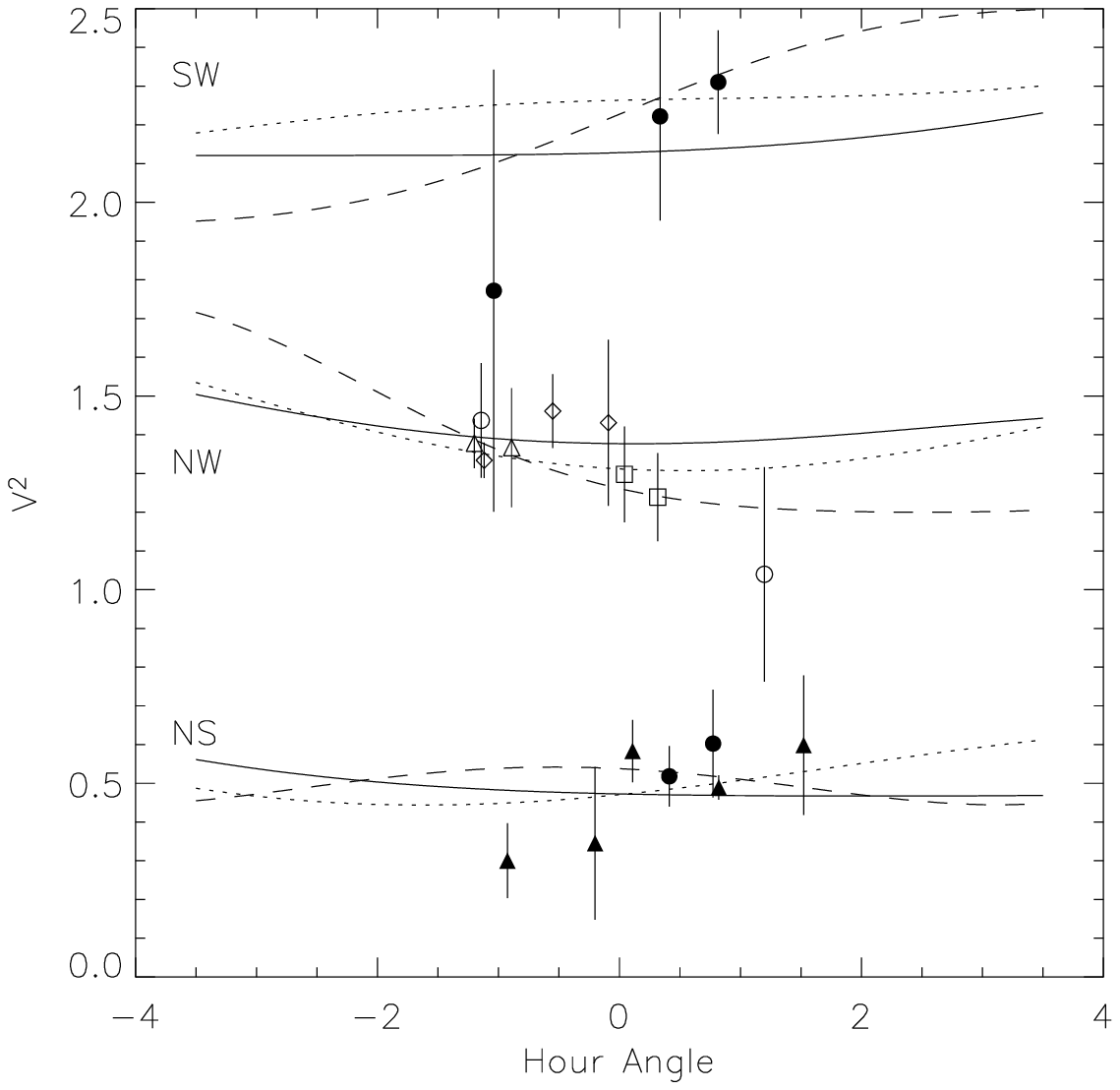}
\caption{$V^2$ PTI data for AS 442, as a function of 
$r_{\rm uv}$ (left panel) and hour angle (right panel).  
Symbols and models are plotted as in Figure \ref{fig:mwc758}.
Inclined disk or binary models 
provide the best fit to the data.
\label{fig:as442}}
\end{figure}

\begin{figure}
\plottwo{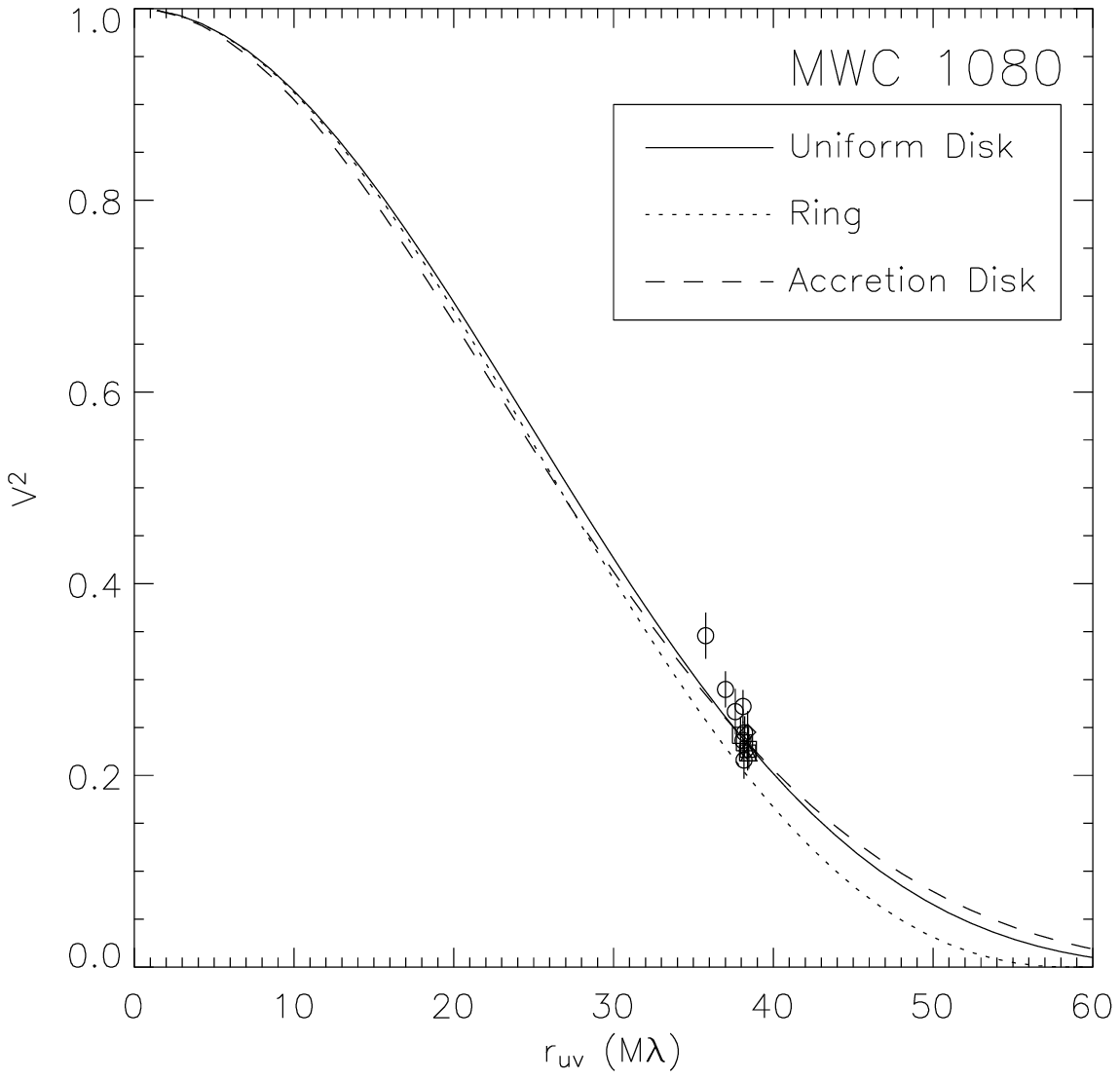}{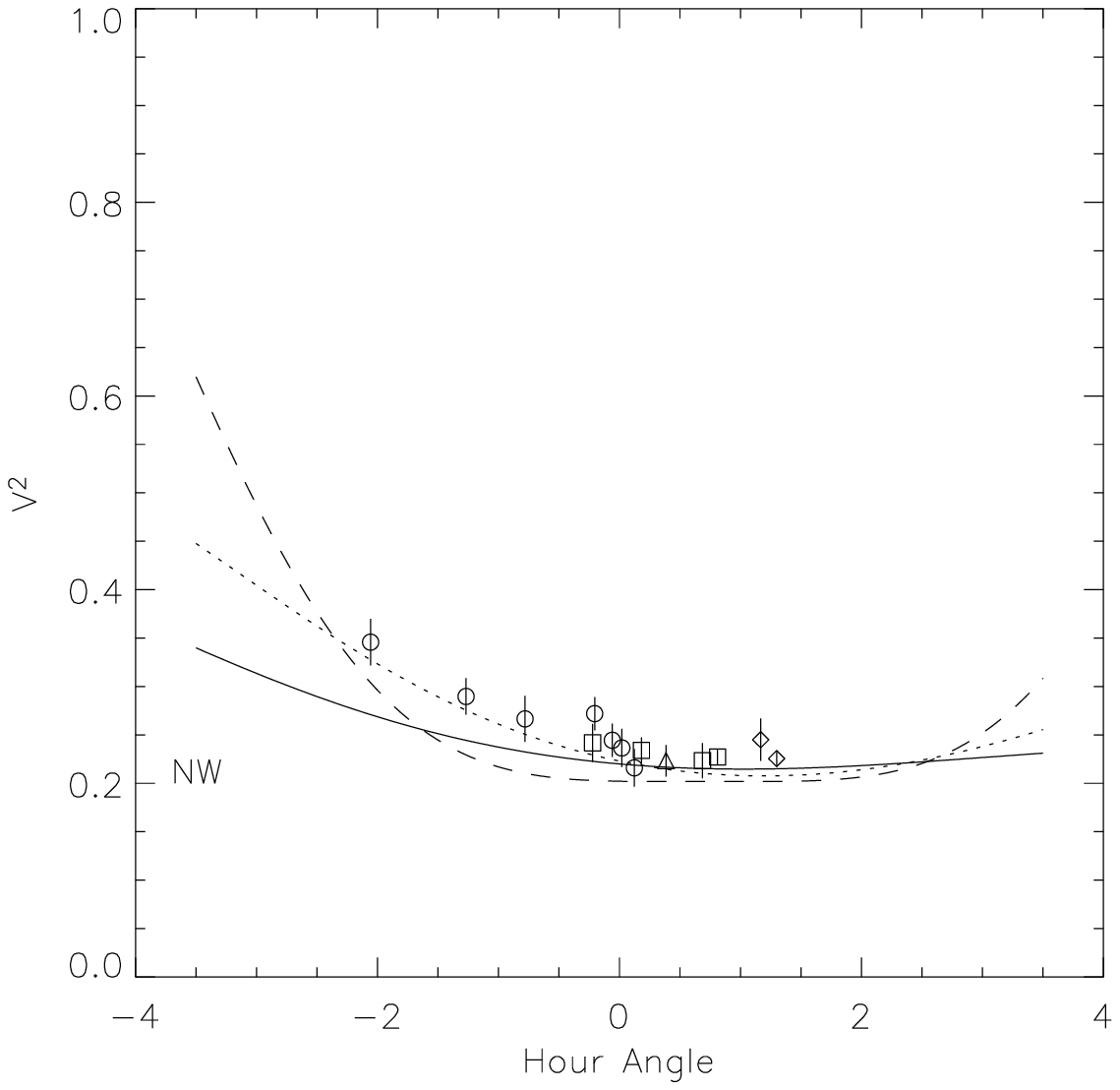}
\caption{$V^2$ PTI data for MWC 1080, as a function of 
$r_{\rm uv}$ (left panel) and hour angle (right panel).  
Symbols and models are plotted as in Figure \ref{fig:abaur},
except that no offset has been applied to the
visibilities in the right panel.
An inclined disk model provides the best fit to the data.
\label{fig:mwc1080}}
\end{figure}

Figures \ref{fig:abaur}--\ref{fig:mwc1080} 
show plots of observed $V^2$
for each source along with the curves predicted by various models.
Inclined disk and binary models are not circularly symmetric, and 
the visibilities are a function of the observed
position angle in addition to the projected baseline 
(Figure \ref{fig:uv}).  Thus, for sources with sufficient data 
to constrain the inclination,  we have plotted $V^2$ as
a function of both $r_{\rm uv}$ and hour angle.
Since we were only able to derive lower limits on the angular size
scales for MWC 297, we do not plot the models for this source here.

AB Aur, VV Ser, V1685 Cyg, AS 442, and MWC 1080
were discussed in ELAHS.  Here, we have obtained additional data 
on additional baselines for all except MWC 1080, and the greatly enhanced 
$u-v$ coverage, shown in Figure \ref{fig:uv},
enables firmer constraints on the models.
We include MWC 1080 in the present discussion largely for 
completeness, since we perform some additional analysis steps that were
absent in the first paper.

\subsubsection{AB Aur}
The PTI visibilities for AB Aur are consistent with a disk-like circumstellar 
distribution that is nearly face-on
(Figure \ref{fig:abaur}).
From Tables \ref{tab:uniforms}-\ref{tab:ddn},  
the best-fit models indicate size 
scales\footnote{ As outlined in \S \ref{sec:gauss}--\ref{sec:ddn},
characteristic size scales for different models measure
different parts of the brightness distributions: 
Gaussian models measure full widths at half
maxima, uniform disk models measure outer diameters, ring models measure inner 
ring diameters, and accretion disk models (flat or flared) measure inner
disk diameters.
The spread in quoted angular sizes for a source is 
mainly due to these differences.} between 2.2 and 5.3 
mas,
and an inclination angle between $8^{\circ}$ and 16$^{\circ}$,
consistent with the values found by ELAHS.  We have, however,
reduced the uncertainties using additional data on a second baseline. 
The data cannot be fit well by a binary model ($\chi_r^2 \sim 117$;
Table \ref{tab:binary}).  

\subsubsection{MWC 480 \label{sec:mwc480}}
The PTI visibilities for MWC 480 are consistent with a disk inclined
by $\sim 30^{\circ}$, at a position angle of $\sim 150^{\circ}$
(Figure \ref{fig:mwc480}).
Specifically, best-fit angular size scales range from 2.0 to 5.0 mas,
inclinations range from $24^{\circ}$ to $32^{\circ}$, and position angles
are between $127^{\circ}$ and $155^{\circ}$ (Tables 
\ref{tab:uniforms}--\ref{tab:ddn}).  Inclined fits give $\chi_r^2 \sim 1.4$,
significantly lower then the $\chi_r^2 \sim 5.0$ values for face-on models.
A binary model can be ruled out with a high degree of confidence
($\chi_r^2 = 13.5$).

\subsubsection{MWC 758 \label{sec:mwc758}}
The angular size scales for best-fit disk
models range from 1.5 to 4.2 mas. Disk inclinations are between
$33^{\circ}$ and $37^{\circ}$, and position angles vary from
$127^{\circ}$ to $130^{\circ}$ (Tables \ref{tab:uniforms}--\ref{tab:ddn}).
For this source, all parameters are firmly constrained because we
obtained data on three baselines.
An inclined disk model clearly fits
the data better than a face-on model (Figure \ref{fig:mwc758}; 
$\chi_r^2<1$ for inclined models,
compared to $\chi_r^2>3$ for face-on models).
A binary model provides a poor fit to the data, with
$\chi_r^2=5.9$.

\subsubsection{CQ Tau \label{sec:cqtau}}
The best-fit angular size scales for CQ Tau are between 1.5 and 4.4 mas.
The best-fit inclination is $48^{\circ}$, and position angles
range from $104^{\circ}$ to $106^{\circ}$ 
(Tables \ref{tab:uniforms}--\ref{tab:ddn}).
All parameters are firmly constrained because we
obtained data on three baselines.  Inclined model fits give
$\chi_r^2 <1$, while face-on fits have much higher $\chi_r^2$ values,
$\ga 5$.  A binary model seems unlikely, with $\chi_r^2=4.4$.

\subsubsection{T Ori and MWC 120 \label{sec:tori}}
Since we obtained data on only one baseline for T Ori and MWC 120, we are
unable to constrain inclinations or position angles.  The best-fit 
angular size scales for T Ori range from 1.1 to 2.7 mas, and the values
for MWC 120 are between 2.1 and 4.9 mas
(Tables \ref{tab:uniforms}--\ref{tab:ddn}).
The limited uv coverage for these sources does not allow us to rule
out (or constrain the parameters of) binary models.

\subsubsection{MWC 297 \label{sec:mwc297}}
This source is extremely resolved, and we are only able to place
lower limits on the angular size scales (corresponding to upper limits on
the visibilities).  We can neither constrain the
exact geometry of the emission, nor rule out a binary model.  For
the face-on disk models discussed above, we find lower-limits on
angular size scales of
2.1 to 5.0 mas.  As we discuss below, this source has also been
resolved by the IOTA interferometer (MST; \S \ref{sec:pti+iota}),
allowing more accurate constraints on angular size.
The puffed-up inner disk model does not fit the
visibility data for this source, since the flared disk is quite
hot even at large radii due to the hotter central star, which leads
to an inner disk radius much larger than allowed by the visibility data.

\subsubsection{VV Ser \label{sec:vvser}}
\begin{figure}
\plottwo{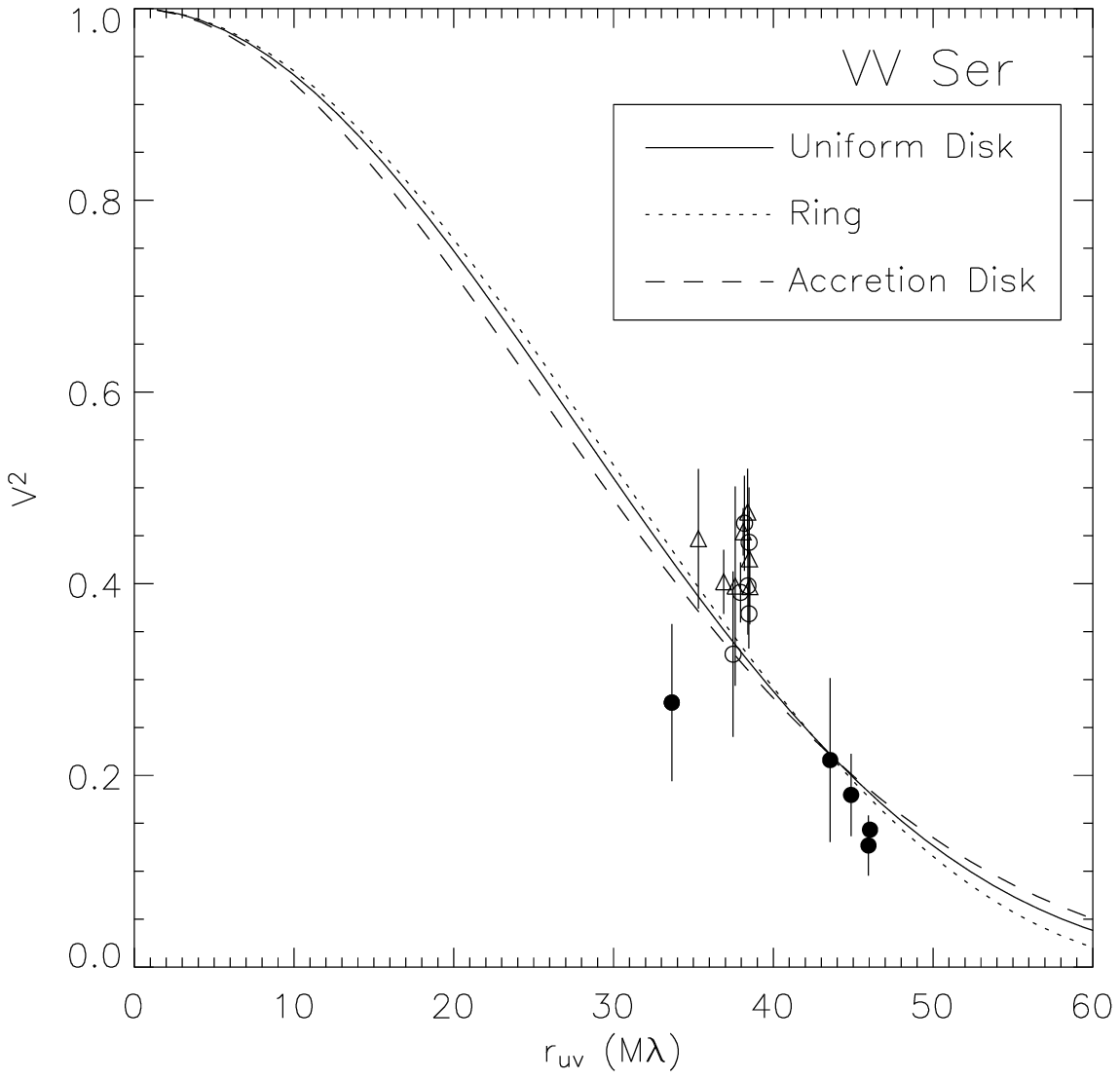}{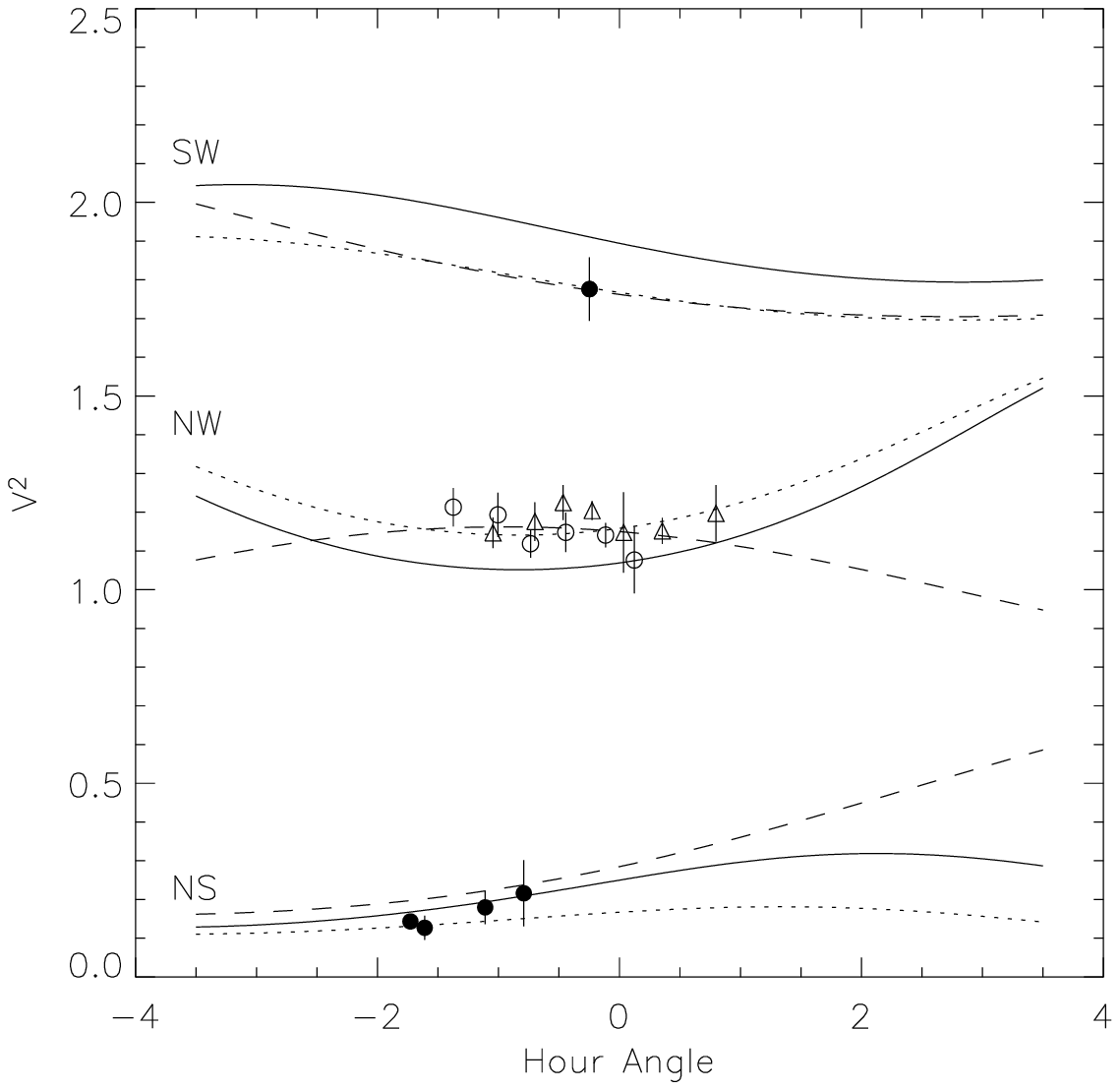}
\caption{$V^2$ PTI data for VV Ser, as a function of 
$r_{\rm uv}$ (left panel) and hour angle (right panel).  
Symbols and models are plotted as in Figure \ref{fig:mwc758}.
Inclined disk or binary models provide the best fit to the 
data for this source.
\label{fig:vvser}}
\end{figure}

The angular size scales for best-fit disk
models range from 1.5 to 4.5 mas, disk inclinations are between
$42^{\circ}$ and $47^{\circ}$, and position angles range from
$165^{\circ}$ to $173^{\circ}$ (Tables \ref{tab:uniforms}--\ref{tab:ddn}).
An inclined disk model clearly fits
the VV Ser data better than a face-on model (Figure \ref{fig:vvser}).
Inclined model fits give $\chi_r^2 < 1$, while face-on model fits
have $\chi_r^2 > 5$ (Table \ref{tab:uniforms}--\ref{tab:ddn}). 
However, as indicated in Figure \ref{fig:uv},
the $u-v$ coverage for this object is somewhat sparse, and 
a binary model cannot be
ruled out ($\chi_r^2 \sim 0.75$; Figure \ref{fig:vvser}).  

While the fitted parameters 
are consistent (within the uncertainties) with those
listed in ELAHS, the inclination determined here is significantly lower
as a result of new data on an additional baseline.  However, we note that
we only have one data point on the SW baseline, which makes it
difficult to estimate a true error bar, and thus the
uncertainties on the fitted parameters may be larger than the
statistical uncertainties quoted in Tables \ref{tab:uniforms}--\ref{tab:ddn}.
We emphasize that if the SW data point is excluded from the fit, the
inclination is closer to edge-on ($80^{\circ}-89^{\circ}$; ELAHS).

\subsubsection{V1295 Aql \label{sec:v1295aql}}
The visibilities for V1295 Aql appear consistent with a disk that is close
to face-on, although we do not rule out a significantly non-zero inclination.
The angular size scales of best-fit models are between 2.4 and 5.6 mas,
and inclinations range from $12^{\circ}$ to $50^{\circ}$.  
While there is no discernible difference in 
the $\chi_r^2$ values for face-on and
edge-on disks, the $u-v$ coverage for this source is sparse
(Figure \ref{fig:uv}), 
and we cannot obtain an accurate estimate of the inclination.
We also cannot rule out a binary model (Figure \ref{fig:v1295aql}).

\subsubsection{V1685 Cyg}
The size scales for V1685 Cyg under the assumptions of various disk models
range from 1.3 to 3.6 mas, the best-fit inclination is
41$^{\circ}$, and the best-fit position angle is 110$^{\circ}$ 
(Tables \ref{tab:uniforms}--\ref{tab:ddn}).
These parameters are consistent with those found in ELAHS, and
we have improved the uncertainties using additional data from
a third baseline.
The visibility data are not fit very well by any model,
although of those considered, inclined disks fit best
($\chi_r^2 \sim 3.9$; Figure \ref{fig:v1685cyg}).
$\chi_r^2 = 15$ for a binary model, making this an unlikely choice.
The flared disk model with a puffed-up inner wall also does not fit the
visibility data for this source, since the flared disk is quite
hot even at large radii due to the hotter central star, which leads
to an inner disk radius much larger than allowed by the visibility data.

Since none of the models provide very good fits to the data,
we attempted to fit several more complex models, including
an inclined disk+point source, binary face-on uniform disks, and
a model consisting of three point sources.  
These models reduce the $\chi_r^2$ (to approximately 2.5), but still
do not appear to accurately fit all of the visibility data.
We speculate that a more complex circumstellar distribution, such as
as a highly non-uniform disk with multiple hot-spots, may be 
necessary to explain the
observations. Complete understanding of this source may have to wait until
multi-baseline interferometers like {\it IOTA-3T}, {\it CHARA}, 
{\it VLTI}, or {\it Keck Interferometer} 
allow synthesis imaging.

\subsubsection{AS 442}
The PTI data for AS 442 generally have low signal-to-noise, and it is 
difficult to distinguish between different models.  We have added 
a third baseline to the dataset used in ELAHS, which allows us
to make estimates of size scales and inclinations (albeit with large
uncertainties).  The size scales for various disk models range from
1.0 to 2.7 mas, the inclination ranges from $46^{\circ}$ to
$48^{\circ}$, and the position angles are between $57^{\circ}$ and
$59^{\circ}$ (Tables \ref{tab:uniforms}--\ref{tab:ddn}).  
While we cannot rule out
a face-on disk model, the $\chi_r^2$ values are somewhat lower for the
inclined disk models: $\sim 0.9$ versus $\sim 1.0$ for face-on models.
We cannot rule out a binary model with these data, and in fact,
the binary model has the lowest $\chi_r^2$ value of all models considered.

\subsubsection{MWC 1080}
The PTI visibilities for MWC 1080 are consistent with a disk inclined
by $\sim 30^{\circ}$ (Figure \ref{fig:mwc1080}).  
The best-fit angular size scales are between 1.7 and 4.1 mas, 
inclination angles range from $28^{\circ}$ to $40^{\circ}$, and 
position angles are between $54^{\circ}$ and $56^{\circ}$.
The $\chi_r^2$ values for inclined models are significantly lower 
than for face-on disk or binary models.  
While the uncertainties
on position angle and inclination
quoted in Tables \ref{tab:uniforms}--\ref{tab:ddn} are somewhat large,
we show below that when IOTA data is included, the
uncertainties are reduced considerably (\S \ref{sec:pti+iota}).
The flared disk model with a puffed-up inner wall does not fit the
visibility data for this source since the flared disk is quite
hot even at large radii due to the hotter central star, which leads
to an inner disk radius much larger than allowed by the visibility data.

\subsection{Comparison with K and H-band IOTA Visibilities 
\label{sec:pti+iota}}
Interferometric observations of AB Aur and 
MWC 1080 at 2.2 $\mu$m have also been obtained with the
21-m and 38-m baselines of the IOTA interferometer \citep{M-G+99,MST01}.
In ELAHS, we combined this 2.2 $\mu$m IOTA data with PTI
data.  However, we misinterpreted the IOTA data
in that analysis (confusing $V$ for $V^2$)
such that the plotted IOTA visibilities appeared closer
to unity than they actually are.  We have rectified that error here.
MST also obtained 1.6 $\mu$m H-band
visibilities for AB Aur, T Ori, MWC 297, V1295 Aql, V1685 Cyg, and MWC 1080.
Based on discussion with R. Millan-Gabet,  
we assign an uncertainty to each IOTA visibility given by the standard
deviation of all data obtained for a given source with a given baseline.

\begin{figure}
\plotone{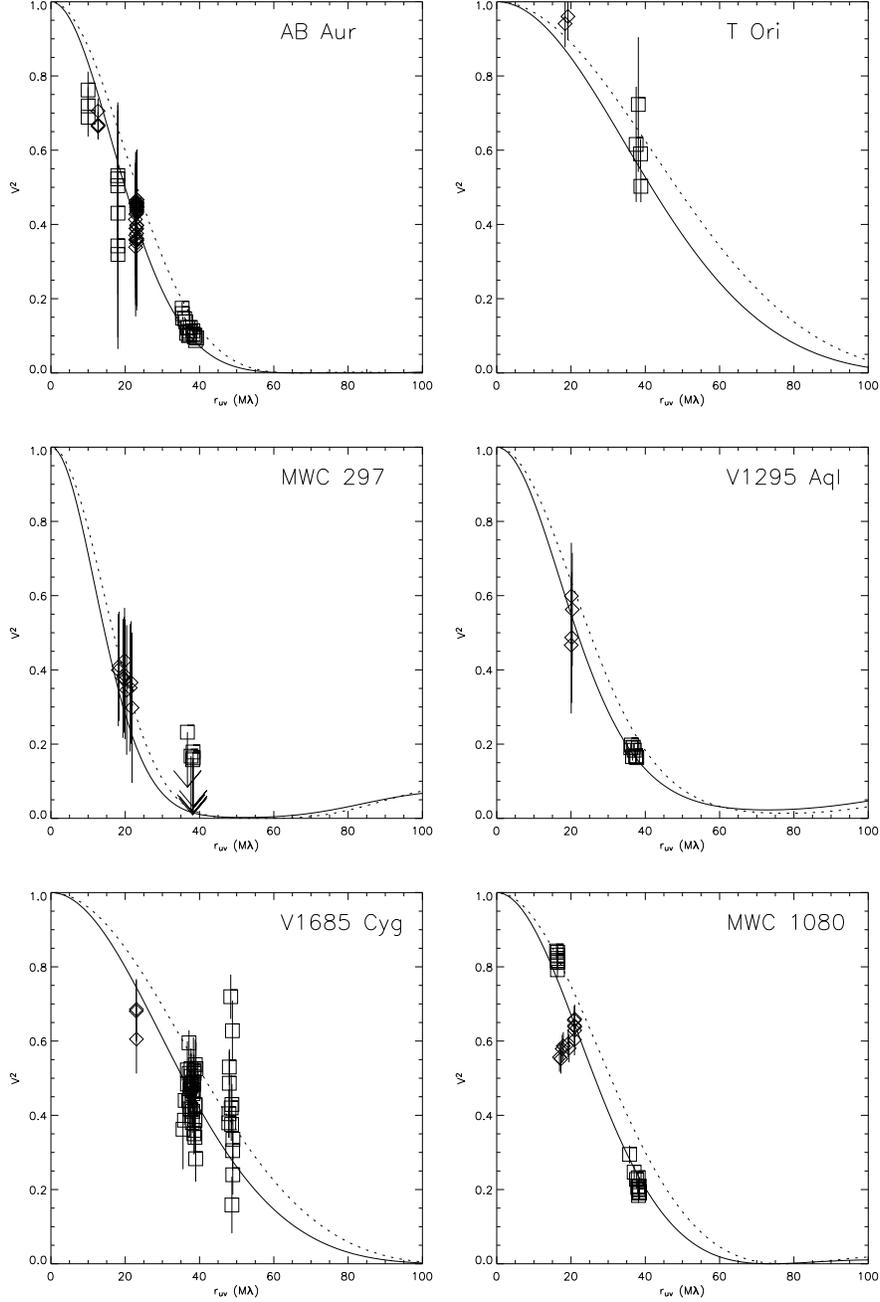}
\caption{PTI and IOTA data for AB Aur, T Ori, MWC 297, V1295 Aql,
V1685 Cyg, and MWC 1080, as a function of $r_{\rm uv} = (u^2 + v^2)^{1/2}$.  
PTI data have $r_{\rm uv}>30$ M$\lambda$, and IOTA data have
$r_{\rm uv}<30$ M$\lambda$. 
K-band data are represented by squares, and H-band data are plotted with
diamonds. 
The best fit of a face-on flat accretion disk model 
to the combined dataset is also
plotted, where the predicted K-band visibilities are indicated by a
solid line and the predicted H-band visibilities are indicated by a dotted 
line.
\label{fig:pti+iota}}
\end{figure}

We compare the visibilities measured by PTI and IOTA for each source by 
fitting a flat accretion disk model (\S \ref{sec:acc})
to the combined H+K-band dataset\footnote{Since we are interested primarily in
the comparison between PTI and IOTA data, the choice of model is
not important; we choose the flat accretion disk model because it is
computationally simple compared to the puffed-up inner disk model.}.
The best-fit face-on flat accretion disk models are plotted in Figure 
\ref{fig:pti+iota}, and the fitted parameters for face-on and
inclined disk models are listed in Table \ref{tab:pti+iota}. Examination
of Figure \ref{fig:pti+iota} and comparison of Tables \ref{tab:acc} and
\ref{tab:pti+iota} show that the PTI and IOTA data are consistent, and
that the combined dataset gives results compatible with those derived
from the PTI data alone.  

For some sources,
there may be a slight trend in the IOTA data toward
lower visibilities than the PTI data: this difference is most 
pronounced for AB Aur, where both the PTI and 
IOTA data sets are consistent with
a nearly face-on disk model, but the IOTA data imply a size that
is $\sim 20\%$ larger.
We speculate that this discrepancy may be due to the $3''$ field of
view of the IOTA interferometer, which is significantly larger than the 
$1''$ field of view
of PTI.  Scattered or thermal emission from dust on large scales ($\ga 1''$)
might contribute incoherent emission to the IOTA visibilities that is 
not present in the PTI data, leading to a larger measured size 
(see \S \ref{sec:extended}).  Including a spatially uniform incoherent 
component via Equation \ref{eq:vscatt}, 
an incoherent flux of $0.6 F_{\ast}=1$ Jy leads to a 
measured size in the IOTA data equal to that measured by PTI.  
Near-IR and optical imaging
have revealed scattered light structures up to $4''-9''$ away from AB Aur 
\citep{FUKAGAWA+04,GRADY+99}, providing some evidence for 
(possibly K-band emitting) dust at large radii.  

The sizes determined when uniform, incoherent emission is included in
the model are $\la 20\%$ larger than the values determined from
the IOTA data when incoherent emission is ignored.  For the PTI data,
this discrepancy is $<1\%$.
Thus, while there may be some uncertainty in the disk sizes determined from 
the IOTA data, the smaller field of view of PTI should lead to uncertainties
of less than a few percent. However, the exact brightness profile of the 
extended emission is unknown, and thus we cannot quantify precisely 
the magnitude of the uncertainty for the PTI data.
Future measurements with interferometers possessing even smaller
fields of view (e.g., {\it Keck Interferometer} with a 50 mas field of view),
when combined with the data discussed here,
will constrain further the effects of extended emission on the
measured visibilities.

For MWC 1080, while the K-band visibilities from IOTA and PTI are compatible
(the fit to the combined dataset gives very similar parameters to those
determined in \S \ref{sec:res}), the H-band data appears to be inconsistent
with a disk model.  This discrepancy was noted by MST, who speculated that the
cause might be a resolved calibrator used in the IOTA observations, or 
possibly increased scattering at shorter wavelengths.  Since the K-band
data are consistent with a disk model while the H-band data are not, we
favor the latter interpretation.

The H-band IOTA visibilities for MWC 297 lead to a size of 3.38 mas. Since our
PTI data presented in \S \ref{sec:res} yielded only an upper limit for
the size of this source,  we will adopt the size measured by IOTA in
the further analysis presented below.

\subsection{Binaries \label{sec:binaries}}
There is always the possibility that the visibilities for
some of the observed HAEBE sources may be (partially) due to
close companions.  For AB Aur, MWC 480, MWC 758, CQ Tau, V1685 Cyg,
and MWC 1080, we can rule out binary models (with separations $\ga 1$ mas) 
with a high degree of confidence based on the near-IR visibility data
(Table \ref{tab:binary}).  The stability of the visibilities
over a long time baseline ($>1$ year) for AB Aur and  V1685 Cyg
supports this conclusion.
However, MWC 1080 is known to be an eclipsing binary with a period of 
$P \approx 2.9$ days \citep{SHEVCHENKO+94,HS99},
and T Ori is an eclipsing spectroscopic binary
with a period of approximately 14 days \citep{SV94}.
These orbital separations are much too small
to be detected by PTI, and the observed visibilities for these sources
are thus probably due to circum-{\it binary} disks.
As yet, the binarity status of MWC 120, MWC 297, V1295 Aql,
AS 442, and VV Ser remains uncertain based on our visibility data.
Radial velocity variations of spectral lines in AS 442
have been attributed to a binary with 
$P \approx 64$ days and $e \approx 0.2$, while a lack of spectral
line variation in V1295 Aql suggests a single source \citep{CL99}.

\subsection{Spectral Energy Distributions \label{sec:seds}}
\begin{figure}
\plotone{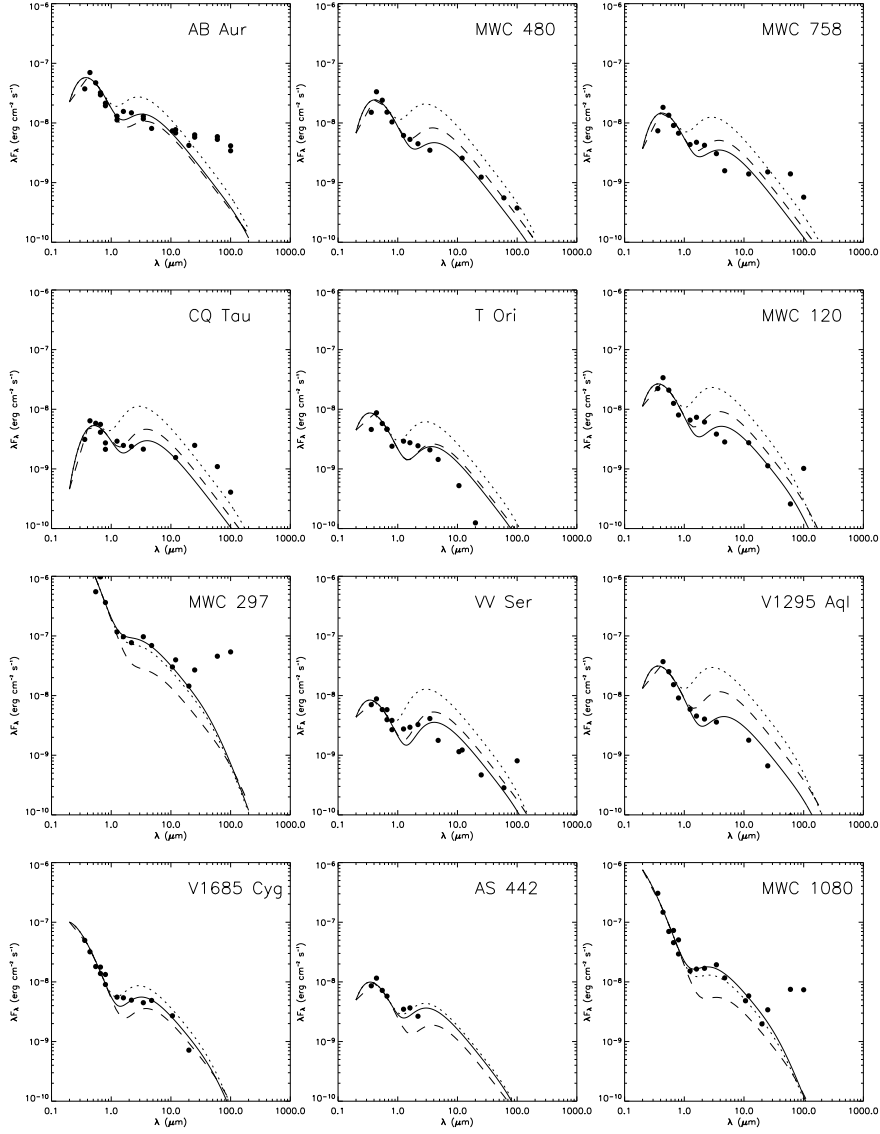}
\caption{Measured SEDs for our sample compiled from the literature 
and new data (points) and predicted SEDs for
geometrically flat accretion disk models (\S \ref{sec:acc}).  
For each source, we 
plot the predicted SED for $T_{\rm in}=1500$ K (dashed lines), 2000 K (dotted
lines), and a value we determined that gives the best fit to the 
near-IR data (solid lines).  The best-fit
$T_{\rm in}$ values are listed in Table \ref{tab:seds}.
These models successfully reproduce the SEDs of early-type sources
(MWC 297, V1685 Cyg, and MWC 1080),
while they are 
less successful than the flared disk models with puffed-up inner 
walls (Figure \ref{fig:ddn-seds}) for later-type sources.
\label{fig:hsvk-seds}}
\end{figure}

\begin{figure}
\plotone{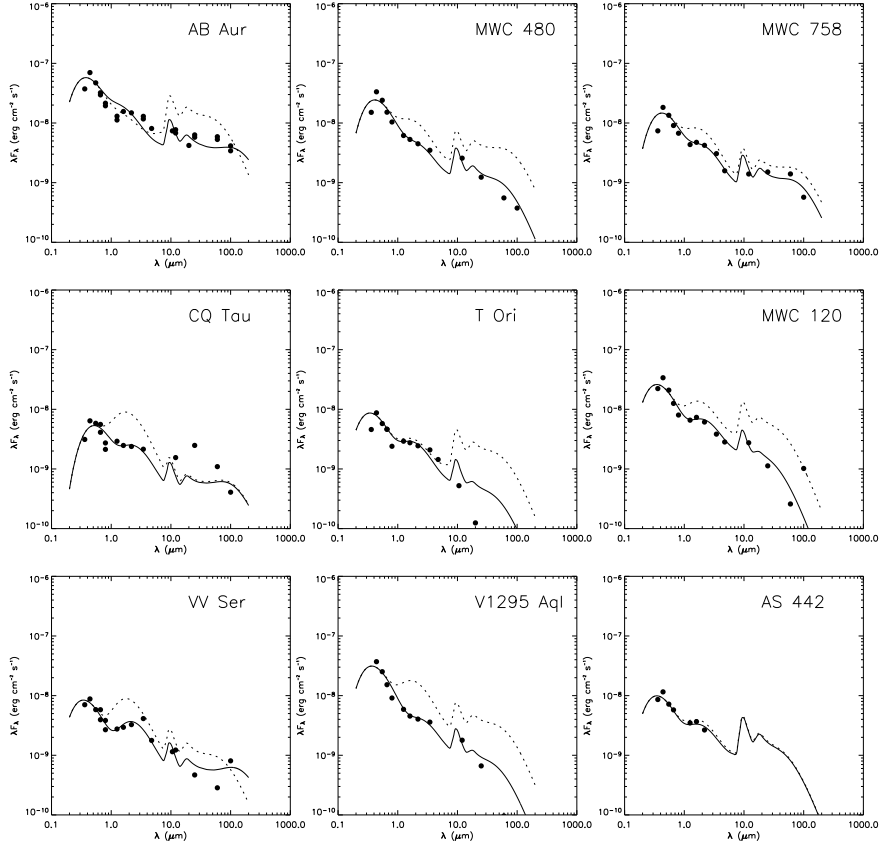}
\caption{Measured SEDs for our sample compiled from the literature 
and new data (points) and predicted SEDs for
flared disk models with puffed-up inner walls (\S \ref{sec:ddn}).  
For each source, we 
plot the predicted SED for the fiducial model ($T_{\rm in}=2000$ K,
$\xi=0.28$, $R_{\rm out}=100$ AU; dotted lines) and for a model
where the values of $T_{\rm in}$, $\xi$, and $R_{\rm out}$ are chosen
to give the best fits to the data (solid lines).  The best-fit parameter
values are listed in Table \ref{tab:seds}.
\label{fig:ddn-seds}}
\end{figure}

The inner disk sizes and inclinations determined from our near-IR 
interferometry data directly constrain common models of the spectral
energy distributions (SEDs) for HAEBEs.  The derived inner radii
constrain the structure
of the inner disks, while the inclination estimates provide constraints
on the structure of the entire disks (assuming the disks are not
significantly warped; see \S \ref{sec:mm}).
For the geometrically flat accretion disk  model discussed in \S \ref{sec:acc}
and the flared disk model with a puffed-up inner wall discussed in
\S \ref{sec:ddn}, we compute the SEDs for our best-fit inner
disk parameters (Tables \ref{tab:acc} and \ref{tab:ddn}; Table 
\ref{tab:pti+iota} for MWC 297). For T Ori, 
MWC 120, and MWC 297,
which do not have inclination estimates, we assume an inclination of zero.  
The model SEDs are plotted in 
Figures \ref{fig:hsvk-seds} and \ref{fig:ddn-seds}.

We compare these predicted SEDs to 
actual measurements compiled from new data (\S \ref{sec:ao}) and
the literature \citep{HS99,HILLENBRAND+92,DEWINTER+01,MBW98,
OUDMAIJER+92,VIEIRA+03}. Although photometry from the literature 
often lacks error bars, typical uncertainties are $\sim 0.05$--0.1 mag for
wavelengths $<10$ $\mu$m and $\sim 10\%$ at longer wavelengths.
Source variability may lead to additional errors since photometric
observations (at different wavelengths) often span several years.
Since we cannot quantify the uncertainties, we do not include them
in our analysis of the SEDs.

For the geometrically flat disk model, the most 
important parameter beside $R_{\rm in}$ and inclination 
is $T_{\rm in}$ (Equation \ref{eq:acc-flux}). As discussed in 
\S \ref{sec:acc}, the fitted inner radius depends very slightly on
$T_{\rm in}$. However, as illustrated by Figure \ref{fig:hsvk-seds}, which
shows the SEDs predicted for different values of $T_{\rm in}$, this 
parameter has
a significant effect on the SED.  The other free parameter in the flat
disk model is $R_{\rm out}$, which is relatively unimportant since
most of the flux at wavelengths $<1$ mm is generated in 
the inner regions of the disk ($<50$ AU).
We vary $T_{\rm in}$ between 1000 K and 2500 K (in increments of 10 K)
and use a least squares technique to determine 
the value that provides the best fit to the measured SED (considering
only the SED long-ward of 1 $\mu$m, since we are interested in the
circumstellar emission).
The best-fit $R_{\rm in}$ and  $T_{\rm in}$ 
values for this model are listed in Table \ref{tab:seds}.

For the flared disk model with a puffed-up inner wall, several parameters
may affect the predicted SED: $T_{\rm in}$, 
$R_{\rm out}$, $\Sigma(R)$, $\kappa(\nu)$, 
and the flaring index, $\xi$ 
(Equations \ref{eq:ddn-fsurf}, \ref{eq:ddn-fint}, and 
\ref{eq:ddn-frim}).  
For simplicity, we will retain our initial assumptions
about $\Sigma$ and $\kappa$, and attempt to find the values of 
$T_{\rm in}$, $\xi$, and $R_{\rm out}$ that give the best fit of the 
model to the observed SEDs (again, considering the SED only long-ward of 
1 $\mu$m).  Specifically, we use a least squares 
fitting method, varying $T_{\rm in}$ from 1000 to 2500 K in 10 K increments,
$\xi$ from 0.10 to 0.28 in increments of 0.02, and  
$R_{\rm out}$ from 30 to 400 AU in 10 AU increments.
We note that varying $\xi$ qualitatively corresponds to changing
the overall flaring of the disk, and while slight modifications
to Equation \ref{eq:ddn-alpha} may be necessary when $\xi$ is changed,
we ignore those here.
Since we can rule out the flared, puffed-up inner disk model for 
MWC 297, V1685 Cyg, and MWC 1080 on the basis of the near-IR visibility 
data alone (\S \ref{sec:models}), 
we do not attempt to fit the SEDs for these sources.
The best-fit $R_{\rm in}$, $T_{\rm in}$, $\xi$, and $R_{\rm out}$ 
values for the remaining sources are listed in Table \ref{tab:seds}.

From Figure \ref{fig:hsvk-seds}, we see that with the parameters derived
from the near-IR interferometry (inner radius and inclination), 
geometrically flat accretion disk models
can fit the SEDs reasonably well for some sources. 
The best fits are
achieved for MWC 297, V1685 Cyg, and MWC 1080, the sources with the
earliest spectral types in our sample.  The far-IR photometry
for these objects appears somewhat inconsistent with the models, perhaps 
due to tenuous dust halos, or possibly due to source
confusion in the large IRAS beams (not unlikely given that these
higher-mass stars are found in small stellar clusters).
For these early-type sources, we also find that the near-IR visibility
data is completely inconsistent with the predictions of flared disk
models with puffed-up inner walls.  

For the other nine sources in our sample, which all have 
spectral types later than B9, the SEDs
are generally fit well by flared disk models with
puffed-up inner walls (Figure \ref{fig:ddn-seds}).  
However, while the puffed-up inner disk wall generally fits the near-IR data
well, for CQ Tau and T Ori the model does not 
fit the long-wavelength data.  
Although this disagreement may be lessened for T Ori
using a non-zero inclination (we assumed a face-on disk since no 
inclination estimate is available from our near-IR interferometry),
the outer disk structure of CQ Tau is inconsistent with the model.
Moreover, many of the sources in our sample require flaring angles smaller
than those predicted by standard flared-disk models (i.e., $\xi<2/7$),
which could support the contention of DDN that 
shadowing by the inner rim plays a prominent role in the outer disk structure 
of HAEBEs.

\begin{figure}
\plotone{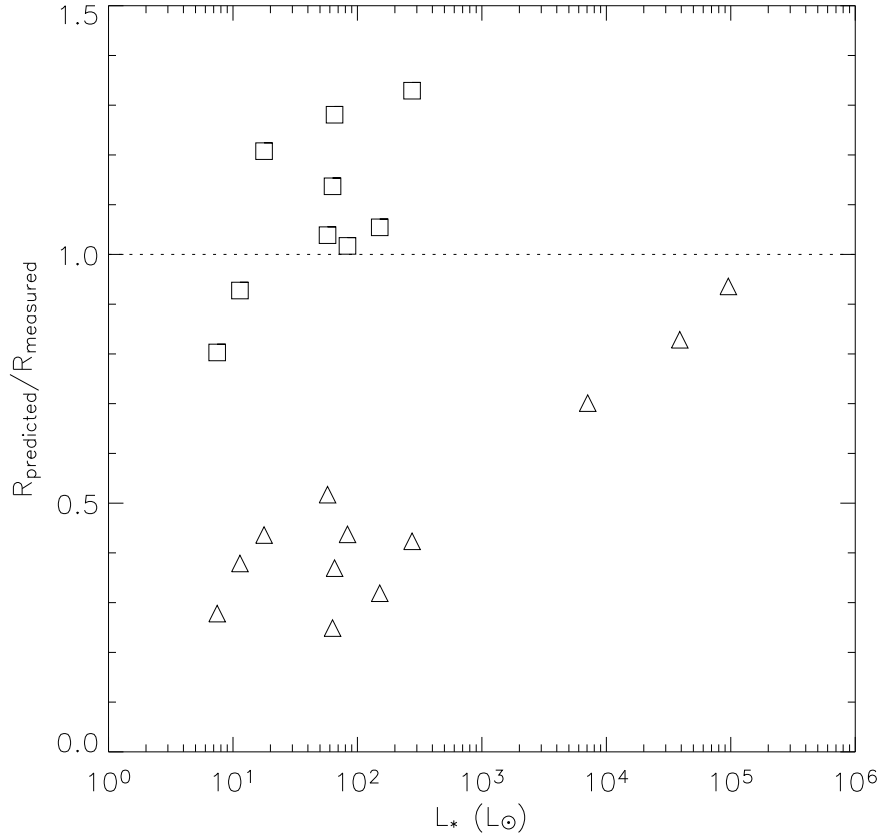}
\caption{The ratio of measured to predicted inner disk sizes, as a function
of stellar luminosity.  The measured sizes are determined from
near-IR interferometry for the geometrically-flat disk model (triangles;
\S \ref{sec:acc}) and the puffed-up inner rim model 
(squares; \S \ref{sec:ddn}), 
and the predicted sizes for each model are computed using the
derived stellar parameters to determine the location in the disk where
$T=T_{\rm in}$.  The ratio of measured to predicted size is independent 
of distance (see ELAHS).  This figure illustrates that for later-type
sources ($L_{\ast} < 10^3$ $L_{\odot}$), the puffed-up inner disk
models predict inner disk sizes within a few tens of percent of the
measured sizes, while the flat disk models predict sizes that are off
by a factor of $\sim 2$.  In contrast, the puffed-up inner disk models
predict sizes much larger than measured for the early-type objects
(not plotted here, since the disagreement is off the scale of the plot), 
while the flat disk model predicts sizes within
$\sim 30\%$ of the measured values.
\label{fig:size-lum}}
\end{figure}

Motivated by Monnier \& Millan-Gabet (2002), 
we explore the apparently different disk structure for early and late-type
HAEBEs from a different perspective in Figure \ref{fig:size-lum},
where we plot the ratio of predicted to measured inner disk size 
as a function of stellar luminosity for the two different
disk models.  The predicted inner disk size is computed by assuming the
disk is heated only by stellar radiation (i.e., no accretion), 
and using the structure equations for flat or puffed-up 
inner disks (see ELAHS) 
to determine the radius where $T=T_{\rm in}$.  
For each source, $T_{\rm in}$ is the value determined from the 
SED (Table \ref{tab:seds}).  The measured size
is calculated from the angular size determined for either
a flat inner disk model (\S \ref{sec:acc}) or a puffed-up inner disk wall
model (\S \ref{sec:ddn}) and the distance assumed
in Table \ref{tab:sources}.  Since the predicted and measured sizes
are both directly proportional to distance (ELAHS), the ratio is independent of
the assumed distance.  
Thus, Figure \ref{fig:size-lum}
is effectively testing whether the inner disk sizes determined from the
near-IR interferometric data are consistent with the disk temperatures
implied by the stellar parameters.  While stellar variability may 
complicate the interpretation of this diagram, the general trends should
be unaffected.

Figure \ref{fig:size-lum} shows that for later-type sources,
the predicted inner disk sizes are within a few tens of percent of the
measured values for puffed-up inner disk models, while flat disk models
are off by a factor of approximately 2.  In contrast, the predictions of
flat models are reasonably close to measured values for early-type sources,
while the puffed-up inner disk models predict inner disk sizes much larger
(factors $\ga 5$) than allowed by the data.
Including accretion luminosity in flat disk models will lead to 
warmer inner disk temperatures and thus larger predicted inner radii,
in better agreement with the measurements.  While this argues for flat
disk models with accretion in early-type Herbig Be stars, the SEDs demonstrate
that puffed-up inner disk models still provide superior fits for
the later-type sources.

\section{Discussion \label{sec:disc}}
Our new PTI results strengthen the arguments supporting the
existence of circumstellar disks around HAEBE stars presented in 
\S \ref{sec:intro}. 
Resolved, small-scale ($\sim 1$ AU)
distributions of dust are found in all observed sources except 
HD 141569 and HD 158352, and the 
non-symmetric intensity distributions of best-fit models for most objects
provide support for inclined disk geometries.  

\subsection{Unresolved Sources}
HD 141569 and HD 158352 show no evidence of near-IR emission in excess
of that expected for the stellar photospheres, and thus it is not
surprising that we did not resolve any circumstellar emission with
PTI.  Rather, our PTI observations imply uniform disk
radii $\la 10$ R$_{\odot}$, consistent with the near-IR
emission arising in the stellar photospheres for these sources.
Although HD 141569 is surrounded by a circumstellar disk,
it appears to have a central gap extending out to
$\sim 17-30$ AU \citep{BRITTAIN+03,MARSH+02,WEINBERGER+99,
AUGEREAU+99,SS96}, which explains the lack of near-IR excess emission.
Moreover, millimeter observations imply a very low dust mass,
suggesting that this system is more evolved than other members
of the HAEBE class \citep{SYLVESTER+01}.
Although HD 158352 was listed as a candidate HAEBE 
by Th\'{e}, de Winter, \& P\'{e}rez. (1994), 
recent observations suggest that 
it may in fact be a more evolved source, such as a shell star or
Vega-like object \citep{GRADY+96,ERITSYAN+02}.  Thus, it appears that
HD 141569 and HD 158352 are probably more evolved than most HAEBE sources
(including the remainder of our sample), which may explain the
lack of near-IR circumstellar emission for these objects.
  
\subsection{Disk Inclinations}
The inclination estimates determined from near-IR interferometric data
are generally compatible with inclinations inferred from other observations.
Our interferometric measurements show that the resolved
circumstellar material around 
MWC 480, MWC 758, CQ Tau, VV Ser, V1685 Cyg, AS 442,
and MWC 1080 is significantly inclined.
Large-amplitude flux and color variability in CQ Tau ($\Delta V \sim 2$ mag), 
VV Ser ($\Delta V \sim 2$ mag), V1685 Cyg ($\Delta V \sim 1$ mag), and AS 442
($\Delta V \sim 1$ mag), which has been attributed to variable obscuration 
from clumps of dust orbiting in inclined circumstellar disks \citep{HS99}, 
provides support for this high inclination distribution.  Furthermore,
high inclinations are suggested by resolved mm emission for MWC 480, 
MWC 758, and CQ Tau (\S \ref{sec:mm}), by optical polarization
measurements for CQ Tau \citep{NW00}, and by ro-vibrational CO emission 
for MWC 480, MWC 758, and VV Ser \citep{BB04}.  However, 
the broad, double-peaked line profile of ro-vibrational CO emission from
VV Ser suggests an inner disk inclination
higher than the $\sim 45^{\circ}$ estimate
presented here. As described in \S \ref{sec:vvser}, the lower 
inclination estimate depends on a single data point, and if this point is 
excluded, we obtain an inclination of $\sim 85^{\circ}$, 
consistent with the CO observations.

The PTI data for AB Aur are consistent with a circumstellar distribution
that is within $20^{\circ}$ of face-on.  
Other data for AB Aur support this contention, including near-IR
interferometric results \citep[][ \S \ref{sec:pti+iota}]{MST01}, 
ro-vibrational CO emission \citep{BB04},
resolved millimeter emission (\S \ref{sec:mm}), 
scattered optical and near-IR light \citep{GRADY+99,FUKAGAWA+04}, 
and small amplitude
optical variability \citep[$\Delta V \sim 0.25$;][]{HS99}.
We also note that the inner disk size for AB Aur determined above
is consistent with recent observations of ro-vibrational emission
from hot CO gas, which show that the CO lies near to, but somewhat
behind the inner disk boundary. Moreover, the CO emission is
found only in the lowest vibrational state, suggesting that the hot gas
is shielded from stellar UV radiation, perhaps by a puffed-up inner
disk wall \citep{BRITTAIN+03,BB04}.

While the V1295 Aql data seems consistent with a nearly face-on 
circumstellar distribution, we cannot rule out a high inclination value for
this source based on the PTI data alone.  The data for T Ori, MWC 120,
and MWC 297 are insufficient to constrain the inclinations.

\subsection{Inner versus Outer Disk Structure \label{sec:mm}}
Our PTI results probe dust in the inner ($< 1$ AU) disk, while
millimeter interferometric observations probe dust and gas in the outer
($\ga 100$ AU) disks of HAEBE stars.  Comparison of these observations 
enables constraints on disk warping.  Of our sample, inclinations are
available from millimeter CO observations for
AB Aur, CQ Tau, MWC 480, and MWC 758
\citep{MS97,MKS97,MS00,SDG01,TESTI+03,CES03}.

As discussed in earlier papers (MST; ELAHS), 
the inclination for AB Aur estimated from the aspect ratio of resolved
millimeter CO emission, $i=76^{\circ}$ \citep{MS97},
is inconsistent with the lower inclinations 
from near-IR interferometry ($i \la 20^{\circ}$; ELAHS)
and modeling of scattered light emission 
\citep[$i < 30^{\circ}-45^{\circ}$;][]{FUKAGAWA+04,GRADY+99}.  
This suggests that the disk around AB Aur may be significantly warped,
which is difficult to explain theoretically.  
However, more detailed kinematic modeling of millimeter
observations with higher angular resolution and sensitivity find an outer 
disk around AB Aur inclined by $\la 30^{\circ}$ \citep{CES03,NATTA+01}, 
compatible with our near-IR results.  

For CQ Tau, the aspect ratio estimated from VLA data at 7mm
implies an inclination angle \citep[$\sim 70^{\circ}$;][]{TESTI+03} somewhat 
larger than that determined from near-IR interferometry 
($\sim 48^{\circ}$; \S \ref{sec:cqtau}).  Kinematic modeling of
more sensitive millimeter observations shows that the outer disk is actually 
inclined by $\sim 45^{\circ}$ \citep{CES03}, consistent with our PTI results.

For MWC 480, there is some variation in the outer disk geometry based
on different observations.  A Keplerian model fit to one set of 
millimeter CO observations yields an
inclination and position angle of $\sim 30^{\circ}$ and
$157^{\circ} \pm 4^{\circ}$, 
respectively \citep{MKS97}, while another gives  $38^{\circ} \pm 
1^{\circ}$ and $148^{\circ} \pm 1^{\circ}$ \citep{SDG01}.  
The millimeter continuum, which traces cool dust, gives yet another estimate
of geometry; 
$i=26^{\circ} \pm 7^{\circ}$, ${\rm pa}=170^{\circ} \pm 11^{\circ}$ 
\citep{SDG01}.  We compare these with estimates of the inner disk geometry
from our PTI results: $i=24^{\circ}-32^{\circ}$, 
${\rm pa} = 149^{\circ}-156^{\circ}$.
Since the outer disk geometry estimates vary considerably, it is difficult
to estimate the true uncertainty in inclination.  However, it seems
reasonable to say that the various observations are consistent
with no offset at all, and that
there is at most a difference of $15^{\circ}$ between
the inner and outer disk. While there is an intriguing hint that
the dust may be somewhat less inclined than the gas, 
we cannot verify this possibility with the current data.

Finally, for MWC 758, the aspect ratio of millimeter CO emission implies a 
disk inclined by $46^{\circ}$ at a position angle of 
$116^{\circ}$ \citep{MS97}.  For comparison, the PTI measurements imply
an inclination of $33^{\circ}-37^{\circ}$ and a position angle of 
$127^{\circ}-130^{\circ}$.
There is some spread between these estimates, but again, it seems
that there is a difference of $\la 10^{\circ}$ between the inner and
outer disk inclinations.

For the 4 sources in our sample that have inner and outer disk
geometry measurements, there is little evidence for inner and outer
disk mis-alignments of more than a few degrees. Thus, the inner disk geometries
derived from our PTI data likely describe the structure of the entire disks.
While we do not rule out small
warps (up to $\sim 10$ degrees), such as those expected from resonant
interactions with giant planets or magnetic warping
\citep[e.g.,][]{MOUILLET+97,LAI99},
our results argue against significant perturbation due to massive 
companions \citep[e.g.,][]{BATE+00}, 
consistent with the lack of binaries in our visibility and
adaptive optics data.

\subsection{Vertical Disk Structure \label{sec:flatvsflared}}
Figures \ref{fig:hsvk-seds} and \ref{fig:ddn-seds}
demonstrate that flared,
puffed-up inner disk models tend to fit the SEDs better than
geometrically flat accretion disk models for later-type HAEBE sources,
while flat disk models fit the visibility and 
SED data well for early-type Herbig Be sources.  Moreover, Figure
\ref{fig:size-lum} clearly demonstrates that puffed-up inner disk
models more accurately predict the inner radii of late-type HAEBEs,
while geometrically flat disk models are far more accurate for
early-type sources.

We have already suggested (ELAHS, see also Vink et al. 2002)
that the apparent difference between inner disk geometries for 
early and late-type HAEBEs may be due do a transition from 
disk accretion \citep{LBP74} in early-type stars to
magneto-spheric accretion \citep{KOENIGL91}
in later-type stars.  There is abundant evidence
for magneto-spheric accretion in T Tauri stars \citep[][and references
therein]{HARTMANN98}, and it is plausible that
the same accretion mechanism would apply to the more massive Herbig Ae
stars.  Magneto-spheric accretion provides a mechanism for truncating
the disks around HAEBE stars, which is necessary to obtain puffed-up
inner disk walls (disk truncation allows direct irradiation of the
inner disk regions, leading to higher temperatures and thus larger
scale heights).
In contrast, for earlier spectral types, such as the Herbig Be
stars, higher accretion rates and/or weaker stellar magnetic fields
may allow the accretion flow to overwhelm the magnetic field,
leading to disk accretion down to the stellar surface.

\section{Summary \label{sec:conc}}
Our new 2.2 $\mu$m observations of 14 HAEBE sources 
have the best $u-v$ coverage of any near-IR interferometric
observations of young stellar objects to date. As a result, we
accurately constrain the sizes and basic geometries of the
material around HAEBEs, providing strong evidence for inner 
circumstellar disks.

We determine the angular size scales and orientations predicted
by uniform disk models, Gaussian models, uniform ring models, 
geometrically flat accretion disk models with inner holes, and flared
passive disk models with puffed-up inner rims.  All except two sources
are resolved, with angular sizes ranging from $\sim 1.0-5.8$ mas.
AB Aur appears to be surrounded by a disk inclined by
$\la 20^{\circ}$, while MWC 480, MWC 758, CQ Tau, VV Ser, V1685 Cyg, AS 442,
and MWC 1080 are associated with more highly inclined circumstellar disks
($\sim 30-50^{\circ}$).
We cannot rule out a high inclination for V1295 Aql, although the data
can be explained by a face-on disk model.
For T Ori, MWC 120, and MWC 297, the $u-v$ coverage is too sparse
to enable constraints on the inclination of the circumstellar material.
There is little supporting evidence that our data
result from binaries, although we cannot rule out binary models
for T Ori, MWC 120, MWC 297, V1295 Aql, AS 442, or VV Ser.

Comparison of our 2.2 $\mu$m PTI visibilities with 1.6 $\mu$m and
2.2 $\mu$m visibilities from the IOTA interferometer (MST) shows that the
two datasets are consistent, and allows firmer constraints on circumstellar
geometry.  Since IOTA has a field of view three times
larger than PTI, we are also able to constrain the degree to which incoherent
emission from extended dust may bias the measured sizes toward larger
values. The smaller field of view of PTI leads to very small biases,
and the resulting uncertainties in the measured sizes are likely
less than a few percent.

We constrain warping of HAEBE disks by comparing 
our near-IR measurements of the inner disks
with resolved millimeter interferometric measurements of the outer
disk geometries (where available).
Our results for 4 sources indicate that the inner and outer
disks of HAEBEs are not significantly mis-aligned. While this
argues against significant perturbations to the disks, we do not rule
out small warps such as those due to interactions between disks and
slightly non-coplanar giant planets.

Our measurements also enable constraints on the vertical structure
of HAEBE disks.  Using the derived inner disk parameters, we compute 
the SEDs for flat accretion disk models with inner holes and 
flared passive disk models with puffed-up inner walls,
and compare these with measured SEDs for our sample.
Geometrically flat disk models fit the data well for the 
early-type Herbig Be stars in our sample, MWC 297, V1685 Cyg and MWC 1080,
while the flared, puffed-up inner disk models cannot fit the data
for these objects.  In contrast, flared disk models with puffed-up inner
rims provide superior fits to the data for the later-type stars in our sample.
The different inner disk geometries may imply a transition from 
magneto-spheric accretion in
late-type HAEBEs to disk accretion in early-type sources.

\noindent{ }

\noindent{\bf Acknowledgments.} The new near-IR interferometry
data presented in this paper
were obtained at the Palomar Observatory using the Palomar Testbed
Interferometer, which is supported by NASA contracts to the Jet Propulsion
Laboratory.  Science operations with PTI are possible through the efforts 
of the PTI Collaboration 
({\tt http://pti.jpl.nasa.gov/ptimembers.html}) and Kevin Rykoski. 
This research made use of software produced by the Michelson Science Center
at the California Institute of Technology.
This publication makes use of data products from the Two Micron
All Sky Survey, which is a joint project of the University of Massachusetts
and the Infrared Processing and Analysis Center, funded by the National
Aeronautics and Space Administration and the National Science Foundation.
2MASS science data and information services were provided by the Infrared
Science Archive (IRSA) at IPAC.
We thank R. Millan-Gabet for providing us with the IOTA
data, for useful discussion, and for a critical review of the manuscript.  
We are also grateful to S. Metchev 
for his assistance with the adaptive optics observing and data reduction.
J.A.E. is supported by a Michelson Graduate Research Fellowship, and
B.F.L. acknowledges support from a Pappalardo
Fellowship in Physics.

\epsscale{0.7}

\epsscale{0.7}

\epsscale{0.7}

\epsscale{0.7}

\epsscale{0.7}

\clearpage
\begin{deluxetable}{lccccccccccc}
\tabletypesize{\footnotesize}
\rotate
\tablewidth{0pt}
\tablecaption{Observed Sources \label{tab:sources}}
\tablehead{\colhead{Source} & \colhead{Alt. Name}  & \colhead{$\alpha$ (J2000)}
&  \colhead{$\delta$ (J2000)} &
\colhead{$d$ (pc)} & \colhead{Sp.Ty.} & \colhead{$V$}
& \colhead{$J$} & \colhead{$H$} 
& \colhead{$K$} & \colhead{$F^{\dag}_{\rm K,\ast}$ (Jy)} &
\colhead{$F_{\rm K,x}^{\dag}$ (Jy)}}
\startdata
AB Aur & HD 31293 &  ${\rm 04^h55^m45.84^s}$ & 
${\rm +30^{\circ}33'04\rlap{.}''3}$ & 140 & A0pe & 7.1 & 6.4 & 5.3 & 4.5 &
1.65 & 9.11 \\
MWC 480 & HD 31648 & ${\rm 04^h58^m46.27^s}$ & 
${\rm +29^{\circ}50'37\rlap{.}''0}$ & 140 & A2/3ep+sh & 7.7 & 7.1 & 6.4 & 
5.7 & 0.81 & 2.45 \\
MWC 758 & HD 36112 &  ${\rm 05^h30^m27.53^s}$ & 
${\rm +25^{\circ}19'57\rlap{.}''1}$ & 150 & A3e & 8.3 & 7.4 & 6.5 & 5.8 & 
0.53 & 2.54 \\
CQ Tau & HD 36910 & ${\rm 05^h35^m58.47^s}$ & 
${\rm +24^{\circ}44'54\rlap{.}''1}$ & 150 & A8V/F2IVea & 10.3 & 8.2 & 7.4 & 
6.5 & 0.35 & 1.38 \\
T Ori & MWC 763 & ${\rm 05^h35^m50.40^s}$ & 
${\rm -05^{\circ}28'35\rlap{.}''0}$ & 450 & A3/5ea & 10.6 & 8.2 & 7.4 & 
6.5 & 0.20 & 1.58 \\
MWC 120 & HD 37806 & ${\rm 05^h41^m02.29^s}$ & 
${\rm -02^{\circ}43'00\rlap{.}''7}$ & 500 & B9Ve+sh & 7.9 & 7.0 & 6.1 &
5.4 & 0.53 & 3.92 \\
HD 141569 & SAO 140789 & ${\rm 15^h49^m57.75^s}$ & 
${\rm -03^{\circ}55'16\rlap{.}''4}$ & 99 & B9/A0V & 7.1 & 6.8 & 6.8 & 
6.7 & 1.29 & 0.02 \\
HD 158352 &  SAO 122418 & ${\rm 17^h28^m49.65^s}$ & 
${\rm +00^{\circ}19'50\rlap{.}''3}$ & 63 & A8V & 5.4 & 5.1 & 4.9 & 4.9 & 
8.36 & -0.98 \\
MWC 297 & NZ Ser & ${\rm 18^h27^m39.60^s}$ & 
${\rm -03^{\circ}49'52\rlap{.}''0}$ & 400 & O9/B1Ve & 12.3 & 
6.0 & 4.5 & 3.3 & 11.65 & 44.44 \\
VV Ser & HBC 282 & ${\rm 18^h28^m47.90^s}$ & 
${\rm +00^{\circ}08'40\rlap{.}''0}$ & 310 & A0Vevp & 11.9 & 8.6 & 7.5 & 
6.3 & 0.21 & 2.16 \\
V1295 Aql & HD 190073 & ${\rm 20^h03^m02.51^s}$ & 
${\rm +05^{\circ}44'16\rlap{.}''7}$ & 290 & B9/A0Vp+sh & 7.7 & 7.1 & 6.6 & 
5.9 & 0.85 & 2.09 \\
V1685 Cyg & BD+40$^{\circ}$4124 & ${\rm 20^h20^m28.25^s}$ & 
${\rm +41^{\circ}21'51\rlap{.}''6}$ & 1000 & B2Ve & 10.7 & 7.9 & 6.8 & 
5.9 & 0.49 & 2.91 \\
AS 442 & V1977 Cyg & ${\rm 20^h47^m37.47^s}$ & 
${\rm +43^{\circ}47'24\rlap{.}''9}$ & 600 & B8Ve & 11.0 & 8.2 & 7.1 & 6.5 &
0.30 & 1.73 \\
MWC 1080 & V628 Cas & ${\rm 23^h17^m26.10^s}$ & 
${\rm +60^{\circ}50'43\rlap{.}''0}$ & 1000 & B0eq & 11.7 & 7.4 & 6.0 &
4.8 & 0.87 & 9.85 
\enddata
\tablerefs{Distances, spectral types and V magnitudes from 
Hillenbrand et al. (1992), 
Th\'{e} et al. (1994), Mora et al. (2001), Strom et al. (1972),
de Lara et al. (1991), Warren \& Hesser (1978), and Bigay \& Garner (1970). 
JHK magnitudes from the present work (\S \ref{sec:ao}).
$\dag$: De-reddened fluxes at K-band.}
\end{deluxetable}

\clearpage
\begin{deluxetable}{lccccc}
\tabletypesize{\footnotesize}
\tablewidth{0pt}
\tablecaption{Summary of Observations \label{tab:obs}}
\tablehead{\colhead{Source} & \colhead{Date (MJD)} &
\colhead{Baseline}  & \colhead{HA$^{\dag}$}
& \colhead{Calibrators (HD)}}
\startdata
AB Aur & 52575 & NW & [1.21,1.85] & 29645 \\
 & 52601 & NW & [-1.80,1.24] & 29645,32301 \\
  & 52602 & NW & [-1.95,1.51] & 29645,32301 \\
  & 52925 & NW & [-2.10,1.23] & 29645,32301 \\
  & 52926 & SW & [-2.86,0.83] & 29645,32301 \\
MWC 480 & 52575 & NW & [-1.61,0.34] & 29645 \\
 & 52601 & NW & [-1.05,0.78] & 29645,32301 \\
  & 52602 & NW & [-0.65,0.63] & 29645,32301 \\
  & 52925 & NW & [-1.91,1.41] & 29645,32301 \\
  & 52926 & SW & [-3.16,1.66] & 29645,32301 \\
MWC 758 & 52575 & NW & [-0.54,0.34] & 29645 \\
 & 52601 & NW & [-2.07,1.37] & 29645,32301 \\
  & 52602 & NW & [-1.61,-1.61] & 29645,32301 \\
  & 52925 & NW & [-1.39,1.09] & 29645,32301 \\
  & 52926 & SW & [-3.28,0.51] & 29645,32301 \\
  & 52977 & NW & [-1.40,-1.40] & 29645,32301 \\
  & 53018 & NS & [-0.52,-0.52] & 29645,32301 \\
  & 53019 & NS & [-1.94,1.11] & 29645,32301 \\
CQ Tau & 52926 & SW & [-1.19,0.66] & 29645,32301 \\
& 52934 & SW & [-1.87,-0.5] & 29645,32301 \\
  & 52977 & NW & [-1.69,0.83] & 29645,32301 \\
  & 53017 & NW & [-2.13,-1.32] & 29645,32301 \\
  & 53018 & NS & [-0.34,-0.34] & 29645,32301 \\
  & 53019 & NS & [-1.79,1.16] & 29645,32301 \\
T Ori & 52977 & NW & [-0.20,0.09] & 33608,38858 \\
 & 53017 & NW & [-1.06,-0.82] & 33608,38858 \\
MWC 120 & 52977 & NW & [0.21,0.21] & 33608,38858 \\
 & 53017 & NW & [-0.70,-0.21] & 33608,38858 \\
HD 141569 & 52780 & NW & [-0.31,0.46] & 139137, 147449 \\
  & 52781 & NW & [-1.01,-1.01] & 139137, 147449 \\
  & 52787 & NW & [-1.22,0.01] & 139137, 147449 \\
HD 158352 & 52775 & NS & [-2.07,-1.38] & 164259,161868 \\
  & 52781 & NW & [0.33,0.33] & 164259,161868 \\
  & 52788 & NW & [-1.37,0.27] & 164259,161868 \\
MWC 297 & 52798 & NW & [-1.91,-0.86] & 164259,171834 \\
  & 52826 & NW & [-1.17,-0.03] & 164259,171834 \\
VV Ser & 52490 & NW & [-0.72,0.54] & 171834 \\
  & 52491 & NS & [-1.55,-0.74] & 171834 \\
  & 52493 & NW & [-1.31,0.20] & 171834 \\
  & 52499 & NW & [-0.96,0.84] & 164259,171834 \\
  & 52869 & SW & [-0.25,-0.25] & 164259,171834 \\
V1295 Aql & 52799 & NW & [-1.28,-0.66] & 187293 \\
& 52827 & NW & [-0.81,-0.42] & 187293,193556 \\
  & 52828 & NW & [-1.55,-0.03] & 187293,193556 \\
  & 52925 & NW & [0.42,0.85] & 187293,193556 \\
  & 52926 & SW & [0.19,0.88] & 187293,193556 \\
V1685 Cyg & 52418 & NW & [-1.10,-1.00] & 192640,192985 \\
& 52475 & NW & [-1.69,1.44] & 192640,192985 \\
  & 52476 & NW & [-1.80,-0.48] & 192640,192985 \\
  & 52490 & NW & [-0.97,1.70] & 192640 \\
  & 52491 & NS & [-1.27,2.38] & 192640 \\
  & 52492 & NW & [-0.90,-0.90] & 192640 \\
  & 52545 & NS & [-1.12,2.48] & 192640,192985 \\
  & 52869 & SW & [-1.24,0.71] & 192640,192985 \\
  & 52878 & SW & [-0.76,1.08] & 192640,192985 \\
  & 52879 & SW & [-0.49,0.23] & 192640,192985 \\
  & 52925 & NW & [1.14,1.97] & 192640,192985 \\
  & 52926 & SW & [0.94,1.19] & 192640 \\
AS 442 & 52475 & NW & [0.21,1.33] & 192640,192985 \\
  & 52476 & NW & [-0.21,-0.21] & 192640,192985 \\
  & 52490 & NW & [-1.11,0.38] & 192640 \\
  & 52491 & NS & [-0.69,2.54] & 192640 \\
  & 52492 & NW & [-1.05,0.00] & 192640 \\
  & 52545 & NS & [-0.87,1.58] & 192640,192985 \\
  & 52878 & SW & [-0.99,0.86] & 192640,192985 \\
MWC 1080 & 52475 & NW & [0.17,0.17] & 219623 \\
  & 52475 & NW & [-1.99,0.52] & 219623 \\
  & 52490 & NW & [-0.14,1.39] & 219623 \\
\enddata
\tablerefs{${\dag}$: Hour angle coverage of the observations.}
\end{deluxetable}

\clearpage
\begin{deluxetable}{lccccccc}
\tabletypesize{\small}
\tablewidth{0pt}
\tablecaption{Properties of Calibrator Sources \label{tab:cals}}
\tablehead{\colhead{Name} & \colhead{$\alpha$ (J2000)}
&  \colhead{$\delta$ (J2000)} & \colhead{Sp.Ty.} & \colhead{$V$}
& \colhead{$K$} & \colhead{Cal. Size (mas)} & \colhead{$\Delta \alpha$
($^{\circ}$)}}
\startdata
HD 29645 & ${\rm 04^h41^m50.26^s}$ & ${\rm +38^{\circ}16'48\rlap{.}''7}$ &
G0V & 6.0 & 4.6 & $0.56 \pm 0.04$ & 
8.2$^{\rm a}$,9.1$^{\rm b}$,16.5$^{\rm c}$,17.7$^{\rm d}$ \\
HD 32301 & ${\rm 05^h03^m05.75^s}$ & ${\rm +21^{\circ}35'23\rlap{.}''9}$ &
A7V & 4.6 & 4.1 & $0.47 \pm 0.11$ & 
9.1$^{\rm a}$,8.3$^{\rm b}$,7.3$^{\rm c}$,8.2$^{\rm d}$ \\
HD 33608 & ${\rm 05^h11^m19.18^s}$ & ${\rm -02^{\circ}29'26\rlap{.}''8}$ &
F5V & 5.9 & 4.8 & $0.47 \pm 0.05$ & 6.5$^{\rm e}$,7.4$^{\rm f}$ \\
HD 38858 & ${\rm 05^h48^m34.94^s}$ & ${\rm -04^{\circ}05'40\rlap{.}''7}$ &
G4V & 6.0 & 4.4 & $0.56 \pm 0.01$ & 3.2$^{\rm e}$,2.3$^{\rm f}$ \\
HD 139137 & ${\rm 15^h36^m33.71^s}$ & ${\rm -00^{\circ}33'41\rlap{.}''5}$ &
G8III & 6.5 & 4.3 & $0.57 \pm 0.08$ & 4.7$^{\rm g}$ \\ 
HD 147449 & ${\rm 16^h22^m04.35^s}$ & ${\rm +01^{\circ}01'44\rlap{.}''5}$ & 
F0V & 4.8 & 4.1 & $0.65 \pm 0.04$ & 9.4$^{\rm g}$ \\
HD 161868 & ${\rm 17^h47^m53.56^s}$ & ${\rm +02^{\circ}42'26\rlap{.}''2}$ & 
A0V & 3.8 & 3.8 & $0.62 \pm 0.12$ & 5.3$^{\rm h}$ \\
HD 164259 & ${\rm 18^h00^m29.01^s}$ & ${\rm -03^{\circ}41'25\rlap{.}''0}$ &
F2IV & 4.6 & 3.7 & $0.77 \pm 0.04$ & 8.9$^{\rm h}$,7.0$^{\rm i}$,7.5$^{\rm j}$ \\
HD 171834 & ${\rm 18^h36^m39.08^s}$ & ${\rm +06^{\circ}40'18\rlap{.}''5}$ &
F3V & 5.4 & 4.5 & $0.54 \pm 0.06$ & 9.1$^{\rm i}$,6.8$^{\rm j}$ \\
HD 187293 & ${\rm 19^h52^m03.44^s}$ & ${\rm +11^{\circ}37'42\rlap{.}''0}$ &
G0V & 6.2 & 4.8 & $0.49 \pm 0.05$ & 6.5$^{\rm k}$ \\
HD 193556 & ${\rm 20^h20^m20.52^s}$ & ${\rm +14^{\circ}34'09\rlap{.}''3}$ &
G8III & 6.2 & 4.0 & $0.82 \pm 0.05$ & 9.8$^{\rm k}$ \\
HD 192640 & ${\rm 20^h14^m32.03^s}$ & ${\rm +36^{\circ}48'22\rlap{.}''7}$ &
A2V & 4.9 & 4.9 & $0.46 \pm 0.01$ & 4.7$^{\rm l}$,9.4$^{\rm m}$ \\ 
HD 192985 & ${\rm 20^h16^m00.62^s}$ & ${\rm +45^{\circ}34'46\rlap{.}''3}$ &
F5V & 5.9 & 4.8 & $0.44 \pm 0.03$ & 4.3$^{\rm l}$,5.9$^{\rm m}$ \\
HD 219623 & ${\rm 23^h16^m42.30^s}$ & ${\rm +53^{\circ}12'48\rlap{.}''5}$ &
F7V & 5.6 & 4.3 & $0.54 \pm 0.02$ & 9.5$^{\rm n}$
\enddata
\tablerefs{$\Delta \alpha$ is the offset on the sky between target and
calibrator.  a: Offset from AB Aur. b: Offset from MWC 480. 
c: Offset from MWC 758. d: Offset from CQ Tau. e: Offset from T Ori.
f: Offset from MWC 120. g: Offset from HD 141569. h: Offset from HD 158352.
i: Offset from MWC 297. j: Offset from VV Ser. k: Offset from V1295 Aql.
l: Offset from V1685 Cyg. m: Offset from AS 442. n: Offset from MWC 1080.}
\end{deluxetable}

\clearpage
\begin{deluxetable}{lcc||ccccccc}
\tablewidth{0pt}
\tablecaption{Uniform Disk Models \label{tab:uniforms}}
\tablecolumns{7}
\tablehead{\colhead{ } & \multicolumn{2}{c}{Face-On Models} & 
\multicolumn{4}{c}{Inclined Models} \\
\colhead{Source} & \colhead{$\chi^2_r$}  & \colhead{$\theta$ (mas)}
& \colhead{$\chi^2_r$} & \colhead{$\theta$ (mas)} & 
\colhead{$\psi$ ($^{\circ}$)} & \colhead{$\phi$ ($^{\circ}$)}}
\startdata
AB Aur & 2.008 & 5.31$^{+0.01}_{-0.01}$ & 2.112 & 5.34$^{+0.06}_{-0.05}$ & 130$^{+50}_{-130}$ & 9$^{+6}_{-9}$ \\
MWC 480 & 5.196 & 4.85$^{+0.01}_{-0.02}$ & 1.543 & 4.99$^{+0.07}_{-0.05}$ & 154$^{+16}_{-13}$ & 26$^{+4}_{-2}$ \\
MWC 758 & 4.695 & 3.69$^{+0.02}_{-0.02}$ & 0.789 & 4.15$^{+0.10}_{-0.10}$ & 128$^{+3}_{-4}$ & 36$^{+2}_{-3}$ \\
CQ Tau & 5.567 & 3.68$^{+0.05}_{-0.05}$ & 0.975 & 4.38$^{+0.18}_{-0.19}$ & 105$^{+5}_{-5}$ & 48$^{+3}_{-5}$ \\
T Ori & 1.001 & 2.71$^{+0.11}_{-0.11}$ &  &  &  &  \\
MWC 120 & 2.736 & 4.94$^{+0.04}_{-0.03}$ &  &  &  &  \\
MWC 297 &  & $>5.02$ &  &  &  &  \\
VV Ser & 6.077 & 3.68$^{+0.03}_{-0.03}$ & 0.816 & 4.49$^{+0.87}_{-0.46}$ & 168$^{+22}_{-12}$ & 43$^{+10}_{-5}$ \\
V1295 Aql & 0.623 & 5.57$^{+0.04}_{-0.04}$ & 0.716 & 5.77$^{+0.62}_{-0.27}$ & 10$^{+170}_{-10}$ & 19$^{+41}_{-19}$ \\
V1685 Cyg & 6.805 & 3.25$^{+0.01}_{-0.02}$ & 3.905 & 3.59$^{+0.07}_{-0.06}$ & 110$^{+3}_{-4}$ & 41$^{+3}_{-2}$ \\
AS 442 & 1.039 & 2.44$^{+0.06}_{-0.06}$ & 0.872 & 2.74$^{+0.26}_{-0.29}$ & 58$^{+59}_{-11}$ & 47$^{+28}_{-33}$ \\
MWC 1080 & 1.251 & 4.09$^{+0.01}_{-0.02}$ & 0.466 & 4.13$^{+0.24}_{-0.05}$ & 55$^{+12}_{-45}$ & 34$^{+23}_{-15}$ \\
\enddata
\tablerefs{Columns 2-3 contain the reduced chi squared and angular size
values for best-fit face-on disk models.  Columns 4-7 list $\chi_r^2$,
angular size, position angle, and inclination for best-fit inclined
disk models. For T Ori, MWC 120, and MWC 297, 
we fit only face-on models, and the
quoted angular size for MWC 297 is a lower limit.}
\end{deluxetable}

\clearpage
\begin{deluxetable}{lcc||ccccccc}
\tablewidth{0pt}
\tablecaption{Gaussian Models \label{tab:gaussians}}
\tablecolumns{7}
\tablehead{\colhead{ } & \multicolumn{2}{c}{Face-On Models} & 
\multicolumn{4}{c}{Inclined Models} \\
\colhead{Source} & \colhead{$\chi^2_r$}  & \colhead{$\theta$ (mas)}
& \colhead{$\chi^2_r$} & \colhead{$\theta$ (mas)} & 
\colhead{$\psi$ ($^{\circ}$)} & \colhead{$\phi$ ($^{\circ}$)}}
\startdata
AB Aur & 4.365 & 3.60$^{+0.01}_{-0.01}$ & 3.279 & 3.69$^{+0.04}_{-0.04}$ & 157$^{+33}_{-157}$ & 16$^{+3}_{-3}$ \\
MWC 480 & 5.860 & 3.21$^{+0.02}_{-0.01}$ & 1.330 & 3.36$^{+0.07}_{-0.05}$ & 149$^{+17}_{-9}$ & 32$^{+4}_{-4}$ \\
MWC 758 & 2.715 & 2.34$^{+0.01}_{-0.01}$ & 0.598 & 2.57$^{+0.08}_{-0.07}$ & 130$^{+6}_{-5}$ & 33$^{+4}_{-4}$ \\
CQ Tau & 4.470 & 2.32$^{+0.03}_{-0.04}$ & 0.871 & 2.75$^{+0.13}_{-0.13}$ & 104$^{+6}_{-6}$ & 48$^{+4}_{-5}$ \\
T Ori & 1.006 & 1.64$^{+0.07}_{-0.07}$ &  &  &  &  \\
MWC 120 & 3.576 & 3.29$^{+0.04}_{-0.03}$ &  &  &  &  \\
MWC 297 &  & $>3.35$ &  &  &  &  \\
VV Ser & 7.963 & 2.33$^{+0.02}_{-0.01}$ & 0.802 & 2.92$^{+0.63}_{-0.29}$ & 173$^{+20}_{-16}$ & 47$^{+9}_{-4}$ \\
V1295 Aql & 0.697 & 3.85$^{+0.04}_{-0.05}$ & 0.666 & 4.29$^{+1.20}_{-0.50}$ & 110$^{+70}_{-110}$ & 50$^{+30}_{-50}$ \\
V1685 Cyg & 6.145 & 2.00$^{+0.02}_{-0.01}$ & 3.895 & 2.21$^{+0.05}_{-0.04}$ & 110$^{+4}_{-4}$ & 41$^{+3}_{-3}$ \\
AS 442 & 1.000 & 1.49$^{+0.04}_{-0.04}$ & 0.868 & 1.66$^{+0.17}_{-0.18}$ & 57$^{+62}_{-12}$ & 47$^{+30}_{-36}$ \\
MWC 1080 & 1.462 & 2.59$^{+0.01}_{-0.01}$ & 0.460 & 2.62$^{+0.16}_{-0.03}$ & 56$^{+10}_{-41}$ & 40$^{+25}_{-17}$ \\
\enddata
\tablerefs{Columns defined as in Table \ref{tab:uniforms}.}
\end{deluxetable}

\clearpage
\begin{deluxetable}{lcc||ccccccc}
\tablewidth{0pt}
\tablecaption{Ring Models \label{tab:rings}}
\tablecolumns{7}
\tablehead{\colhead{ } & \multicolumn{2}{c}{Face-On Models} & 
\multicolumn{4}{c}{Inclined Models} \\
\colhead{Source} & \colhead{$\chi^2_r$}  & \colhead{$\theta$ (mas)}
& \colhead{$\chi^2_r$} & \colhead{$\theta$ (mas)} & 
\colhead{$\psi$ ($^{\circ}$)} & \colhead{$\phi$ ($^{\circ}$)}}
\startdata
AB Aur & 2.263 & 3.23$^{+0.01}_{-0.01}$ & 2.016 & 3.25$^{+0.03}_{-0.02}$ & 105$^{+75}_{-105}$ & 9$^{+5}_{-9}$ \\
MWC 480 & 4.879 & 2.97$^{+0.01}_{-0.01}$ & 1.390 & 3.05$^{+0.03}_{-0.03}$ & 155$^{+17}_{-14}$ & 24$^{+3}_{-2}$ \\
MWC 758 & 6.149 & 2.31$^{+0.01}_{-0.01}$ & 0.938 & 2.62$^{+0.05}_{-0.05}$ & 127$^{+3}_{-4}$ & 37$^{+2}_{-2}$ \\
CQ Tau & 6.254 & 2.30$^{+0.03}_{-0.02}$ & 1.033 & 2.75$^{+0.10}_{-0.11}$ & 106$^{+4}_{-5}$ & 48$^{+3}_{-4}$ \\
T Ori & 1.000 & 1.73$^{+0.06}_{-0.06}$ &  &  &  &  \\
MWC 120 & 2.339 & 3.03$^{+0.02}_{-0.02}$ &  &  &  &  \\
MWC 297 &  & $>2.95$ &  &  &  &  \\
VV Ser & 5.194 & 2.30$^{+0.01}_{-0.01}$ & 0.822 & 2.80$^{+0.50}_{-0.31}$ & 165$^{+13}_{-10}$ & 42$^{+10}_{-7}$ \\
V1295 Aql & 0.647 & 3.37$^{+0.02}_{-0.01}$ & 0.748 & 3.76$^{+0.23}_{-0.42}$ & 16$^{+164}_{-0}$ & 33$^{+23}_{-33}$ \\
V1685 Cyg & 7.198 & 1.96$^{+0.01}_{-0.01}$ & 3.914 & 2.18$^{+0.04}_{-0.04}$ & 110$^{+3}_{-4}$ & 41$^{+3}_{-2}$ \\
AS 442 & 1.062 & 1.55$^{+0.04}_{-0.04}$ & 0.874 & 1.74$^{+0.17}_{-0.17}$ & 59$^{+57}_{-9}$ & 46$^{+28}_{-30}$ \\
MWC 1080 & 0.706 & 2.55$^{+0.01}_{-0.01}$ & 0.344 & 2.57$^{+0.24}_{-0.02}$ & 56$^{+123}_{-56}$ & 28$^{+21}_{-18}$ \\
\enddata
\tablerefs{Columns defined as in Table \ref{tab:uniforms}.}
\end{deluxetable}

\clearpage
\begin{deluxetable}{lcc||ccccccc}
\tabletypesize{\small}
\tablewidth{0pt}
\tablecaption{Geometrically Flat Accretion Disk Models \label{tab:acc}}
\tablecolumns{7}
\tablehead{\colhead{ } & \multicolumn{2}{c}{Face-On Models} & 
\multicolumn{4}{c}{Inclined Models} \\
\colhead{Source} & \colhead{$\chi^2_r$}  & \colhead{$\theta$ (mas)}
& \colhead{$\chi^2_r$} & \colhead{$\theta$ (mas)} & 
\colhead{$\psi$ ($^{\circ}$)} & \colhead{$\phi$ ($^{\circ}$)}}
\startdata
\multicolumn{7}{c}{$T_{\rm in}=2000$ K} \\ \hline
AB Aur & 2.549 & 2.24$^{+0.01}_{-0.01}$ & 2.465 & 2.26$^{+0.03}_{-0.01}$ & 148$^{+32}_{-148}$ & 10$^{+6}_{-4}$ \\
MWC 480 & 5.479 & 2.03$^{+0.01}_{-0.01}$ & 1.333 & 2.10$^{+0.03}_{-0.02}$ & 153$^{+15}_{-12}$ & 28$^{+5}_{-2}$ \\
MWC 758 & 3.395 & 1.51$^{+0.01}_{-0.01}$ & 0.660 & 1.69$^{+0.03}_{-0.05}$ & 129$^{+5}_{-4}$ & 35$^{+3}_{-4}$ \\
CQ Tau & 4.846 & 1.50$^{+0.02}_{-0.02}$ & 0.898 & 1.78$^{+0.08}_{-0.08}$ & 105$^{+5}_{-6}$ & 48$^{+4}_{-5}$ \\
T Ori & 1.007 & 1.08$^{+0.04}_{-0.04}$ &  &  &  &  \\
MWC 120 & 3.057 & 2.09$^{+0.02}_{-0.02}$ &  &  &  &  \\
MWC 297 &  & $>2.12$ &  &  &  &  \\
VV Ser & 7.195 & 1.52$^{+0.01}_{-0.01}$ & 0.804 & 1.86$^{+0.40}_{-0.18}$ & 172$^{+20}_{-15}$ & 45$^{+10}_{-4}$ \\
V1295 Aql & 0.631 & 2.37$^{+0.02}_{-0.01}$ & 0.694 & 2.39$^{+0.30}_{-0.05}$ & 135$^{+45}_{-135}$ & 12$^{+22}_{-12}$ \\
V1685 Cyg & 6.330 & 1.32$^{+0.01}_{-0.01}$ & 3.896 & 1.46$^{+0.03}_{-0.03}$ & 110$^{+4}_{-4}$ & 41$^{+3}_{-2}$ \\
AS 442 & 1.004 & 0.98$^{+0.02}_{-0.03}$ & 0.869 & 1.10$^{+0.11}_{-0.12}$ & 57$^{+62}_{-12}$ & 48$^{+29}_{-35}$ \\
MWC 1080 & 1.372 & 1.69$^{+0.01}_{-0.01}$ & 0.461 & 1.71$^{+0.10}_{-0.02}$ & 56$^{+9}_{-42}$ & 38$^{+24}_{-17}$ \\
\hline \multicolumn{7}{c}{$T_{\rm in}=1500$ K} \\ \hline
AB Aur & 2.071 & 2.49$^{+0.01}_{-0.01}$ & 2.205 & 2.50$^{+0.03}_{-0.01}$ & 136$^{+44}_{-136}$ & 8$^{+7}_{-8}$ \\
MWC 480 & 5.262 & 2.27$^{+0.01}_{-0.01}$ & 1.342 & 2.34$^{+0.03}_{-0.02}$ & 152$^{+18}_{-11}$ & 27$^{+4}_{-3}$ \\
MWC 758 & 4.233 & 1.71$^{+0.01}_{-0.01}$ & 0.738 & 1.93$^{+0.04}_{-0.05}$ & 128$^{+4}_{-4}$ & 36$^{+2}_{-4}$ \\
CQ Tau & 5.309 & 1.71$^{+0.02}_{-0.03}$ & 0.944 & 2.03$^{+0.09}_{-0.08}$ & 105$^{+5}_{-5}$ & 48$^{+4}_{-5}$ \\
T Ori & 1.003 & 1.25$^{+0.05}_{-0.05}$ &  &  &  &  \\
MWC 120 & 2.795 & 2.33$^{+0.01}_{-0.02}$ &  &  &  &  \\
MWC 297 &  & $>2.36$ &  &  &  &  \\
VV Ser & 6.434 & 1.72$^{+0.01}_{-0.01}$ & 0.810 & 2.11$^{+0.42}_{-0.22}$ & 169$^{+22}_{-9}$ & 44$^{+10}_{-5}$ \\
V1295 Aql & 0.623 & 2.62$^{+0.02}_{-0.01}$ & 0.712 & 2.67$^{+0.32}_{-0.08}$ & 3$^{+177}_{-3}$ & 14$^{+25}_{-14}$ \\
V1685 Cyg & 6.631 & 1.51$^{+0.01}_{-0.01}$ & 3.903 & 1.67$^{+0.04}_{-0.02}$ & 110$^{+3}_{-4}$ & 41$^{+3}_{-2}$ \\
AS 442 & 1.024 & 1.13$^{+0.03}_{-0.03}$ & 0.871 & 1.26$^{+0.13}_{-0.13}$ & 58$^{+60}_{-8}$ & 46$^{+18}_{-33}$ \\
MWC 1080 & 1.300 & 1.92$^{+0.01}_{-0.01}$ & 0.471 & 1.94$^{+0.0.}_{-0.02}$ & 54$^{+13}_{-43}$ & 35$^{+19}_{-16}$ \\
\enddata
\tablerefs{Columns defined as in Table \ref{tab:uniforms}.}
\end{deluxetable}

\clearpage
\begin{deluxetable}{lcc||ccccccc}
\tablewidth{0pt}
\tablecaption{Flared Disk Models with Puffed-Up Inner Walls \label{tab:ddn}}
\tablecolumns{7}
\tablehead{\colhead{ } & \multicolumn{2}{c}{Face-On Models} & 
\multicolumn{4}{c}{Inclined Models} \\
\colhead{Source} & \colhead{$\chi^2_r$}  & \colhead{$\theta$ (mas)}
& \colhead{$\chi^2_r$} & \colhead{$\theta$ (mas)} & 
\colhead{$\psi$ ($^{\circ}$)} & \colhead{$\phi$ ($^{\circ}$)}}
\startdata
\multicolumn{7}{c}{$T_{\rm in}=2000$ K} \\ \hline
AB Aur & 1.994 & 3.35$^{+0.01}_{-0.01}$ & 2.163 & 3.37$^{+0.05}_{-0.03}$ & 130$^{+50}_{-130}$ & 8$^{+7}_{-8}$ \\
MWC 480 & 4.919 & 3.22$^{+0.01}_{-0.01}$ & 1.387 & 3.31$^{+0.03}_{-0.03}$ & 156$^{+16}_{-15}$ & 24$^{+4}_{-2}$ \\
MWC 758 & 5.858 & 2.49$^{+0.01}_{-0.01}$ & 0.921 & 2.85$^{+0.06}_{-0.05}$ & 127$^{+3}_{-4}$ & 37$^{+2}_{-2}$ \\
CQ Tau & 6.254 & 2.51$^{+0.04}_{-0.03}$ & 1.034 & 3.01$^{+0.10}_{-0.13}$ & 106$^{+4}_{-5}$ & 48$^{+3}_{-4}$ \\
T Ori & 1.006 & 1.78$^{+0.09}_{-0.10}$ &  &  &  &  \\
MWC 120 & 2.340 & 3.31$^{+0.02}_{-0.02}$ &  &  &  &  \\
VV Ser & 5.287 & 2.50$^{+0.01}_{-0.02}$ & 0.820 & 3.05$^{+0.31}_{-0.21}$ & 165$^{+15}_{-5}$ & 42$^{+6}_{-2}$ \\
V1295 Aql & 0.646 & 3.67$^{+0.02}_{-0.02}$ & 0.765 & 3.84$^{+0.47}_{-0.20}$ & 14$^{+166}_{-14}$ & 21$^{+19}_{-21}$ \\
AS 442 & 0.990 & 1.57$^{+0.06}_{-0.07}$ & 0.871 & 1.83$^{+0.19}_{-0.27}$ & 57$^{+59}_{-11}$ & 48$^{+24}_{-38}$ \\
\hline \multicolumn{7}{c}{$T_{\rm in}=1500$ K} \\ \hline
AB Aur & 4.320 & 3.13$^{+0.01}_{-0.01}$ & 3.404 & 3.15$^{+0.06}_{-0.01}$ & 155$^{+25}_{-155}$ & 16$^{+3}_{-3}$ \\
MWC 480 & 5.052 & 3.16$^{+0.01}_{-0.02}$ & 1.474 & 3.27$^{+0.03}_{-0.02}$ & 145$^{+9}_{-6}$ & 28$^{+2}_{-1}$ \\
MWC 758 & 5.033 & 2.42$^{+0.01}_{-0.02}$ & 0.867 & 2.78$^{+0.08}_{-0.06}$ & 127$^{+4}_{-3}$ & 36$^{+3}_{-2}$ \\
CQ Tau & 6.211 & 2.50$^{+0.04}_{-0.03}$ & 1.032 & 3.00$^{+0.11}_{-0.12}$ & 106$^{+4}_{-5}$ & 48$^{+3}_{-4}$ \\
T Ori & 2.246 & 1.41$^{+0.23}_{-0.07}$ &  &  &  &  \\
MWC 120 & 2.348 & 3.30$^{+0.02}_{-0.02}$ &  &  &  &  \\
VV Ser & 5.585 & 2.47$^{+0.01}_{-0.02}$ & 0.818 & 3.02$^{+0.32}_{-0.24}$ & 166$^{+17}_{-6}$ & 42$^{+6}_{-2}$ \\
V1295 Aql & 0.643 & 3.65$^{+0.02}_{-0.02}$ & 0.751 & 3.85$^{+0.38}_{-0.24}$ & 14$^{+166}_{-14}$ & 23$^{+15}_{-23}$ \\
AS 442 & 1.628 & 1.36$^{+0.02}_{-0.05}$ & 0.863 & 1.49$^{+0.32}_{-0.31}$ & 53$^{+58}_{-11}$ & 58$^{+16}_{-40}$ \\
\enddata
\tablerefs{Columns defined as in Table \ref{tab:uniforms}. MWC 297,
V1685 Cyg, and MWC 1080 are excluded from this table because the
puffed-up inner disk wall model cannot fit the data for these sources.}
\end{deluxetable}

\clearpage
\begin{deluxetable}{lccccccc}
\tablewidth{0pt}
\tablecaption{Binary Models \label{tab:binary}}
\tablecolumns{7}
\tablehead{\colhead{Source} & \colhead{$\chi^2_r$}  & \colhead{$\theta$ (mas)}
& \colhead{$\psi$ ($^{\circ}$)} & \colhead{$R$}}
\startdata
AB Aur & 117.08 & 4.06$^{+0.01}_{-0.01}$ & 28$^{+1}_{-1}$ & 0.71$^{+0.01}_{-0.01}$ \\
MWC 480 & 13.46 & 3.24$^{+0.03}_{-0.03}$ & 127$^{+1}_{-1}$ & 0.43$^{+0.01}_{-0.01}$ \\
MWC 758 & 14.50 & 2.69$^{+0.02}_{-0.03}$ & 26$^{+1}_{-1}$ & 0.48$^{+0.01}_{-0.02}$ \\
CQ Tau & 4.408 & 2.95$^{+0.05}_{-0.06}$ & 32$^{+1}_{-1}$ & 0.48$^{+0.03}_{-0.04}$ \\
VV Ser & 0.750 & 9.27$^{+3.44}_{-2.74}$ & 175$^{+1}_{-172}$ & 0.55$^{+0.16}_{-0.03}$ \\
V1295 Aql & 0.692 & 3.09$^{+2.52}_{-0.20}$ & 109$^{+25}_{-102}$ & 0.43$^{+0.33}_{-0.02}$ \\
V1685 Cyg & 14.90 & 3.61$^{+0.04}_{-0.05}$ & 135$^{+1}_{-1}$ & 0.28$^{+0.01}_{-0.01}$ \\
AS 442 & 0.864 & 2.69$^{+0.21}_{-0.69}$ & 30$^{+19}_{-8}$ & 0.20$^{+0.07}_{-0.01}$ \\
MWC 1080 & 0.774 & 2.55$^{+0.11}_{-0.31}$ & 63$^{+5}_{-5}$ & 0.38$^{+0.05}_{-0.01}$ \\
\enddata
\tablerefs{Columns 2-5 list the reduced chi squared values, angular 
separations, position angles, and brightness ratios of best-fit binary models.
T Ori, MWC 120, and MWC 297 are excluded from this table because the
limited $u-v$ coverage for these sources is insufficient to constrain the 
parameters of the model.}
\end{deluxetable}

\clearpage
\begin{deluxetable}{lcc||ccccccc}
\tablewidth{0pt}
\tablecaption{Accretion Disk Models for PTI+IOTA Visibilities 
\label{tab:pti+iota}}
\tablecolumns{7}
\tablehead{\colhead{ } & \multicolumn{2}{c}{Face-On Models} & 
\multicolumn{4}{c}{Inclined Models} \\
\colhead{Source} & \colhead{$\chi^2_r$}  & \colhead{$\theta$ (mas)}
& \colhead{$\chi^2_r$} & \colhead{$\theta$ (mas)} & 
\colhead{$\psi$ ($^{\circ}$)} & \colhead{$\phi$ ($^{\circ}$)}} 
\startdata
\multicolumn{7}{c}{$T_{\rm in}=2000$ K} \\ \hline
AB Aur & 2.495 & 2.26$^{+0.01}_{-0.01}$ & 2.889 & 2.27$^{+0.02}_{-0.01}$ & 170$^{+0}_{-170}$ & 11$^{+4}_{-3}$ \\
T Ori & 0.834 & 1.08$^{+0.04}_{-0.05}$ &  &  &  &  \\
MWC 297 & 0.040 & 3.38$^{+0.24}_{-0.22}$ &  &  &  &  \\
V1295 Aql & 0.604 & 2.38$^{+0.02}_{-0.02}$ & 0.343 & 2.68$^{+0.11}_{-0.33}$ & 15$^{+165}_{-15}$ & 35$^{+4}_{-35}$ \\
V1685 Cyg & 6.141 & 1.32$^{+0.01}_{-0.01}$ & 2.021 & 1.46$^{+0.03}_{-0.02}$ & 111$^{+3}_{-4}$ & 41$^{+3}_{-3}$ \\
MWC 1080 & 9.355 & 1.69$^{+0.01}_{-0.01}$ & 6.408 & 1.76$^{+0.03}_{-0.03}$ & 43$^{+6}_{-4}$ & 37$^{+5}_{-5}$ \\
\enddata
\tablerefs{Columns defined as in Table \ref{tab:uniforms}.}
\end{deluxetable}

\clearpage
\begin{deluxetable}{lcc||cccc}
\tablewidth{0pt}
\tablecaption{Disk Parameters from Near-IR Interferometry and SEDs 
\label{tab:seds}}
\tablecolumns{7}
\tablehead{\colhead{ } & \multicolumn{2}{c}{Flat Disks} & 
\multicolumn{4}{c}{Flared, Puffed-Up Inner Disks} \\
\colhead{Source} & \colhead{$R_{\rm in}$} & 
\colhead{$T_{\rm in}$} &  
\colhead{$R_{\rm in}$} & \colhead{$T_{\rm in}$} & \colhead{$\xi$}
& \colhead{$R_{\rm out}$} \\
& \colhead{(AU)} & \colhead{(K)} & \colhead{(AU)} & \colhead{(K)} & & 
\colhead{(AU)}}
\startdata
AB Aur & 0.17 & 1690 & 0.25 & 2230 & 0.12 & 360 \\
MWC 480 & 0.16 & 1370 & 0.23 & 1580 & 0.16 & 30 \\
MWC 758 & 0.14 & 1450 & 0.21 & 1700 & 0.24 & 70 \\
CQ Tau & 0.15 & 1430 & 0.23 & 1470 & 0.28 & 100 \\
T Ori & 0.28 & 1570 & 0.40 & 1950 & 0.10 & 30 \\
MWC 120 & 0.58 & 1370 & 0.83 & 1690 & 0.14 & 30 \\
MWC 297 & 0.68 & 2140 \\
VV Ser & 0.33 & 1450 & 0.47 & 1630 & 0.20 & 400 \\
V1295 Aql & 0.39 & 1240 & 0.55 & 1390 & 0.12 & 30 \\
V1685 Cyg & 0.73 & 1790 \\
AS 442 & 0.33 & 1910 & 0.55 & 1910 & 0.28 & 100 \\
MWC 1080 & 0.86 & 2170 \\
\enddata
\tablerefs{Columns 2-3 list the inner disk radii and temperatures for
geometrically flat disk models (\S \ref{sec:acc}) determined from
near-IR interferometry and SEDs. Inner radii are computed from inner disk 
angular sizes (Table \ref{tab:acc}, and Table \ref{tab:pti+iota} for MWC 297), 
using the distances assumed in Table \ref{tab:sources}.
Columns 4-7 list the inner radii,
inner disk temperatures, flaring indexes, and outer radii determined
for flared disk models with puffed-up inner walls (\S \ref{sec:ddn}). 
The inner radii are computed from angular sizes in Table \ref{tab:ddn} and 
distances listed in Table \ref{tab:sources}. Puffed-up inner disk
wall models are ruled out for MWC 297, V1685 Cyg, and MWC 1080, 
and thus no parameter values are listed.}
\end{deluxetable}

\end{document}